\documentclass[final,3p,times]{elsarticle}

\usepackage{graphicx}

\usepackage{amssymb}
\usepackage{amsthm}
\usepackage{cancel}

\usepackage{hyperref}
\usepackage{verbatim}

 \biboptions{square,sort&compress,comma}

\journal{XXX}

\usepackage{mathrsfs}
\usepackage{amssymb}
\usepackage{amsmath}
\usepackage{bm}
\usepackage{graphicx}
\usepackage{color}

\begin{document}

\begin{frontmatter}



\title{A Unified Numerically Solvable Framework for Complicated Kinetic Plasma Dispersion Relations}

\date{\today}


\author[label1,label2]{Hua-sheng XIE}
\ead{huashengxie@gmail.com, xiehuasheng@enn.cn}

\address[label1]{Hebei Key Laboratory of Compact Fusion, Langfang 065001, China}
\address[label2]{ENN Science and Technology Development Co., Ltd., Langfang 065001, China}

\begin{abstract}
A unified numerically solvable framework for dispersion relations with arbitrary number of species drifting at arbitrary directions and with Krook collision is derived for linear uniform/homogenous kinetic plasma, which largely extended the standard one [say, T. Stix, {\em Waves in Plasmas}, AIP Press, 1992]. The purpose of this work is to provide a kinetic plasma dispersion relation tool not only the physical model but also the numerical approach be as general/powerful as possible. As a very general application example, we give the final dispersion relations which assume further the equilibrium distribution function be bi-Maxwellian and including parallel drift, two directions of perpendicular drift (i.e., drift across magnetic field), ring beam and loss-cone.  Both electromagnetic and electrostatic versions are provided, with also the Darwin (a.k.a., magnetoinductive or magnetostatic) version. The species can be treated either magnetized or unmagnetized. Later, the equations are transformed to the matrix form be solvable by using the powerful matrix algorithm [H. S. Xie and Y. Xiao, Plasma Science and Technology, 18, 2, 97, 2016], which is the first approach can give all the important solutions of a linear kinetic plasma system without requiring initial guess for root finding and thus can be extremely useful to the community. To our knowledge, the present model would be the most comprehensive one in literature for the distribution function constructed bases on Maxwellian, which thus can be applied widely for study waves and instabilities in space, astrophysics, fusion and laser plasma. We limit the present work to non-relativistic case.
\end{abstract}


\end{frontmatter}


\section{Introduction}\label{sec:intro}

Due to the complicated evolution of charged particles and electromagnetic field, one of the most important feature of plasma is the numerous waves and instabilities. 
The fundamental features of linear waves and instabilities in uniform/homogenous plasma can be described by dispersion relation and are discussed by many authors in monographes \cite{Stix1992,Gary1993,Swanson2003}  and textbooks (cf. \cite{Gurnett2005}). In standard treatment of kinetic plasma dispersion relation, the velocity space equilibrium distribution function is assumed to be $f_{s0}=f_{s0}(v_\parallel,v_\perp)$, thus can not treat the cases with drift across magnetic field. To including the arbitrary directions of drift and collision would largely extend the application range of the dispersion relation, with the instabilities in the non-uniform shock \cite{Muschietti2017, Umeda2018} be one of numerous of them.

In this work, we try to provide a most comprehensive linear kinetic plasma dispersion relation tool to the community, which includes lots of new capabilities in both the physical model and algorithm. The tool is largely benefit from the powerful matrix approach of PDRK solver \cite{Xie2016}, which is the first algorithm can yield all the important kinetic solutions without initial guess for root finding. The new tool is named as PASS (Plasma wAve and inStability analySis), which includes the present work be kinetic version PASS-K succeed from PDRK \cite{Xie2016} (hereafter 'PASS-K' is refered equivalent to 'PDRK'), the fluid version PASS-F succeed from PDRF \cite{Xie2014}, and possibly more.

In the following sections, we firstly derive the most general kinetic dispersion relation equation with drift across the magnetic field and Krook collision in section \ref{sec:general_dr}. In section \ref{sec:maxwell_dr}, we assume a very general extended Maxwellian equilibrium distribution function for all species and derive the corresponding final dispersion relation. In section \ref{sec:passk_dr}, the corresponding equations suitable be solved by PASS-K matrix algorithm are derived. In sections \ref{sec:benchmark}, we give some benchmark results. In section \ref{sec:discussion}, we give a summary and some discussions.

\section{The General Non-relativistic Dispersion Relation}\label{sec:general_dr}

Considering that the widely used magnetized kinetic dispersion relation in literature such as in Ref.\cite{Stix1992} does not including the drift across magnetic field, we firstly derive our new models. A similar magnetized model is derived only recently in Ref.\cite{Umeda2018}, but which is still not as general as the present one. We limit our study to non-relativistic model. To help the reader, we will give detailed steps of our derivations.

\subsection{Starting equations}
We consider only the collisionless case or with a Krook collision, the kinetic equation for each species is the Vlasov equation with Krook collision at the right hand side
 \begin{equation}
\frac{\partial F_s}{\partial t}+\bm v\cdot\frac{\partial F_s}{\partial \bm r}+\Big[\bm a_{s}+\frac{q_s}{m_s}(\bm E+\bm v\times\bm B)\Big]\cdot\frac{\partial F_s}{\partial \bm v}=-\nu_s(F_s-F_{s0}),
\end{equation}
where the distribution function $F_s(\bm r, \bm v, t)=n_{s0}f_s(\bm r, \bm v, t)$, and $q_s$, $m_s$ and $n_{s0}$ are the charge, mass and the number density of species $s$, respectively. When the collision frequency $\nu_s=0$, the equation reduces to a collisionless case. The Maxwell equations for fields can be either 
\begin{eqnarray}
  & \partial_t {\bm E} = c^2\nabla\times{\bm B}-{\bm J}/\epsilon_0,\\
  & \partial_t {\bm B} = -\nabla\times{\bm E},
\end{eqnarray}
for electromagnetic case, or
\begin{eqnarray}
  & \nabla \cdot \bm E=\rho/\epsilon_0,\\
  & \bm E = -\nabla\Phi,
\end{eqnarray}
for electrostatic case, where
\begin{eqnarray}
   \bm J&=&\sum \bm J_s=\sum_s q_s n_{s0}\int dv^3 \bm vf_s,\\
   \rho&=&\sum \rho_s= \sum_s q_sn_{s0}\int dv^3 f_s,
\end{eqnarray}
$d\bm v=dv^3=dv_xdv_ydv_z$ and $c=1/\sqrt{\epsilon_0\mu_0}$ is the speed of light. The accelerate term $\bm a_s$ can be caused by other external forces such as the gravity and other (magneto-)hydro-dynamic forces due to spatial inhomogeneity, which then would cause drift motions across the magnetic field. A typical example is the low hybrid drift instability (LHDI) in space (e.g., in current sheet and shock cases) and fusion \cite{Steinhauer2011}  plasma (e.g., in mirror and field-reversed configuration).

To extend the application range, we will also given the Darwin model \cite{Xie2014} version. In Darwin mode, all field variables can be divided into two parts: the transverse (T, divergence free) part and the longitudinal (L, curl free) part, i.e.,
\begin{eqnarray}
  &\bm E = \bm E_L +\bm E_T,~~ \nabla\cdot\bm E_T=0,~~\nabla\times\bm E_L=0,\\
  & \bm B = \bm B_T,~~\nabla\cdot\bm B_T=0,\\
  &\bm J = \bm J_L +\bm J_T,~~ \nabla\cdot\bm J_T=0,~~\nabla\times\bm J_L=0.
\end{eqnarray}
The corresponding field equations are
\begin{eqnarray}
  & \nabla \cdot \bm E_L=\rho/\epsilon_0,\\
  & \nabla \cdot \bm B=0,\\
  & \partial_t {\bm E_L} +\underbrace{ \cancel{\partial_t {\bm E_T}}}_{\rm droped~in~Darwin~model} = c^2\nabla\times{\bm B}-{\bm J}/\epsilon_0,\\
  & \partial_t {\bm B} = -\nabla\times{\bm E}~~({\rm or}, \partial_t {\bm B} = -\nabla\times{\bm E_L}),
\end{eqnarray}
where for our usage we need only the last two of them. The only difference from full electromagnetic model is that the term $\partial_t {\bm E_T}$ is dropped in the Darwin model. The Darwin model which eliminates the high frequency electromagnetic wave $\omega^2\sim k^2c^2$, is particular interesting in theoretical study and kinetic particle-in-cell and Vlasov simulations, i.e., which can use large time steps in simulation and saves the computation resource a lot \cite{Busnardo-Neto1977}.

We assume the zero-order term be a homogenous system with $f_s(\bm v)=f_{s0}(\bm v)+f_{s1}(\bm v)$, $\int dv^3 f_{s0}=1$, $\bm a_s=\bm a_{s0}$, $\bm E=\bm E_0+\bm E_1$ and $\bm B=\bm B_0+\bm B_1$. The conventional derivation is further assume $\bm a_{s0}=0$ and $\bm E_0=0$, whereas we treat also $\bm a_{s0}\neq0$ and $\bm E_0\neq0$ in this work.

Without loss of generality, we assume the background magnetic field in $z$ direction, i.e., $\bm B_0=B_0{\bm \hat{z}}=(0,0,B_0)$. The zero-order equations
\begin{eqnarray}\label{eq:eqfs0}
   && \Big[\bm a_{s0}+\frac{q_s}{m_s}(\bm E_0+\bm v\times\bm B_0)\Big]\cdot\frac{\partial f_{s0}}{\partial \bm v}=0,\\
   && \bm J_0= \sum_s q_sn_{s0}\bm v_{ds}=0,\\
   && \rho_0= \sum_s q_sn_{s0}=0,
\end{eqnarray}
where $\bm v_{ds}=\int dv^3 \bm vf_{s0}$. We introduce a cylindrical velocity coordinates\footnote{When $\bm v_{ds}\neq0$, we have assumed the system to be Galilean invariant. It is not sure how much influence to the result yet. This issue is subtle, because the Vlasov equation is Galilean invariant and the electromagnetic field equations are Lorentz invariant and the total system is neither Galilean nor Lorentz invariant.} $(v'_\perp,\phi',v'_\parallel)$ with $v'_\perp=\sqrt{(v_x-v_{dsx})^2+(v_y-v_{dsy})^2}$ and $v'_\parallel=v_z$, where $v_{dsx}=(\frac{m_s}{q_s}a_{s0y}+E_{0y})/B_0$ and $v_{dsy}=-(\frac{m_s}{q_s}a_{s0x}+E_{0x})/B_0$. We have $v_x=v'_\perp\cos\phi'+v_{dsx}$, $v_y=v'_\perp\sin\phi'+v_{dsy}$ and $v_z=v'_\parallel$. We have also $\frac{\partial v_x}{\partial\phi'}=-v'_\perp\sin\phi'$, $\frac{\partial v_y}{\partial\phi'}=v'_\perp\cos\phi'$, {\color{red}$\frac{\partial v'_\perp}{\partial v_x}=\cos\phi'$ and $\frac{\partial v'_\perp}{\partial v_y}=\sin\phi'$}.
Eq.(\ref{eq:eqfs0}) is
\begin{eqnarray}\label{eq:eqfs0_1}\nonumber
   0&=& \Big[\bm a_{s0}+\frac{q_s}{m_s}(\bm E_0+\bm v\times\bm B_0)\Big]\cdot\frac{\partial f_{s0}}{\partial \bm v}\\\nonumber
   &=&\frac{q_s}{m_s}\Big[\Big(\frac{m_s}{q_s}a_{s0x}+E_{0x}+v_yB_0\Big)\frac{\partial f_{s0}}{\partial v_x}+\Big(\frac{m_s}{q_s}a_{s0y}+E_{0y}-v_xB_0\Big)\frac{\partial f_{s0}}{\partial v_y}+\Big(\frac{m_s}{q_s}a_{s0z}+E_{0z}\Big)\frac{\partial f_{s0}}{\partial v_z}\Big]\\\nonumber
   &=&\frac{q_sB_0}{m_s}\Big[(-v_{dsy}+v_y)\frac{\partial f_{s0}}{\partial v_x}+(v_{dsx}-v_x)\frac{\partial f_{s0}}{\partial v_y}\Big]=\frac{q_sB_0}{m_s}\Big(v'_\perp\sin\phi'\frac{\partial f_{s0}}{\partial v_x}-v'_\perp\cos\phi'\frac{\partial f_{s0}}{\partial v_y}\Big)\\
   &=&  -\frac{q_sB_0}{m_s}\Big(\frac{\partial v_x}{\partial\phi'}\frac{\partial f_{s0}}{\partial v_x}+\frac{\partial v_y}{\partial\phi'}\frac{\partial f_{s0}}{\partial v_y}\Big)= -\frac{q_sB_0}{m_s}\frac{\partial f_{s0}}{\partial \phi'}.
\end{eqnarray}
Thus we have $\frac{\partial f_{s0}}{\partial \phi'}=0$ if $B_0\neq0$, i.e., $f_{s0}=f_{s0}(v'_\perp,v'_\parallel)$. And, we have assume the force balance in the z-direction, i.e., $\frac{m_s}{q_s}a_{s0z}+E_{0z}=0$. 

The first order kinetic equation is (we have dropped the '1' subscript)
\begin{equation}\label{eq:eqfs1_0}
   \frac{\partial f_s}{\partial t}+\bm v\cdot\frac{\partial f_s}{\partial \bm r}+\frac{q_s}{m_s}(\bm v\times\bm B_0)\cdot\frac{\partial f_s}{\partial \bm v}+\frac{q_s}{m_s}(\bm E+\bm v\times\bm B)\cdot\frac{\partial f_{s0}}{\partial \bm v}=-\nu_sf_s,
\end{equation}
and Fourier transform the equation using $\partial/\partial\bm r\to i\bm k$, $\partial/\partial t\to -i\omega$ and also $\partial_t {\bm B} = -\nabla\times{\bm E}$, we obtain
\begin{equation}\label{eq:eqfs1}
  i(\omega -\bm k\cdot\bm v +i\nu_s)f_s+\omega_{cs}\frac{\partial f_s}{\partial \phi'}=\frac{q_s}{m_s}\Big[\bm E+\bm v\times\Big(\frac{\bm k\times\bm E}{\omega}\Big)\Big]\cdot\frac{\partial f_{s0}}{\partial \bm v},
\end{equation}
where $\omega_{cs}=\frac{q_sB_0}{m_s}$ is the cyclone frequency. Without loss of generality, we can assume\footnote{Ref.\cite{Umeda2018} assumed the across magnetic field drift $\bm v_{ds}$ at $x$ direction but $\bm k=(k_x,k_y,k_z)$, which will limit that all species can only drift at $x-z$ direction. Later, they limit their discussion also to only $k_y=0$. And thus their result is our result at $v_{dsy}=0$ and $\nu_s=0$. } the wave vector $\bm k=(k_x,0,k_z)=(k\sin\theta,0,k\cos\theta)$, which gives $k_\perp=k_x$ and $k_\parallel=k_z$. We study four cases:
\begin{enumerate}
\item Electromagnetic or Darwin, magnetized: $B_0\neq0$, $\bm B_1\neq0$.
\item Electromagnetic or Darwin, unmagnetized: $B_0=0$ ($\omega_{cs}=0$), $\bm B_1\neq0$.
\item Electrostatic, magnetized: $B_0\neq0$, $\bm B_1=0$ ($\bm k\times\bm E_1=0$).
\item Electrostatic, unmagnetized: $B_0=0$  ($\omega_{cs}=0$), $\bm B_1=0$ ($\bm k\times\bm E_1=0$).
\end{enumerate}

For convenient to theoretical study, we let the user to choose whether a species is magnetized or unmagnetized\footnote{A typical case is in shock study, say Ref.\cite{Muschietti2017}, where electron are magnetized and drift ions are treated unmagnetized.}, i.e., say for a electromagnetic case, the different species can be either magnetized (labeled as 'm') or unmagnetized (labeled as 'u').

For unmagnetized case, Eq.(\ref{eq:eqfs1}) reduces to
\begin{equation}\label{eq:eqfsu1}
  i(\omega -\bm k\cdot\bm v +i\nu_s)f_s^u=\frac{q_s}{m_s}\Big[\bm E+\bm v\times\Big(\frac{\bm k\times\bm E}{\omega}\Big)\Big]\cdot\frac{\partial f_{s0}}{\partial \bm v},
\end{equation}
where $\bm k\cdot\bm v=k_xv_x+k_zv_z$ or $\bm k\cdot\bm v=k_zv'_\parallel+k_xv'_\perp\cos\phi'+{\color{red}k_xv_{dsx}}$, which gives the solution for $f_s^u$ is
\begin{equation}\label{eq:fsu1}
 f_s^u=\frac{\frac{q_s}{m_s}\Big[\bm E+\bm v\times\Big(\frac{\bm k\times\bm E}{\omega}\Big)\Big]\cdot\frac{\partial f_{s0}}{\partial \bm v}}{i(\omega-\bm k\cdot\bm v +i\nu_s)}.
\end{equation}

For magnetized case, Eq.(\ref{eq:eqfs1}) reduces to
\begin{equation}\label{eq:eqfsm1}
   \frac{\partial f_s}{\partial \phi'}-i(x_s+y_s\cos\phi')f_s=\frac{q_s}{m_s\omega_{cs}}\Big[\bm E+\bm v\times\Big(\frac{\bm k\times\bm E}{\omega}\Big)\Big]\cdot\frac{\partial f_{s0}}{\partial \bm v},
\end{equation}
where $x_s=-\frac{\omega-k_\parallel v'_\parallel {-\color{red}k_xv_{dsx}+i\nu_s}}{\omega_{cs}}$ and $y_s=\frac{k_\perp v'_\perp}{\omega_{cs}}$.
The solution for magnetized $f_s^m$ is much complicated. Instead of the approaches using the method of characteristics in Refs.\cite{Stix1992} and \cite{Umeda2018}, we using the approach similar to Ref.\cite{Gurnett2005}, which solves the differential equation directly.
That is to say,  Eq.(\ref{eq:eqfsm1}) is a first order linear differential equation of the form
\begin{equation}
   \frac{df}{dx}+P(x)f=Q(x),
\end{equation}
which has a general solution
\begin{equation}
   f=e^{-\int^xP(x')dx'}\Big[\int^xQ(x')e^{\int^{x'}P(x'')dx''}dx'\Big],
\end{equation}
where the integration factor term $\int P(x)dx$ in our case is
\begin{equation}
   \int^{\phi'} P(\phi')d\phi''=-i\int^{\phi'}(x_s+y_s\cos\phi'')d\phi''=-i(x_s\phi'+y_s\sin\phi'),
\end{equation}
and thus
\begin{equation}\label{eq:fsm1}
   f_s^m=e^{i(x_s\phi'+y_s\sin\phi')}\Big[\int^{\phi'}Q(\phi'')e^{-i(x_s\phi''+y_s\sin\phi'')}d\phi''\Big],
\end{equation}
with
\begin{equation}
   Q=\frac{q_s}{m_s\omega_{cs}}\Big[\bm E+\bm v\times\Big(\frac{\bm k\times\bm E}{\omega}\Big)\Big]\cdot\frac{\partial f_{s0}}{\partial \bm v}.
\end{equation}

We notice that Eq.(\ref{eq:fsm1}) {\color{red}can not} reduce to (\ref{eq:fsu1}) by set $\omega_{cs}=0$, i.e., {\color{red}the magnetized version can not reduce to unmagnetized version by set $\omega_{cs}=0$}. Thus we should treat them separately.

\subsection{Electrostatic case}
For electrostatic case, $\bm k\times\bm E=0$, $\bm E=-i\bm k\Phi$, and only Poisson equation for field is required. We have
\begin{eqnarray}\label{eq:poisson1}
   &&k^2\Phi=\rho/\epsilon_0,\\
   &&\rho=\rho^u+\rho^m=\sum \rho_s^u+\sum \rho_s^m= \sum_{s=u,m} q_sn_{s0}\int dv^3 f_s.
\end{eqnarray}
The unmagnetized species 
\begin{equation}
 f_s^u=\frac{-\frac{q_s\Phi}{m_s}\bm k\cdot\frac{\partial f_{s0}}{\partial \bm v}}{(\omega-\bm k\cdot\bm v +i\nu_s)},
\end{equation}
where $\bm k\cdot\frac{\partial f_{s0}}{\partial \bm v}=k_x\frac{\partial f_{s0}}{\partial v_x}+k_z\frac{\partial f_{s0}}{\partial v_z}$ or $\bm k\cdot\frac{\partial f_{s0}}{\partial \bm v}=k_x\cos\phi'\frac{\partial f_{s0}}{\partial v'_\perp}+k_z\frac{\partial f_{s0}}{\partial v'_\parallel}$.
The corresponding charge density is
\begin{eqnarray}
   \rho_s^u= q_sn_{s0}\int dv^3 f_s^u= -\frac{q_s^2n_{s0}\Phi}{m_s}\int dv^3 \frac{\bm k\cdot\frac{\partial f_{s0}}{\partial \bm v}}{(\omega-\bm k\cdot\bm v +i\nu_s)}.
\end{eqnarray}
The velocity space integral may not be easy for complicated $f_{s0}$, and probably better calculate at $(v_x,v_y,v_z)$ coordinates instead of $(v'_\perp,\phi',v'_\parallel)$.

For magnetized species
\begin{equation}
  f_s^m=\frac{-i\Phi q_s}{m_s\omega_{cs}}e^{i(x_s\phi'+y_s\sin\phi')}\Big[\int^{\phi'}\Big(\bm k\cdot\frac{\partial f_{s0}}{\partial \bm v}\Big)e^{-i(x_s\phi''+y_s\sin\phi'')}d\phi''\Big],
\end{equation}
and
\begin{equation}\label{eq:esrhosm}
  \rho_s^m= q_sn_{s0}\int dv^3 f_s^m=\frac{-iq_s^2n_{s0}\Phi }{m_s\omega_{cs}}\int e^{i(x_s\phi'+y_s\sin\phi')}\Big[\int^{\phi'}\Big(\bm k\cdot\frac{\partial f_{s0}}{\partial \bm v}\Big)e^{-i(x_s\phi''+y_s\sin\phi'')}d\phi''\Big] d\phi'v'_\perp dv'_\perp dv'_\parallel,
\end{equation}
which is not yet in a useful form. Let us do some further calculations step by step. We have used cylindrical coordinates $(v'_\perp,\phi',v'_\parallel)$ with $dv^3=v'_\perp d\phi'dv'_\perp dv'_\parallel$, and note $\partial f_{s0}/\partial \phi'=0$. Use $\cos\phi'=(e^{i\phi'}+e^{-i\phi'})/2$, we have
\begin{eqnarray}
  && \int^{\phi'}\Big(\bm k\cdot\frac{\partial f_{s0}}{\partial \bm v}\Big)e^{-i(x_s\phi''+y_s\sin\phi'')}d\phi''=\int^{\phi'}\Big(k_x\cos\phi''\frac{\partial f_{s0}}{\partial v'_\perp}+k_z\frac{\partial f_{s0}}{\partial v'_\parallel}\Big)e^{-i(x_s\phi''+y_s\sin\phi'')}d\phi''\\\nonumber
   &&=k_z\frac{\partial f_{s0}}{\partial v'_\parallel}\int^{\phi'}e^{-i(x_s\phi''+y_s\sin\phi'')}d\phi''+k_x\frac{\partial f_{s0}}{\partial v'_\perp}\int^{\phi'}\frac{1}{2}(e^{i\phi''}+e^{-i\phi''})e^{-i(x_s\phi''+y_s\sin\phi'')}d\phi''.
\end{eqnarray}
Now, we use the following expansion
\begin{equation}
  e^{-iy_s\sin\phi''}=\sum_{n=-\infty}^{\infty}J_n(y_s)e^{-in\phi''},
\end{equation}
where $J_n(y_s)$ it the $n$th order Bessel function. And we have
\begin{eqnarray}
  \int^{\phi'}e^{-i(x_s\phi''+y_s\sin\phi'')}d\phi''=\sum_nJ_n(y_s)\int^{\phi'}e^{-i(x_s+n)\phi''}d\phi''=i\sum_n\frac{J_n(y_s)}{x_s+n}e^{-i(x_s+n)\phi'},
\end{eqnarray}
and
\begin{eqnarray}
  \int^{\phi'}(e^{i\phi''}+e^{-i\phi''})e^{-i(x_s\phi''+y_s\sin\phi'')}d\phi''&=&\sum_nJ_n(y_s)\int^{\phi'}[e^{-i(x_s-1+n)\phi''}+e^{-i(x_s+1+n)\phi''}]d\phi''\\
  &=&i\sum_nJ_n(y_s)\Big[\frac{e^{-i(x_s-1+n)\phi'}}{x_s-1+n}+\frac{e^{-i(x_s+1+n)\phi'}}{x_s+1+n}\Big].
\end{eqnarray}
We further use
\begin{equation}
  e^{i(x_s\phi'+y_s\sin\phi')}=\sum_{m=-\infty}^{\infty}J_m(y_s)e^{i(x_s+m)\phi'},
\end{equation}
and thus in Eq.(\ref{eq:esrhosm})
\begin{eqnarray}
  &&e^{i(x_s\phi'+y_s\sin\phi')}\Big[\int^{\phi'}\cdots d\phi''\Big] \\\nonumber
  &&  =ik_z\frac{\partial f_{s0}}{\partial v'_\parallel}\sum_{m,n}\frac{J_nJ_m}{x_s+n}e^{i(m-n)\phi'}+
  \frac{i}{2}k_x\frac{\partial f_{s0}}{\partial v'_\perp}\sum_{m,n}J_nJ_m\Big[\frac{e^{i(m-n+1)\phi'}}{x_s+n-1}+\frac{e^{i(m-n-1)\phi'}}{x_s+n+1}\Big]\\\nonumber
  &&  =ik_z\frac{\partial f_{s0}}{\partial v'_\parallel}\sum_{m,n}\frac{J_nJ_m}{x_s+n}e^{i(m-n)\phi'}+
  \frac{i}{2}k_x\frac{\partial f_{s0}}{\partial v'_\perp}\sum_{m,n}J_m\Big[J_{n+1}\frac{e^{i(m-n)\phi'}}{x_s+n}+J_{n-1}\frac{e^{i(m-n)\phi'}}{x_s+n}\Big]\\\nonumber
  &&  =i\sum_{m,n}J_mJ_n\Big(k_z\frac{\partial f_{s0}}{\partial v'_\parallel}+
  \frac{n\omega_{cs}}{v'_\perp}\frac{\partial f_{s0}}{\partial v'_\perp}\Big)\frac{e^{i(m-n)\phi'}}{x_s+n},
\end{eqnarray}
where we have used $(J_{n+1}+J_{n-1})=(2n/y_s)J_n$. Further integral out $\int d\phi'$ in  Eq.(\ref{eq:esrhosm}) gives
\begin{eqnarray}
  &&\int e^{i(x_s\phi'+y_s\sin\phi')}\Big[\int^{\phi'}\cdots d\phi''\Big]d\phi'   =i\int \sum_{m,n}J_mJ_n\Big(k_z\frac{\partial f_{s0}}{\partial v'_\parallel}+
  \frac{n\omega_{cs}}{v'_\perp}\frac{\partial f_{s0}}{\partial v'_\perp}\Big)\frac{e^{i(m-n)\phi'}}{x_s+n}d\phi' \\\nonumber
    &&  =i2\pi\sum_{n}\frac{J_n^2}{x_s+n}\Big(k_z\frac{\partial f_{s0}}{\partial v'_\parallel}+
  \frac{n\omega_{cs}}{v'_\perp}\frac{\partial f_{s0}}{\partial v'_\perp}\Big)\\\nonumber
    && =-i2\pi\sum_{n}\frac{\omega_{cs}J_n^2(y_s)}{\omega-k_\parallel v'_\parallel+{\color{red}i\nu_s-k_xv_{dsx}}-n\omega_{cs}}\Big(k_z\frac{\partial f_{s0}}{\partial v'_\parallel}+
  \frac{n\omega_{cs}}{v'_\perp}\frac{\partial f_{s0}}{\partial v'_\perp}\Big),
\end{eqnarray}
where because $\int_0^{2\pi}e^{i(m-n)\phi'}d\phi'=2\pi\delta_{m,n}$, and $\delta_{m,n}$ is Kronecker delta.
Thus we find the final form is very similar to the standard Harris dispersion relation form, except the term $x_s$.

Combine the magnetized and unmagnetized species $\rho_s$ and substitute them to the Poisson equation (\ref{eq:poisson1}), we obtain the final electrostatic dispersion relation
\begin{eqnarray}\label{eq:esdr0}\nonumber
  D(\omega,\bm k)&=&1+\sum_{s=m}\frac{\omega_{ps}^2}{k^2}\int_{-\infty}^{\infty}\int_{0}^{\infty} \sum_{n=-\infty}^{\infty}\frac{J_n^2(y_s)\Big(k_\parallel \frac{\partial f_{s0}}{\partial v'_\parallel}+
  \frac{n\omega_{cs}}{v'_\perp}\frac{\partial f_{s0}}{\partial v'_\perp}\Big)}{\omega-k_\parallel v'_\parallel+{\color{red}i\nu_s-k_xv_{dsx}}-n\omega_{cs}}2\pi v'_\perp dv'_\perp dv'_\parallel\\
  &&+\sum_{s=u} \frac{\omega_{ps}^2}{k^2}\int dv^3 \frac{\bm k\cdot\frac{\partial f_{s0}}{\partial \bm v}}{(\omega-\bm k\cdot\bm v +i\nu_s)}=0,
\end{eqnarray}
where $\omega_{ps}^2=\frac{n_{s0}q_s^2}{\epsilon_0 m_s}$.

\subsection{Electromagnetic  case}
For electromagnetic case, $\bm B=(\bm k\times\bm E)/\omega$, the field equations we needed are
\begin{eqnarray}\label{eq:em1}
   &&-i\omega \bm E= ic^2\bm k\times\Big(\frac{\bm k\times\bm E}{\omega}\Big)-{\bm J}/\epsilon_0,\\\label{eq:em1_J}
   &&\bm J=\bm J^u+\bm J^m=\sum \bm J_s^u+\sum \bm J_s^m= \sum_{s=u,m} q_sn_{s0}\int dv^3\bm v f_s.
\end{eqnarray}

If we assume the relation between current $\bm J$ and electric field $\bm E$ be 
\begin{equation}
\bm J=\bm\sigma\cdot\bm E.
\end{equation}
which will be calculated later via the kinetic equation, we obtain from Eq.(\ref{eq:em1})
\begin{equation}
\bm D(\omega,\bm k)\cdot\bm E=0,
\end{equation}
where $\bm D$ can be expressed in terms of the dielectric tensor $\bm K(\omega,\bm k)$ and gives the dispersion relation
\begin{equation}\label{eq:emdr0}
|\bm D(\omega,\bm k)|=|\bm K(\omega,\bm k)+(\bm k\bm k-k^2\bm I )\frac{c^2}{\omega^2}|=0,
\end{equation}
where $\bm I$ is the unit tensor, and we have used $\bm k\times(\bm k\times\bm E)=\bm k(\bm k\cdot\bm  E)-(\bm k\cdot\bm k)\bm E=(\bm k\bm k-k^2\bm I )\cdot\bm E$. And the relation to conductivity tensor $\bm\sigma$ is 
\begin{equation}
\bm K=\bm I+\bm Q=\bm I-\bm\sigma/(i\omega\epsilon_0),
\end{equation}
with $\bm Q=-\bm\sigma/(i\omega\epsilon_0)$.

Now, we calculate $\bm J$. 

For unmagnetized species
\begin{eqnarray}
\bm J_s^u&=&q_sn_{s0}\int dv^3\bm v f_s^u=\epsilon_0\omega_{ps}^2\int dv^3\bm v \frac{\Big[\bm E+\bm v\times\Big(\frac{\bm k\times\bm E}{\omega}\Big)\Big]\cdot\frac{\partial f_{s0}}{\partial \bm v}}{i(\omega-\bm k\cdot\bm v +i\nu_s)}\\\nonumber
&=&\frac{\epsilon_0\omega_{ps}^2}{i\omega}\int dv^3\bm v \frac{\Big[(\omega-\bm k\cdot\bm v+i\nu_s)\bm E-i\nu_s\bm E+\bm k(\bm v\cdot\bm E)\Big]\cdot\frac{\partial f_{s0}}{\partial \bm v}}{(\omega-\bm k\cdot\bm v +i\nu_s)}\\\nonumber
&=&-\frac{\epsilon_0\omega_{ps}^2}{i\omega}\bm I\cdot\bm E+\frac{\epsilon_0\omega_{ps}^2}{i\omega}\int dv^3 \frac{\Big[-i\nu_s(\bm v \frac{\partial f_{s0}}{\partial \bm v})+(\bm k\cdot\frac{\partial f_{s0}}{\partial \bm v})(\bm v\bm v)\Big]}{(\omega-\bm k\cdot\bm v +i\nu_s)}\cdot\bm E,
\end{eqnarray}
where we have used $\bm v\times(\bm k\times\bm E)=\bm k(\bm v\cdot\bm  E)-(\bm v\cdot\bm k)\bm E$, and $\bm v\bm v$, $\bm v \frac{\partial f_{s0}}{\partial \bm v}$ are dyadic product tensors. The result is similar to the one in Ref.\cite{Muschietti2017}, except our new $i\nu_s$ term. Note also that the tensor $\bm v \frac{\partial f_{s0}}{\partial \bm v}\neq \frac{\partial f_{s0}}{\partial \bm v} \bm v $.

For magnetized species
\begin{eqnarray}\label{eq:emJm}
  \bm J_s^m&=& q_sn_{s0}\int dv^3\bm v f_s^m\\\nonumber
  &=&\frac{q_s^2n_{s0} }{m_s\omega_{cs}}\int e^{i(x_s\phi'+y_s\sin\phi')}\Big\{\int^{\phi'}\Big[\bm E+\bm v\times\Big(\frac{\bm k\times\bm E}{\omega}\Big)\Big]\cdot\frac{\partial f_{s0}}{\partial \bm v}e^{-i(x_s\phi''+y_s\sin\phi'')}d\phi''\Big\} d\phi'v'_\perp dv'_\perp dv'_\parallel\\\nonumber
  &=&\frac{q_s^2n_{s0} }{m_s\omega_{cs}\omega}\int e^{i(x_s\phi'+y_s\sin\phi')}\Big\{\int^{\phi'}\Big[(\omega-\bm k\cdot\bm v)\bm E+\bm k(\bm v\cdot\bm  E)\Big]\cdot\frac{\partial f_{s0}}{\partial \bm v}e^{-i(x_s\phi''+y_s\sin\phi'')}d\phi''\Big\} d\phi'v'_\perp dv'_\perp dv'_\parallel.
\end{eqnarray}
We calculate firstly
\begin{eqnarray}\label{eq:evke}\nonumber
 &&\Big[(\omega-\bm k\cdot\bm v)\bm E+\bm k(\bm v\cdot\bm  E)\Big]\cdot\frac{\partial f_{s0}}{\partial \bm v}=\Big[(\omega-k_zv'_\parallel-k_xv'_\perp\cos\phi'-{\color{red}k_xv_{dsx}})\bm E+\bm k(\bm v\cdot\bm  E)\Big]\cdot\frac{\partial f_{s0}}{\partial \bm v}\\\nonumber
 &=&(\omega-k_zv'_\parallel-k_xv'_\perp\cos\phi'-{\color{red}k_xv_{dsx}})[\frac{\partial f_{s0}}{\partial v'_\perp}(E_x\cos\phi'+E_y\sin\phi')+\frac{\partial f_{s0}}{\partial v'_\parallel}E_z]\\\nonumber
 &&+[v'_\perp (E_x\cos\phi'+E_y\sin\phi')+v'_\parallel E_z+v_{dsx}E_x+v_{dsy}E_y](k_x\cos\phi'\frac{\partial f_{s0}}{\partial v'_\perp}+k_z\frac{\partial f_{s0}}{\partial v'_\parallel})\\\nonumber
 &=&\Big\{(\omega-{\color{red}k_xv_{dsx}})E_z+k_zv_{dsx}E_x+k_zv_{dsy}E_y+(k_zv'_\perp E_x-k_xv'_\perp E_z)\cos\phi'+k_zv'_\perp E_y\sin\phi'\Big\}\frac{\partial f_{s0}}{\partial v'_\parallel}\\\nonumber
 &&+\Big\{(\omega-k_zv'_\parallel-{\color{red}k_xv_{dsx}})E_y\sin\phi'+[(\omega-k_zv'_\parallel)E_x+k_xv_{dsy}E_y+k_xv'_\parallel E_z]\cos\phi'\Big\}\frac{\partial f_{s0}}{\partial v'_\perp}\\\nonumber
 &=&\Big\{(k_zv'_\perp E_x-k_xv'_\perp E_z)\frac{\partial f_{s0}}{\partial v'_\parallel}+[(\omega-k_zv'_\parallel)E_x+k_xv_{dsy}E_y+k_xv'_\parallel E_z]\frac{\partial f_{s0}}{\partial v'_\perp}\Big\}\cos\phi'\\\nonumber
 &&+\Big\{(\omega-k_zv'_\parallel-{\color{red}k_xv_{dsx}})E_y\frac{\partial f_{s0}}{\partial v'_\perp}+k_zv'_\perp E_y\frac{\partial f_{s0}}{\partial v'_\parallel}\Big\}\sin\phi'+[(\omega-{\color{red}k_xv_{dsx}})E_z+k_zv_{dsx}E_x+k_zv_{dsy}E_y]\frac{\partial f_{s0}}{\partial v'_\parallel}\\
 &=&(U_{s1}E_x+U_{s2}E_y +U_{s3}E_z)\cos\phi'+U_{s4}E_y\sin\phi'+(U_{s5}E_x+U_{s6}E_y+U_{s7}E_z),
\end{eqnarray}
with
\begin{eqnarray}\nonumber
 &&U_{s1}=[(\omega-k_zv'_\parallel)\frac{\partial f_{s0}}{\partial v'_\perp}+k_zv'_\perp \frac{\partial f_{s0}}{\partial v'_\parallel}],~~~U_{s2}=k_xv_{dsy}\frac{\partial f_{s0}}{\partial v'_\perp},~~~U_{s3}= [k_xv'_\parallel \frac{\partial f_{s0}}{\partial v'_\perp}-k_xv'_\perp \frac{\partial f_{s0}}{\partial v'_\parallel}],\\\nonumber
 &&U_{s4}=[(\omega-k_zv'_\parallel-{\color{red}k_xv_{dsx}})\frac{\partial f_{s0}}{\partial v'_\perp}+k_zv'_\perp \frac{\partial f_{s0}}{\partial v'_\parallel}],~~~U_{s5}=k_zv_{dsx}\frac{\partial f_{s0}}{\partial v'_\parallel},\\
 &&U_{s6}=k_zv_{dsy}\frac{\partial f_{s0}}{\partial v'_\parallel},~~~U_{s7}=(\omega-{\color{red}k_xv_{dsx}})\frac{\partial f_{s0}}{\partial v'_\parallel},
 \end{eqnarray}
where we have used $\bm E\cdot\frac{\partial f_{s0}}{\partial \bm v}=E_x\frac{\partial f_{s0}}{\partial v_x}+E_y\frac{\partial f_{s0}}{\partial v_y}+E_z\frac{\partial f_{s0}}{\partial v_z}=\frac{\partial f_{s0}}{\partial v'_\perp}(E_x\cos\phi'+E_y\sin\phi')+\frac{\partial f_{s0}}{\partial v'_\parallel}E_z$, $\bm v\cdot\bm E=v_xE_x+v_yE_y+v_zE_z=v'_\perp (E_x\cos\phi'+E_y\sin\phi')+v'_\parallel E_z+v_{dsx}E_x+v_{dsy}E_y$, $\bm k\cdot\frac{\partial f_{s0}}{\partial \bm v}=k_x\frac{\partial f_{s0}}{\partial v_x}+k_z\frac{\partial f_{s0}}{\partial v_z}=k_x\cos\phi'\frac{\partial f_{s0}}{\partial v'_\perp}+k_z\frac{\partial f_{s0}}{\partial v'_\parallel}$.

$\frac{\partial f_{s0}}{\partial v'_\perp}$ terms: $(\omega-k_zv'_\parallel-k_xv'_\perp\cos\phi'-{\color{red}k_xv_{dsx}})(E_x\cos\phi'+E_y\sin\phi')+[v'_\perp (E_x\cos\phi'+E_y\sin\phi')+v'_\parallel E_z+v_{dsx}E_x+v_{dsy}E_y]k_x\cos\phi'=(\omega-k_zv'_\parallel-{\color{red}k_xv_{dsx}})(E_x\cos\phi'+E_y\sin\phi')+(v'_\parallel E_z+v_{dsx}E_x+v_{dsy}E_y)k_x\cos\phi'=(\omega-k_zv'_\parallel-{\color{red}k_xv_{dsx}})E_y\sin\phi'+[(\omega-k_zv'_\parallel)E_x+k_xv_{dsy}E_y+k_xv'_\parallel E_z]\cos\phi'$.

$\frac{\partial f_{s0}}{\partial v'_\parallel}$ terms: $(\omega-k_zv'_\parallel-k_xv'_\perp\cos\phi'-{\color{red}k_xv_{dsx}})E_z+[v'_\perp (E_x\cos\phi'+E_y\sin\phi')+v'_\parallel E_z+v_{dsx}E_x+v_{dsy}E_y]k_z=(\omega-{\color{red}k_xv_{dsx}})E_z+k_zv_{dsx}E_x+k_zv_{dsy}E_y+(k_zv'_\perp E_x-k_xv'_\perp E_z)\cos\phi'+k_zv'_\perp E_y\sin\phi'$.

Our Eq.(\ref{eq:evke}) is the same as in Ref.\cite{Umeda2018} when set our $v_{dsy}=0$ and $\nu_s=0$, and their $k_y=0$.

Thus, using $\sin\phi=-i(e^{i\phi}-e^{-i\phi})/2$ and $J_{n+1}-J_{n-1}=-2J'_n$,
\begin{eqnarray}\nonumber
 f_s^{m}&=&\frac{q_s }{m_s\omega_{cs}\omega} e^{i(x_s\phi'+y_s\sin\phi')}\Big\{\int^{\phi'}\Big[(U_{s1}E_x+U_{s2}E_y +U_{s3}E_z)\frac{1}{2}(e^{i\phi''}+e^{-i\phi''})+U_{s4}E_y\frac{-i}{2}(e^{i\phi''}-e^{-i\phi''})\\\nonumber
 &&+(U_{s5}E_x+U_{s6}E_y+U_{s7}E_z)\Big]e^{-i(x_s\phi''+y_s\sin\phi'')}d\phi''\Big\}\\\nonumber
 &=&\frac{q_s }{m_s\omega_{cs}\omega}  \sum_mJ_me^{i(x_s+m)\phi'}\Big\{\int^{\phi'}\Big[(U_{s1}E_x+U_{s2}E_y +U_{s3}E_z)\frac{1}{2}(e^{i\phi''}+e^{-i\phi''})+U_{s4}E_y\frac{-i}{2}(e^{i\phi''}-e^{-i\phi''})\\\nonumber
 &&+(U_{s5}E_x+U_{s6}E_y+U_{s7}E_z)\Big]\sum_nJ_ne^{-i(x_s+n)\phi''}d\phi''\Big\}\\\nonumber
 &=&\frac{q_s }{m_s\omega_{cs}\omega}  \sum_mJ_me^{i(x_s+m)\phi'}\sum_nJ_n\Big\{(U_{s1}E_x+U_{s2}E_y +U_{s3}E_z)\frac{1}{2}\Big[\frac{ie^{-i(x_s+n-1)\phi'}}{(x_s+n-1)}+\frac{ie^{-i(x_s+n+1)\phi'}}{(x_s+n+1)}\Big]\\\nonumber
 &&+U_{s4}E_y\frac{-i}{2}\Big[\frac{ie^{-i(x_s+n-1)\phi'}}{(x_s+n-1)}-\frac{ie^{-i(x_s+n+1)\phi'}}{(x_s+n+1)}\Big]+(U_{s5}E_x+U_{s6}E_y+U_{s7}E_z)\frac{ie^{-i(x_s+n)\phi'}}{(x_s+n)}\Big\}\\\nonumber
 &=&\frac{iq_s }{m_s\omega_{cs}\omega}  \sum_{m,n}J_m\frac{e^{-i(n-m)\phi'}}{(x_s+n)}\Big\{(U_{s1}E_x+U_{s2}E_y +U_{s3}E_z)\frac{n}{y_s}J_n+U_{s4}E_yiJ'_n+\\\nonumber
 &&(U_{s5}E_x+U_{s6}E_y+U_{s7}E_z)J_n\Big\},
\end{eqnarray}
And thus we have
\begin{eqnarray}\label{eq:emJm}\nonumber
  \bm J_s^m&=& q_sn_{s0}\int dv^3\bm v f_s^m= q_sn_{s0}\int dv^3\left( \begin{array}{c}
    v'_\perp(e^{i\phi'}+e^{-i\phi'})/2+v_{dsx}  \\
    -iv'_\perp(e^{i\phi'}-e^{-i\phi'})/2+v_{dsy}   \\
    v'_\parallel
    \end{array}\right) f_s^m\\\nonumber
   &=&\frac{iq_s^2n_{s0} }{m_s\omega_{cs}\omega}\sum_{m,n}\int dv^3 \frac{e^{-i(n-m)\phi'}}{(x_s+n)}\Big\{(U_{s1}E_x+U_{s2}E_y +U_{s3}E_z)\frac{n}{y_s}J_n+U_{s4}E_yiJ'_n+\\\nonumber
 &&(U_{s5}E_x+U_{s6}E_y+U_{s7}E_z)J_n\Big\}\left( \begin{array}{c}
    v'_\perp mJ_m/y_s+v_{dsx}J_m  \\
    -iv'_\perp J'_m+v_{dsy}J_m   \\
    v'_\parallel J_m
    \end{array}\right)\\\nonumber
   &=&\frac{iq_s^2n_{s0} }{m_s\omega_{cs}\omega}\sum_{n}\int_{-\infty}^{\infty}\int_0^{\infty}  \frac{2\pi v'_\perp dv'_\perp dv'_\parallel}{(x_s+n)}\Big\{(\frac{n}{y_s}U_{s1}+U_{s5})J_nE_x+(\frac{n}{y_s}J_nU_{s2}+iJ'_nU_{s4}+J_nU_{s6})E_y +\\\nonumber
 &&(\frac{n}{y_s}U_{s3}+U_{s7})J_nE_z\Big\}\left( \begin{array}{c}
    (v'_\perp \frac{n}{y_s}+v_{dsx})J_n  \\
    (-iv'_\perp J'_n+v_{dsy}J_n)   \\
    v'_\parallel J_n
    \end{array}\right)\\
   &=&\frac{iq_s^2n_{s0} }{m_s\omega_{cs}\omega}\sum_{n}\int_{-\infty}^{\infty}\int_0^{\infty}  \frac{2\pi v'_\perp dv'_\perp dv'_\parallel}{(x_s+n)}\bm \Pi_s\cdot\bm E ,
\end{eqnarray}
with
{\scriptsize
\begin{equation}\label{eq:Pis}
    {\bm \Pi_s}=\left( \begin{array}{ccc}
  (\frac{n}{y_s}U_{s1}+U_{s5})(v'_\perp \frac{n}{y_s}+v_{dsx})J_n^2  &  (\frac{n}{y_s}J_nU_{s2}+iJ'_nU_{s4}+J_nU_{s6})(v'_\perp \frac{n}{y_s}+v_{dsx})J_n &  (\frac{n}{y_s}U_{s3}+U_{s7})(v'_\perp \frac{n}{y_s}+v_{dsx})J_n^2\\
  (\frac{n}{y_s}U_{s1}+U_{s5})(-iv'_\perp J'_n+v_{dsy}J_n)J_n  & (\frac{n}{y_s}J_nU_{s2}+iJ'_nU_{s4}+J_nU_{s6})(-iv'_\perp J'_n+v_{dsy}J_n) & (\frac{n}{y_s}U_{s3}+U_{s7})(-iv'_\perp J'_n+v_{dsy}J_n)J_n \\
  (\frac{n}{y_s}U_{s1}+U_{s5})v'_\parallel J_n^2  & (\frac{n}{y_s}J_nU_{s2}+iJ'_nU_{s4}+J_nU_{s6})v'_\parallel J_n & (\frac{n}{y_s}U_{s3}+U_{s7})v'_\parallel J_n^2
    \end{array}\right).
\end{equation}
}
Write out each terms:
\begin{itemize}

\item  $\Pi_{11}=(\frac{n}{y_s}U_{s1}+U_{s5})(v'_\perp \frac{n}{y_s}+v_{dsx})J_n^2=(\frac{n}{y_s}[(\omega-k_zv'_\parallel)\frac{\partial f_{s0}}{\partial v'_\perp}+k_zv'_\perp \frac{\partial f_{s0}}{\partial v'_\parallel}]+k_zv_{dsx}\frac{\partial f_{s0}}{\partial v'_\parallel})(v'_\perp \frac{n}{y_s}+v_{dsx})J_n^2 \\
= J_n^2[\frac{n\omega_{cs}}{k_x v'_\perp}(\omega-k_zv'_\parallel)\frac{\partial f_{s0}}{\partial v'_\perp}+k_z(\frac{n\omega_{cs}}{k_x } +v_{dsx})\frac{\partial f_{s0}}{\partial v'_\parallel}]( \frac{n\omega_{cs}}{k_x }+v_{dsx})$. [agree with Umeda18, except that his four terms can cancel]

\item $\Pi_{12}=(\frac{n}{y_s}J_nU_{s2}+iJ'_nU_{s4}+J_nU_{s6})(v'_\perp \frac{n}{y_s}+v_{dsx})J_n=(\frac{n}{y_s}J_nk_xv_{dsy}\frac{\partial f_{s0}}{\partial v'_\perp}+iJ'_n[(\omega-k_zv'_\parallel-{\color{red}k_xv_{dsx}})\frac{\partial f_{s0}}{\partial v'_\perp}+k_zv'_\perp \frac{\partial f_{s0}}{\partial v'_\parallel}]+J_nk_zv_{dsy}\frac{\partial f_{s0}}{\partial v'_\parallel})(v'_\perp \frac{n}{y_s}+v_{dsx})J_n \\
= J_n^2v_{dsy}( \frac{n\omega_{cs}}{k_x }+v_{dsx})(\frac{n\omega_{cs}}{ v'_\perp}\frac{\partial f_{s0}}{\partial v'_\perp}+k_z\frac{\partial f_{s0}}{\partial v'_\parallel})+iJ_nJ'_n( \frac{n\omega_{cs}}{k_x }+v_{dsx})[(\omega-k_zv'_\parallel-{\color{red}k_xv_{dsx}})\frac{\partial f_{s0}}{\partial v'_\perp}+k_zv'_\perp \frac{\partial f_{s0}}{\partial v'_\parallel}]$. [agree with Umeda18]

\item $\Pi_{21}=(\frac{n}{y_s}U_{s1}+U_{s5})(-iv'_\perp J'_n+v_{dsy}J_n)J_n=(\frac{n}{y_s}[(\omega-k_zv'_\parallel)\frac{\partial f_{s0}}{\partial v'_\perp}+k_zv'_\perp \frac{\partial f_{s0}}{\partial v'_\parallel}]+k_zv_{dsx}\frac{\partial f_{s0}}{\partial v'_\parallel})(-iv'_\perp J'_n+v_{dsy}J_n)J_n \\
= (-iv'_\perp J_nJ'_n+v_{dsy}J_n^2)[\frac{n\omega_{cs}}{k_x v'_\perp}(\omega-k_zv'_\parallel)\frac{\partial f_{s0}}{\partial v'_\perp}+\frac{n\omega_{cs}}{k_x }k_z \frac{\partial f_{s0}}{\partial v'_\parallel}+k_zv_{dsx}\frac{\partial f_{s0}}{\partial v'_\parallel}]$.  [agree with Umeda18, except that his two terms can cancel]

\item $\Pi_{22}=(\frac{n}{y_s}J_nU_{s2}+iJ'_nU_{s4}+J_nU_{s6})(-iv'_\perp J'_n+v_{dsy}J_n)=(\frac{n}{y_s}J_nk_xv_{dsy}\frac{\partial f_{s0}}{\partial v'_\perp}+iJ'_n[(\omega-k_zv'_\parallel-{\color{red}k_xv_{dsx}})\frac{\partial f_{s0}}{\partial v'_\perp}+k_zv'_\perp \frac{\partial f_{s0}}{\partial v'_\parallel}]+J_nk_zv_{dsy}\frac{\partial f_{s0}}{\partial v'_\parallel})(-iv'_\perp J'_n+v_{dsy}J_n) \\
= J_nv_{dsy}(-iv'_\perp J'_n+v_{dsy}J_n)[\frac{n\omega_{cs}}{v'_\perp}\frac{\partial f_{s0}}{\partial v'_\perp}+k_z\frac{\partial f_{s0}}{\partial v'_\parallel}]+iJ'_n(-iv'_\perp J'_n+v_{dsy}J_n)[(\omega-k_zv'_\parallel-{\color{red}k_xv_{dsx}})\frac{\partial f_{s0}}{\partial v'_\perp}+k_zv'_\perp \frac{\partial f_{s0}}{\partial v'_\parallel}]$. [agree with Umeda18]

\item $\Pi_{13}=(\frac{n}{y_s}U_{s3}+U_{s7})(v'_\perp \frac{n}{y_s}+v_{dsx})J_n^2=(\frac{n}{y_s} [k_xv'_\parallel \frac{\partial f_{s0}}{\partial v'_\perp}-k_xv'_\perp \frac{\partial f_{s0}}{\partial v'_\parallel}]+(\omega-{\color{red}k_xv_{dsx}})\frac{\partial f_{s0}}{\partial v'_\parallel})(v'_\perp \frac{n}{y_s}+v_{dsx})J_n^2 \\
=J_n^2( \frac{n\omega_{cs}}{k_x }+v_{dsx})[n\omega_{cs} \frac{v'_\parallel}{v'_\perp} \frac{\partial f_{s0}}{\partial v'_\perp}+(\omega-n\omega_{cs} -{\color{red}k_xv_{dsx}})\frac{\partial f_{s0}}{\partial v'_\parallel}]$. [agree with Umeda18]

\item $\Pi_{31}= (\frac{n}{y_s}U_{s1}+U_{s5})v'_\parallel J_n^2= (\frac{n}{y_s}[(\omega-k_zv'_\parallel)\frac{\partial f_{s0}}{\partial v'_\perp}+k_zv'_\perp \frac{\partial f_{s0}}{\partial v'_\parallel}]+k_zv_{dsx}\frac{\partial f_{s0}}{\partial v'_\parallel})v'_\parallel J_n^2 \\
= v'_\parallel J_n^2[\frac{n\omega_{cs}}{k_x v'_\perp}(\omega-k_zv'_\parallel)\frac{\partial f_{s0}}{\partial v'_\perp}+\frac{n\omega_{cs}}{k_x }k_z \frac{\partial f_{s0}}{\partial v'_\parallel}+k_zv_{dsx}\frac{\partial f_{s0}}{\partial v'_\parallel}]$.   [agree with Umeda18, except that his two terms can cancel]

\item $\Pi_{23}=(\frac{n}{y_s}U_{s3}+U_{s7})J_n(-iv'_\perp J'_n+v_{dsy}J_n)=(\frac{n}{y_s} [k_xv'_\parallel \frac{\partial f_{s0}}{\partial v'_\perp}-k_xv'_\perp \frac{\partial f_{s0}}{\partial v'_\parallel}]+(\omega-{\color{red}k_xv_{dsx}})\frac{\partial f_{s0}}{\partial v'_\parallel})J_n(-iv'_\perp J'_n+v_{dsy}J_n) \\
=(v_{dsy}J_n^2-iv'_\perp J_nJ'_n)[n\omega_{cs}\frac{v'_\parallel }{v'_\perp}\frac{\partial f_{s0}}{\partial v'_\perp}+(\omega-n\omega_{cs}-{\color{red}k_xv_{dsx}})\frac{\partial f_{s0}}{\partial v'_\parallel}]$. [agree with Umeda18]

\item $\Pi_{32}=(\frac{n}{y_s}J_nU_{s2}+iJ'_nU_{s4}+J_nU_{s6})v'_\parallel J_n=(\frac{n}{y_s}J_nk_xv_{dsy}\frac{\partial f_{s0}}{\partial v'_\perp}+iJ'_n[(\omega-k_zv'_\parallel-{\color{red}k_xv_{dsx}})\frac{\partial f_{s0}}{\partial v'_\perp}+k_zv'_\perp \frac{\partial f_{s0}}{\partial v'_\parallel}]+J_nk_zv_{dsy}\frac{\partial f_{s0}}{\partial v'_\parallel})v'_\parallel J_n  \\
=v'_\parallel J_n^2(\frac{n\omega_{cs}}{v'_\perp}\frac{\partial f_{s0}}{\partial v'_\perp}+k_z\frac{\partial f_{s0}}{\partial v'_\parallel})v_{dsy}+iv'_\parallel J_nJ'_n[(\omega-k_zv'_\parallel-{\color{red}k_xv_{dsx}})\frac{\partial f_{s0}}{\partial v'_\perp}+k_zv'_\perp \frac{\partial f_{s0}}{\partial v'_\parallel}]$. [agree with Umeda18]

\item $\Pi_{33}=(\frac{n}{y_s}U_{s3}+U_{s7})v'_\parallel J_n^2=(\frac{n}{y_s} [k_xv'_\parallel \frac{\partial f_{s0}}{\partial v'_\perp}-k_xv'_\perp \frac{\partial f_{s0}}{\partial v'_\parallel}]+(\omega-{\color{red}k_xv_{dsx}})\frac{\partial f_{s0}}{\partial v'_\parallel})v'_\parallel J_n^2  \\
=v'_\parallel J_n^2[n\omega_{cs} \frac{v'_\parallel}{v'_\perp}\frac{\partial f_{s0}}{\partial v'_\perp}+(\omega-n\omega_{cs}-{\color{red}k_xv_{dsx}})\frac{\partial f_{s0}}{\partial v'_\parallel}]$. [agree with Umeda18]

\end{itemize}

Note: $x_s=-\frac{\omega-k_\parallel v'_\parallel {-\color{red}k_xv_{dsx}+i\nu_s}}{\omega_{cs}}$ and $y_s=\frac{k_x v'_\perp}{\omega_{cs}}$. 

At $v_{dsy}=0$ and $\nu_s=0$ limit, our result reduces to exactly the same result as the Eq.(25) of Ref.\cite{Umeda2018}. By further set $v_{dsx}=0$, our result reduces to the standard without across magnetic field drifts one in Ref.\cite{Gurnett2005} and the non-relativistic case in Ref.\cite{Stix1992}. We should also note that when $v_{dsx,y}\neq0$, the matrix elements $\Pi_{ij}$ are not symmetric or antisymmetric any more.

The $\bm Q$ in electromagnetic dispersion relation is
\begin{eqnarray}\label{eq:emdrQ}\nonumber
   \bm Q &=& -\frac{\bm \sigma}{i\omega\epsilon_0}\\\nonumber
   &=&\sum_{s=u}\Big\{-\frac{\omega_{ps}^2}{\omega^2}\bm I +\frac{\omega_{ps}^2}{\omega^2}\int dv^3 \frac{\Big[-i\nu_s(\bm v \frac{\partial f_{s0}}{\partial \bm v})+(\bm k\cdot\frac{\partial f_{s0}}{\partial \bm v})(\bm v\bm v)\Big]}{(\omega-\bm k\cdot\bm v +i\nu_s)}\Big\}\\
   &&+\sum_{s=m}\frac{\omega_{ps}^2 }{\omega^2}\sum_{n=-\infty}^{\infty}\int_{-\infty}^{\infty}\int_0^{\infty}  \frac{2\pi v'_\perp dv'_\perp dv'_\parallel}{(\omega-k_\parallel v'_\parallel {-\color{red}k_xv_{dsx}+i\nu_s}-n\omega_{cs})}\bm \Pi_s.
\end{eqnarray}

\subsection{Darwin model  case}
For Darwin model, we still have $\bm B=(\bm k\times\bm E)/\omega$, and thus the calculation of current $\bm J=\bm J^u+\bm J^m$ is exactly the same as in Eq.(\ref{eq:em1_J}), i.e., the kinetic solutions to the distribution function $f_{s}$ and current $\bm J$ do not need change as in the above electromagnetic model, which simplify our derivation a lot.

We discuss the change of the field equation here. In Fourier space, we have
\begin{eqnarray}
   \bm E_T=\bm E-\frac{\bm k (\bm k\cdot\bm E)}{k^2}=(\bm I -\frac{\bm k\bm k}{k^2})\cdot\bm E,~~~\bm E_L=\frac{\bm k (\bm k\cdot\bm E)}{k^2}=(\frac{\bm k\bm k}{k^2})\cdot\bm E,
\end{eqnarray}
which can satisfied $\bm k\cdot \bm E_T=0$, $\bm k\times \bm E_L=0$ and $\bm E=\bm E_T+\bm E_L$. We have used the tensor relation $(\bm U\bm V)\cdot\bm W=\bm U(\bm V\cdot\bm W)$ and $\bm W\cdot(\bm U\bm V)=(\bm W\cdot\bm U)\bm V$, where $\bm U$, $\bm V$ and $\bm W$ are vectors.
The field equations we needed are
\begin{eqnarray}
   &&-i\omega \bm E_L= ic^2\bm k\times\Big(\frac{\bm k\times\bm E}{\omega}\Big)-{\bm J}/\epsilon_0,
\end{eqnarray}
i.e.,
\begin{eqnarray}
   &&{\color{red}\underbrace{(\frac{\bm k\bm k}{k^2})\cdot \bm E}_{\bm I\cdot\bm E {\rm ~in~electromagnetic~model}}}+ \frac{c^2}{\omega^2}(\bm k\bm k-k^2\bm I )\cdot\bm E+\bm Q\cdot\bm E=0,
\end{eqnarray}
or
\begin{eqnarray}
   &&\bm I\cdot\bm E+ {\color{red}\underbrace{\Big(\frac{k^2c^2}{\omega^2}+1\Big)}_{\frac{k^2c^2}{\omega^2} {\rm ~in~electromagnetic~model}}} \Big(\frac{\bm k\bm k}{k^2}-\bm I \Big)\cdot\bm E+\bm Q\cdot\bm E=0,
\end{eqnarray}
where the only change from the electromagnetic model is that the $\bm I\cdot\bm E$ term is changed to be $(\frac{\bm k\bm k}{k^2})\cdot \bm E$, or $\frac{k^2c^2}{\omega^2}$ term is changed to be $\frac{k^2c^2}{\omega^2}+1$. And thus the Darwin dispersion relation is
\begin{equation}\label{eq:darwindr0}
|\bm D(\omega,\bm k)|=|\bm K(\omega,\bm k)+{\color{red}\underbrace{\Big(\frac{k^2c^2}{\omega^2}+1\Big)}_{\frac{k^2c^2}{\omega^2} {\rm ~in~electromagnetic~model}}} \Big(\frac{\bm k\bm k}{k^2}-\bm I \Big)|=0,
\end{equation}

\vspace{20pt}
Eqs.(\ref{eq:esdr0}), (\ref{eq:emdr0}) and (\ref{eq:darwindr0}) with $\bm Q$ in (\ref{eq:emdrQ}) are our starting electrostatic, electromagnetic and Darwin dispersion relations with drift across magnetic field. The above dispersion relations are valid for arbitrary non-relativistic distribution functions. Later, we will limit our study to treat a special case of the distribution function $f_{s0}$.

\section{The Dispersion Relation for Extend Maxwellian Distribution}\label{sec:maxwell_dr}
The extend Maxwellian distribution here means bi-Maxwellian distribution with loss cone, parallel and perpendicular drifts and ring beam. We model it use the following equilibrium distribution function.

\subsection{Equilibrium distribution function}
We assume equilibrium distribution function $F_{s0}(v'_\parallel,v'_\perp)=n_{s0}f_{s0}(v'_\parallel,v'_\perp)$, with $v'_\parallel=v_z$, $v'_\perp=\sqrt{(v_x-v_{dsx})^2+(v_y-v_{dsy})^2}$, and
\begin{eqnarray}\label{eq:f0s}
   &&f_{s0}(v'_\parallel,v'_\perp)=f_{s0z}(v'_\parallel )f_{s0\perp}(v'_\perp)
   \\\nonumber
   &=&\frac{1}{\pi^{3/2}v_{zts}v_{\perp ts}^2}\exp\Big[-\frac{(v'_\parallel-v_{dsz})^2}{v_{zts}^2}\Big]\Big\{\frac{r_{sa}}{A_{sa}}\exp\Big[-\frac{(v'_\perp-v_{dsr})^2}{v_{\perp ts}^2}\Big]+\frac{r_{sb}}{\alpha_sA_{sb}}\exp\Big[-\frac{(v'_\perp-v_{dsr})^2}{\alpha_sv_{\perp ts}^2}\Big]\Big\},
\end{eqnarray}
where $r_{sa}=\Big(\frac{1-\alpha_s\Delta_s}{1-\alpha_s}\Big)$ and $r_{sb}=\Big(\frac{-\alpha_s+\alpha_s\Delta_s}{1-\alpha_s}\Big)$, and
\begin{eqnarray}\label{eq:Asab}
   A_{s\sigma}&=&\exp\Big(-\frac{v_{dsr}^2}{v_{\perp ts\sigma}^2}\Big)+\frac{\sqrt{\pi}v_{dsr}}{v_{\perp ts\sigma}}{\rm erfc}\Big(-\frac{v_{dsr}}{v_{\perp ts\sigma}}\Big),~~~{\rm for}~~ \sigma=a,b,
\end{eqnarray}
with $v_{\perp tsa}=v_{\perp ts}$ and $v_{\perp tsb}=\sqrt{\alpha_s}v_{\perp ts}$.

That is, the equilibrium distribution function $f_{s0}$ is separated to two sub-distributions: $f_{s0}=\sum_{\sigma=a,b}r_{s\sigma}f_{s0\sigma}=f_{s0z}(v'_\parallel)\sum_{\sigma=a,b}r_{s\sigma}f_{s0\perp\sigma}$, $f_{s0\sigma}=f_{s0z}(v'_\parallel)f_{s0\perp\sigma}(v'_\perp)$, $f_{s0\perp}=\sum_{\sigma=a,b}r_{s\sigma}f_{s0\perp\sigma}$,
\begin{equation}\label{eq:f0z}
   f_{s0z}(v'_\parallel)=\frac{1}{\pi^{1/2}v_{zts}}\exp\Big[-\frac{(v'_\parallel-v_{dsz})^2}{v_{zts}^2}\Big]=\frac{1}{\pi^{1/2}v_{zts}}\exp\Big[-\frac{(v_z-v_{dsz})^2}{v_{zts}^2}\Big],
\end{equation}
and
\begin{eqnarray}\label{eq:f0p}\nonumber
   f_{s0\perp\sigma}(v'_\perp)&=&\frac{1}{\pi A_{s\sigma}v_{\perp ts\sigma}^2}\exp\Big[-\frac{(v'_\perp-v_{dsr})^2}{v_{\perp ts\sigma}^2}\Big]\\
   &=&\frac{1}{\pi A_{s\sigma}v_{\perp ts\sigma}^2}\exp\Big[-\frac{(\sqrt{(v_x-v_{dsx})^2+(v_y-v_{dsy})^2}-v_{dsr})^2}{v_{\perp ts\sigma}^2}\Big],
\end{eqnarray}
where 
$v_{dsx}$, $v_{dsy}$, $v_{dsz}$ are the drift velocities in $x$ (perpendicular 1), $y$  (perpendicular 2) and $z$  (parallel) directions, respectively. And, $v_{dsr}$ is the perpendicular ring beam velocity, and ${\rm erfc}(-x)=1-{\rm erf}(-x)=1+{\rm erf}(x)$ is the complementary error function, with ${\rm erfc}(x)\equiv\frac{2}{\sqrt{\pi}}\int_{x}^{\infty}e^{-t^2}dt$ and ${\rm erf}(x)\equiv\frac{2}{\sqrt{\pi}}\int_{0}^{x}e^{-t^2}dt$. The $v_{zts}$ and $v_{\perp ts\sigma}$ are the parallel and perpendicular thermal velocities and corresponding temperatures are $T_{zs}=\frac{1}{2}k_Bm_sv_{zts}^2$ and $T_{\perp s\sigma}=\frac{1}{2}k_Bm_sv_{\perp ts\sigma}^2$. We define the temperature anisotropic $\lambda_{Ts\sigma}=T_{zs}/T_{\perp s\sigma}$. 
The parameters $\Delta_s$ and $\alpha_s$ determine the depth and size of the loss-cone\footnote{Actually, the user can also use two different specieses to represent the loss cone distribution, with one of them has negative density. Then, they can also have different $v_{dsx}$, $v_{dsy}$ and $v_{dsr}$. Many complicated distribution function can be constructed based on our model, which we leave it to the user. For example, Ref.\cite{Min2015} constructed the shell distribution based on ring beam model.}. Here, $\Delta_s\in[0,1]$, for max loss cone and no loss cone. If $\Delta_s=1$ or $\alpha_s=1$, i.e., $r_{sa}=1$ and $r_{sb}=0$, the above equation reduced to no loss cone case.
Note $\int f_{s0}d\bm v=\int f_{s0z}d\bm v=\int f_{s0\perp}d\bm v=\int f_{s0\sigma}d\bm v=\int f_{s0\perp\sigma}d\bm v=1$.

We will use the same distribution in magnetized and unmagnetized versions to simplified the notations. {\color{red}For unmagnetized version, to remove the trouble of integral, we study only $v_{dsr}=0$.} How to justify the assumed distribution to the realistic physics problem is leave to the user. For example, the linearized equation (i.e., the dispersion relations) can also run when the zero order current and charge density are not zero. Thus, the user should justify whether it is reasonable.

Several $\partial f_{s0}/\partial \bm v$ terms are particular useful for further discussions:
\begin{eqnarray}\nonumber
   \frac{\partial f_{s0}}{\partial v'_\parallel}&=&-\frac{2(v'_\parallel-v_{dsz})}{v_{zts}^2}f_{s0}=-2(v'_\parallel-v_{dsz})\sum_{\sigma=a,b}\frac{r_{s\sigma}f_{s0\sigma}}{v_{zts}^2},\\\nonumber
   \frac{\partial f_{s0}}{\partial v'_\perp}&=&-2(v'_\perp-v_{dsr})\sum_{\sigma=a,b}\frac{r_{s\sigma}f_{s0\sigma}}{v_{\perp ts\sigma}^2},\\\nonumber
   \frac{\partial f_{s0}}{\partial v_x}&=&-2\Big[1-\frac{v_{dsr}}{\sqrt{(v_x-v_{dsx})^2+(v_y-v_{dsy})^2}}\Big](v_x-v_{dsx})\sum_{\sigma=a,b}\frac{r_{s\sigma}f_{s0\sigma}}{v_{\perp ts\sigma}^2},\\\nonumber
   \frac{\partial f_{s0}}{\partial v_y}&=&-2\Big[1-\frac{v_{dsr}}{\sqrt{(v_x-v_{dsx})^2+(v_y-v_{dsy})^2}}\Big](v_y-v_{dsx})\sum_{\sigma=a,b}\frac{r_{s\sigma}f_{s0\sigma}}{v_{\perp ts\sigma}^2},\\\nonumber
   \frac{\partial f_{s0}}{\partial v_z}&=&-\frac{2(v_z-v_{dsz})}{v_{zts}^2}f_{s0}=-2(v_z-v_{dsz})\sum_{\sigma=a,b}\frac{r_{s\sigma}f_{s0\sigma}}{v_{zts}^2},
\end{eqnarray}
where we have used that $f_{s0}=\sum_{\sigma=a,b}r_{s\sigma}f_{s0\sigma}$.

\subsection{Notations}

Note the definition of $v_{ts}$, i.e., $v_{ts}=\sqrt{\frac{2k_BT_s}{m_s}}$, not $v_{ts}=\sqrt{\frac{k_BT_s}{m_s}}$ as in Ref.\cite{Umeda2012}. Other notations:
$\omega_{ps}^2=\frac{n_{s0}q_s^2}{\epsilon_0 m_s}$, $\omega_p^2=\sum_s\omega_{ps}^2$, $\Omega_{s}=\frac{q_sB_0}{m_s}$,
$\lambda_{Ds}^2=\frac{\epsilon_0k_BT_{zs}}{n_{s0}q_s^2}$,
$a_{s\sigma}={\color{red}\sqrt{2}}k_\perp\rho_{cs\sigma}$, $b_{s\sigma}=\frac{v_{dsr}}{v_{\perp ts\sigma}}$,
$\rho_{cs\sigma}=\sqrt{\frac{k_BT_{s\perp\sigma}}{m_s}}\frac{1}{\Omega_s}=\frac{ v_{\perp ts\sigma}}{\sqrt{2}\Omega_s}$. Note: $\Omega_{s}<0$ for electron ($q_s<0$), and thus also $\rho_{cs\sigma}<0$ and $a_{s\sigma}<0$.
[Note the definition in previous bi-Maxwellian version: $a_s=k_\perp\rho_{cs}$, $b_s=k^2_\perp\rho^2_{cs}$,
$\rho_{cs}=\sqrt{\frac{k_BT_{s\perp}}{m_s}}\frac{1}{\Omega_s}=\frac{ v_{\perp ts}}{\sqrt{2}\Omega_s}$, $\Gamma_n(b)=I_n(b)e^{-b}$, $I_n$ is the modified
Bessel function.]

We use 'ES3D' or 'ES', 'EM3D' or 'EM', 'Darwin' to represent the electrostatic, electromagnetic and Darwin version, respectively. And with suffix '-U' and '-M' to represent the corresponding unmagnetized and magnetized species.

\subsection{Some integrals and functions}
Here, we also clarify the correctly treat of the plasma dispersion function for $k_z\leq0$.
The standard definition of plasma dispersion function $Z(\zeta)$ is
\begin{eqnarray*}
  &&Z(\zeta)=\frac{1}{\sqrt{\pi}}\int_C\frac{1}{x-\zeta}e^{-x^2}dx, \\
  &&\frac{dZ}{d\zeta} = -2(1+\zeta Z),
\end{eqnarray*}
where $C$ is the Landau contour to analytic continuation from $Im(\zeta)>0$ to $Im(\zeta)\leq0$. However, for our usage the above definition to plasma dispersion function with $\zeta_s=\frac{\omega-k_{cs}}{k_{ts}}$ is only correct when $k_{ts}>0$, where for example $k_{ts}=k_\parallel v_{zts}$. 
To correctly capture the physics, one should be careful of the analytic continuation for both $k_{ts}>0$ and $k_{ts}\leq0$. From standard derivation of the dispersion relations based on Laplacian transformation instead of Fourier transformation which considered the  causality, say Ref.\cite{Gurnett2005}, we obtain the correct analytic continuation one should be
\begin{equation}
    Z(\zeta_s)=\left\{ \begin{array}{ccc}
    i\sqrt{\pi}e^{-\zeta_s^2}+\frac{1}{\sqrt{\pi}}P\int_{-\infty}^{\infty}\frac{1}{x-\zeta_s}e^{-x^2}dx, && k_{ts}>0,\\
    -\frac{1}{\zeta_s}, && k_{ts}=0,\\    
    -i\sqrt{\pi}e^{-\zeta^2}+\frac{1}{\sqrt{\pi}}P\int_{-\infty}^{\infty}\frac{1}{x-\zeta_s}e^{-x^2}dx,&& k_{ts}<0,
    \end{array}\right.
\end{equation}
where $P$ refers to the principal value integral.
The corresponding $J$-pole expansion should also be modified for $k_{ts}\leq0$, which will be discussed later. We can find easily that for $k_{ts}<0$, one can use $Z(\zeta_s)=-Z(-\zeta_s)$, which will simplify our later usage.

To simplify the notation, we define also the function $Z_p(\zeta_s,k_{ts})$, and have
\begin{eqnarray*}
  &&Z_p(\zeta_s,k_{ts})=\frac{1}{\sqrt{\pi}}\int_C\frac{x^p}{ x- \zeta_s}e^{-x^2}dx, \\
  &&Z_0(\zeta_s,k_{ts}) =Z(\zeta_s)\\
  &&Z_1(\zeta_s,k_{ts}) = [1+\zeta_s Z_0(\zeta_s)],\\
  &&Z_2(\zeta_s,k_{ts}) = \zeta_s[1+\zeta_s Z_0(\zeta_s)],\\
  &&Z_3(\zeta_s,k_{ts}) =\frac{1}{2}+\zeta_s^2[1+\zeta_s Z_0(\zeta_s)],
\end{eqnarray*}
where we have used $\int_{-\infty}^{\infty}x^2e^{-x^2}dx=\frac{\sqrt{\pi}}{2}$. Later, we will find by using $Z_{0,1,2,3}$ will simplify the notations a lot, and also will make the matrix linear transformation quite straightforward, say,
\begin{eqnarray}
&&\int_{-\infty}^{\infty}\frac{f_{sz}}{\omega_{sn}-k_z (v_\parallel-v_{dsz})}dv_\parallel=-\frac{Z(\zeta_{sn})}{k_zv_{zts}}=-\frac{Z_0(\zeta_{sn})}{k_zv_{zts}},\\
&&\int_{-\infty}^{\infty}\frac{k_z(v_\parallel-v_{dsz})f_{sz}}{\omega_{sn}-k_z (v_\parallel-v_{dsz})}dv_\parallel=-[1+\zeta_{sn}Z(\zeta_{sn})]=-Z_1(\zeta_{sn}),\\
&&\int_{-\infty}^{\infty}\frac{[k_z(v_\parallel-v_{dsz})]^2f_{sz}}{\omega_{sn}-k_z (v_\parallel-v_{dsz})}dv_\parallel=-\omega_{sn}[1+\zeta_{sn}Z(\zeta_{sn})]=-k_zv_{zts}Z_2(\zeta_{sn}),\\
&&\int_{-\infty}^{\infty}\frac{[k_z(v_\parallel-v_{dsz})]^3f_{sz}}{\omega_{sn}-k_z (v_\parallel-v_{dsz})}dv_\parallel=-\frac{k_z^2v_{zts}^2}{2}-\omega_{sn}^2[1+\zeta_{sn}Z(\zeta_{sn})]=-k_z^2v_{zts}^2Z_3(\zeta_{sn}).
\end{eqnarray}

For Bessel function, we have (p256 of Ref.\cite{Stix1992}): $J'_n(x)=[J_{n-1}(x)-J_{n+1}(x)]/2$, ~$nJ_n(x)/x=[J_{n-1}(x)+J_{n+1}(x)]/2$, ~$\sum_{n=-
\infty}^{\infty}J_n^2=1$, ~$\sum_{n=-\infty}^{\infty}J_nJ'_n=0$, ~$\sum_{n=-\infty}^{\infty}nJ_n^2=0$, ~$\sum_{n=-\infty}^{\infty}(J'_n)^2=\frac{1}{2}$, ~$\sum_{n=-\infty}^{\infty}\frac{n^2J_n^2(x)}{x^2}=\frac{1}{2}$, ~$\sum_{n=-
\infty}^{\infty}nJ_nJ'_n=0$, ~$J_n(-x)=J_{-n}(x)=(-1)^nJ_n(x)$, and $\sum_{m\neq0,n=-\infty}^{\infty}J_nJ_{n+m}=0$.

We define
\begin{eqnarray*}
  A_n(a,b,c)&\equiv&\int_0^{\infty}J_n^2(ay)e^{-(y-b)^2}(y-c)dy,\\
  B_n(a,b,c)&\equiv&\int_0^{\infty}J_n(ay)J'_n(ay)e^{-(y-b)^2}y(y-c)dy,\\
  C_n(a,b,c)&\equiv&\int_0^{\infty}J'_n{}^2(ay)e^{-(y-b)^2}y^2(y-c)dy.
\end{eqnarray*}
For our usage $a=\sqrt{2}k_\perp\rho_{cs}$, $b=\frac{v_{dsr}}{v_{\perp ts}}$ and $c=b$ or $0$. Here, we calculate $A_n$, $B_n$ and $C_n$ using numerical integral. When $b=c=0$, the above integrals reduce to the conventional Maxwellian form with modified Bessel function $I_n$ and $I'_n$, i.e.,
\begin{eqnarray*}
  A_n(a,0,0)&=&\frac{1}{2}e^{-\frac{a^2}{2}}I_n(\frac{a^2}{2})=\frac{1}{2}\Gamma_n(\frac{a^2}{2}),\\
  B_n(a,0,0)&=&\frac{a}{4}e^{-\frac{a^2}{2}}[I'_n(\frac{a^2}{2})-I_n(\frac{a^2}{2})]=\frac{a}{4}\Gamma'_n(\frac{a^2}{2}),\\
  C_n(a,0,0)&=&e^{-\frac{a^2}{2}}\Big[\frac{n^2}{2a^2}I_n(\frac{a^2}{2})-\frac{a^2}{4}I'_n(\frac{a^2}{2})+\frac{a^2}{4}I_n(\frac{a^2}{2})\Big]=\frac{n^2}{2a^2}\Gamma_n(\frac{a^2}{2})-\frac{a^2}{4}\Gamma'_n(\frac{a^2}{2}).
\end{eqnarray*}
where $\Gamma_n(b)=e^{-b}I_n(b)$. And $\Gamma'_n(b)=(I'_n-I_n)e^{-b}$, $I'_n(b)=(I_{n+1}+I_{n-1})/2$, $I_{-n}=I_{n}$. Thus, for species with $v_{dsr}=0$, we will still use the Bessel function form $\Gamma_n$ and $\Gamma'_n$.

Note, we have
\begin{eqnarray}
&&\int_0^{\infty}J_n^2(y_s)f_{s0\perp\sigma}v'_\perp dv'_\perp=\frac{1}{\pi A_{s\sigma}}A_n(a_{s\sigma},b_{s\sigma},0), \\
&& \int_0^{\infty}J_n^2(y_s)f_{s0\perp\sigma}(v'_\perp-v_{dsr})dv'_\perp=\frac{1}{\pi A_{s\sigma}}A_n(a_{s\sigma},b_{s\sigma},b_{s\sigma}),\\
&&\int_0^{\infty}J_n(y_s)J'_n(y_s)f_{s0\perp\sigma}v'^2_\perp dv'_\perp=\frac{v_{\perp ts\sigma}}{\pi A_{s\sigma}}B_n(a_{s\sigma},b_{s\sigma},0), \\
&& \int_0^{\infty}J_n(y_s)J'_n(y_s)f_{s0\perp\sigma}v'_\perp(v'_\perp-v_{dsr})dv'_\perp=\frac{v_{\perp ts\sigma}}{\pi A_{s\sigma}}B_n(a_{s\sigma},b_{s\sigma},b_{s\sigma}),\\
&&\int_0^{\infty}J'_n(y_s)J'_n(y_s)f_{s0\perp\sigma}v'^3_\perp dv'_\perp=\frac{v_{\perp ts\sigma}^2}{\pi A_{s\sigma}}C_n(a_{s\sigma},b_{s\sigma},0), \\
&& \int_0^{\infty}J'_n(y_s)J'_n(y_s)f_{s0\perp\sigma}v'^2_\perp(v'_\perp-v_{dsr})dv'_\perp=\frac{v_{\perp ts\sigma}^2}{\pi A_{s\sigma}}C_n(a_{s\sigma},b_{s\sigma},b_{s\sigma}).
\end{eqnarray}

We have checked in Matlab, the speed of numerical integral is also fast, compared to using the Bessel function. To short the notation, we would also use such as $A_n(0)=A_n(a_{s\sigma},b_{s\sigma},0)$ and $A_n(b_{s\sigma})=A_n(a_{s\sigma},b_{s\sigma},b_{s\sigma})$.

For short the notations using $A_{nbs\sigma}=\frac{4}{A_{s\sigma}}A_n(a_{s\sigma},b_{s\sigma},b_{s\sigma})$,  $B_{nbs\sigma}=\frac{4}{A_{s\sigma}}B_n(a_{s\sigma},b_{s\sigma},b_{s\sigma})$, $C_{nbs\sigma}=\frac{4}{A_{s\sigma}}C_n(a_{s\sigma},b_{s\sigma},b_{s\sigma})$, $A_{n0\sigma}=\frac{4}{A_{s\sigma}}A_n(a_{s\sigma},b_{s\sigma},0)$,  $B_{n0\sigma}=\frac{4}{A_{s\sigma}}B_n(a_{s\sigma},b_{s\sigma},0)$ and  $C_{n0\sigma}=\frac{4}{A_{s\sigma}}C_n(a_{s\sigma},b_{s\sigma},0)$. 

Note also: $\frac{2}{A_s}\int_0^{\infty}y^2e^{-(y-b_s)^2}(y-b_s)dy=1$ and $\frac{2}{A_s}\int_0^{\infty}ye^{-(y-b_s)^2}dy=1$.

Expansion $A_n$ and $B_n$ at $a\to0$ would be useful. For $|y|\ll1$ and $n\geq0$, $J_n(y)=\sum_{m=0}^{\infty}\frac{(-1)^m}{m!\Gamma(m+n+1)}(\frac{y}{2})^{2m+n}$, with $\Gamma(n+1)=n!$  be the Euler $\Gamma$ function. Thus, we have $J_n(ay)\sim \frac{1}{n!}(\frac{ay}{2})^{n}-\frac{1}{(n+1)!}(\frac{ay}{2})^{n+2}$ and $J'_{n\geq1}(ay)\sim \frac{1}{2(n-1)!}(\frac{ay}{2})^{n-1}-\frac{(n+2)}{2(n+1)!}(\frac{ay}{2})^{n+1}$. To $O(a^2)$, we have
\begin{eqnarray*}
  A_n(a,b,b)&=&\int_0^{\infty}J_n^2(ay)e^{-(y-b)^2}(y-b)dy\sim\left\{ \begin{array}{ccc}
  \frac{B_s}{2}- \frac{A_s}{4}a^2, && n=0,\\
    \frac{A_s}{2}a^2, && n=\pm1,\\    
    0,&& {\rm others},
    \end{array}\right.\\
  B_n(a,b,b)&\sim&\int_0^{\infty}J_n(ay)J'_n(ay)e^{-(y-b)^2}y(y-b)dy\sim\left\{ \begin{array}{ccc}
   -\frac{A_s}{4}a, && n=0,\\
    \frac{A_s}{8}a, && n=\pm1,\\    
    0,&& {\rm others},
    \end{array}\right.\\
  A_n(a,b,0)&=&\int_0^{\infty}J_n^2(ay)e^{-(y-b)^2}ydy\sim\left\{ \begin{array}{ccc}
 \frac{1}{2}A_s- \frac{2b^2+3}{8}a^2A_s+ \frac{1}{8}a^2B_s, && n=0,\\
   \frac{2b^2+3}{16}a^2A_s-\frac{1}{16}a^2B_s, && n=\pm1,\\    
    0,&& {\rm others},
    \end{array}\right.\\
  B_n(a,b,0)&\sim&\int_0^{\infty}J_n(ay)J'_n(ay)e^{-(y-b)^2}y^2dy\sim\left\{ \begin{array}{ccc}
  - \frac{2b^2+3}{8}aA_s+ \frac{1}{8}aB_s, && n=0,\\
     \frac{2b^2+3}{16}aA_s- \frac{1}{16}aB_s, && n=\pm1,\\    
    0,&& {\rm others},
    \end{array}\right.\\
\end{eqnarray*}
where we have used $J_0(ay)\sim1-(\frac{ay}{2})^2$, $J_1(ay)\sim \frac{ay}{2}$, $J_2(ay)\sim \frac{1}{2}(\frac{ay}{2})^2$, $J_{n\geq3}\sim0$,
$J'_0(ay)\sim-(\frac{ay}{2})$, $J'_1(ay)\sim \frac{1}{2}-\frac{3}{4}(\frac{ay}{2})^2$,  $J'_2(ay)\sim \frac{1}{2}(\frac{ay}{2})$, $J'_3(ay)\sim \frac{1}{4}(\frac{ay}{2})^2$, $J'_{n\geq4}\sim0$,
and defined $A_s=2\int_0^{\infty}e^{-(y-b)^2}ydy=e^{-b^2}+\sqrt{\pi}b[{\rm erf}(b)+1]$, $B_s=2\int_0^{\infty}e^{-(y-b)^2}(y-b)dy=e^{-b^2}$.

\subsection{Electrostatic dispersion relation}
We derive the electrostatic dispersion relation in this subsection base on Eq.(\ref{eq:esdr0}) and the distribution function (\ref{eq:f0s}).
The term
\begin{eqnarray}\nonumber
   k_\parallel \frac{\partial f_{s0}}{\partial v'_\parallel}+
  \frac{n\omega_{cs}}{v'_\perp}\frac{\partial f_{s0}}{\partial v'_\perp}=-2(v'_\parallel-v_{dsz})k_z\sum_{\sigma=a,b}\frac{r_{s\sigma}f_{s0\sigma}}{v_{zts}^2}
 -2(v'_\perp-v_{dsr}) \frac{n\omega_{cs}}{v'_\perp}\sum_{\sigma=a,b}\frac{r_{s\sigma}f_{s0\sigma}}{v_{\perp tsb}^2},
\end{eqnarray}
and thus
\begin{eqnarray}\nonumber
   \frac{v'_\perp\Big(k_\parallel \frac{\partial f_{s0}}{\partial v'_\parallel}+
  \frac{n\omega_{cs}}{v'_\perp}\frac{\partial f_{s0}}{\partial v'_\perp}\Big)}{\omega-k_\parallel v'_\parallel+{\color{red}i\nu_s-k_xv_{dsx}}-n\omega_{cs}} = \sum_{\sigma=a,b}\frac{2r_{s\sigma}f_{s0\sigma}}{v_{zts}^2}\frac{-(v'_\parallel-v_{dsz})k_zv'_\perp
 -(v'_\perp-v_{dsr})  n\omega_{cs} \lambda_{Ts\sigma}}{\omega-k_z(v'_\parallel-v_{dsz})-k_zv_{dsz}+{\color{red}i\nu_s-k_xv_{dsx}}-n\omega_{cs}},
\end{eqnarray}
and thus the magnetized $\sum_{s=m}$ term in Eq.(\ref{eq:esdr0}) is
\begin{eqnarray}\nonumber
  \sum_{s=m}\cdots&=& \sum_{s=m}\frac{\omega_{ps}^2}{k^2}\int_{-\infty}^{\infty}\int_{0}^{\infty} \sum_{n=-\infty}^{\infty}J_n^2(y_s) \sum_{\sigma=a,b}\frac{4\pi r_{s\sigma}f'_{s0\sigma}}{v_{zts}^2}\frac{-v'_\parallel k_zv'_\perp
 -(v'_\perp-v_{dsr})  n\omega_{cs} \lambda_{Ts\sigma}}{\omega-k_z v'_\parallel-k_zv_{dsz}+{\color{red}i\nu_s-k_xv_{dsx}}-n\omega_{cs}}dv'_\perp dv'_\parallel\\\nonumber
 &=& -\sum_{s=m}\frac{\omega_{ps}^2}{k^2}\int_{-\infty}^{\infty}\int_{0}^{\infty} \sum_{n=-\infty}^{\infty}J_n^2(y_s) \sum_{\sigma=a,b}\frac{4\pi r_{s\sigma}f'_{s0\sigma}}{v_{zts}^2}\frac{v'_\parallel k_zv'_\perp
 +(v'_\perp-v_{dsr})  n\omega_{cs} \lambda_{Ts\sigma}}{\omega_{sn}-k_z v'_\parallel}dv'_\perp dv'_\parallel\\\nonumber
 &=& \sum_{s=m}\frac{\omega_{ps}^2}{k^2}\int_{0}^{\infty} \sum_{n=-\infty}^{\infty}J_n^2(y_s) \sum_{\sigma=a,b}\frac{4\pi r_{s\sigma}f_{s0\perp\sigma}}{v_{zts}^2}\Big\{v'_\perp[1+\zeta_{sn}Z(\zeta_{sn})]+(v'_\perp-v_{dsr})  n\omega_{cs} \lambda_{Ts\sigma}\frac{Z(\zeta_{sn})}{k_zv_{zts}}\Big\}dv'_\perp \\\nonumber
 &=& \sum_{s=m}\frac{\omega_{ps}^2}{k^2v_{zts}^2}\sum_{n=-\infty}^{\infty} \sum_{\sigma=a,b}\frac{4 r_{s\sigma}}{A_{s\sigma}}\Big\{Z_1(\zeta_{sn})A_n(a_{s\sigma},b_{s\sigma},0) + \frac{n\omega_{cs} \lambda_{Ts\sigma}}{k_zv_{zts}}Z_0(\zeta_{sn})A_n(a_{s\sigma},b_{s\sigma},b_{s\sigma})\Big\}\\
 &=& \sum_{s=m}\frac{\omega_{ps}^2}{k^2v_{zts}^2}\sum_{n=-\infty}^{\infty} \sum_{\sigma=a,b}r_{s\sigma}\Big\{Z_1(\zeta_{sn})A_{n0\sigma} + \frac{n\omega_{cs} \lambda_{Ts\sigma}}{k_zv_{zts}}Z_0(\zeta_{sn})A_{nbs\sigma}\Big\} ,
\end{eqnarray}
We find the above result is the same as in Ref.\cite{Umeda2007} ring-beam case, except that our new (1) $\omega_{sn}=\omega-k_zv_{dsz}-n\omega_{cs}{\color{red}-k_xv_{dsx}+i\nu_s}$, and (2) the {\color{red}summation for loss cone}.

For the unmagnetized species
\begin{eqnarray}\nonumber
  \bm k\cdot\frac{\partial f_{s0}}{\partial \bm v}&=&-2\Big[1-\frac{v_{dsr}}{\sqrt{(v_x-v_{dsx})^2+(v_y-v_{dsy})^2}}\Big]k_x(v_x-v_{dsx})\sum_{\sigma=a,b}\frac{r_{s\sigma}f_{s0\sigma}}{v_{\perp ts\sigma}^2}-2k_z(v_z-v_{dsz})\sum_{\sigma=a,b}\frac{r_{s\sigma}f_{s0\sigma}}{v_{zts}^2}\\\nonumber
 &=&-2\sum_{\sigma=a,b}\Big[\Big(1-\frac{b_{s\sigma}}{\sqrt{x'^2+y'^2}}\Big)\frac{k_xx'}{v_{\perp ts\sigma}}+\frac{k_zz'}{v_{zts}} \Big]r_{s\sigma}f_{s0\sigma}\\\nonumber
 &=&-2\sum_{\sigma=a,b}\Big[\Big(1-\frac{b_{s\sigma}}{\sqrt{x'^2+y'^2}}\Big)\frac{k_xx'}{v_{\perp ts\sigma}}+\frac{k_zz'}{v_{zts}} \Big]\frac{r_{s\sigma}}{A_{s\sigma}}\frac{1}{\pi^{3/2}v_{zts}v_{\perp ts\sigma}^2}e^{-z'^2-(\sqrt{x'^2+y'^2}-b_{s\sigma})^2},
\end{eqnarray}
where we have chosen the transformation
\begin{equation}
    \left\{ \begin{array}{l}
    x'=\frac{(v_x-v_{dsx})}{v_{\perp ts\sigma}} \\
    y'=\frac{(v_y-v_{dsy})}{v_{\perp ts\sigma}} \\
    z'=\frac{(v_z-v_{dsz})}{v_{zts}}
      \end{array}\right.,~~~~~~~
          \left\{ \begin{array}{l}
    v_x=v_{\perp ts\sigma}x'+v_{dsx} \\
    v_y=v_{\perp ts\sigma}y'+v_{dsy} \\
    v_z=v_{zts}z'+v_{dsz} 
      \end{array}\right..
\end{equation}
Further consider 
\begin{eqnarray}\nonumber
 \omega-\bm k\cdot\bm v +i\nu_s&=&\omega-k_xv_x-k_zv_z +i\nu_s\\\nonumber
 &=&\omega-k_xv_{\perp ts\sigma}x'-k_zv_{zts}z' -k_xv_{dsx}-k_zv_{dsz}+i\nu_s\\
 &=&\omega-kv_{ts\sigma}x -k_xv_{dsx}-k_zv_{dsz}+i\nu_s,
\end{eqnarray}
we need a new transformation to let the two variables $k_xv_{\perp ts\sigma}x'+k_zv_{zts}z'$ change to a single variable $kv_{ts\sigma}x$, which is
\begin{equation}
    \left\{ \begin{array}{l}
    x=\frac{k_xv_{\perp ts\sigma}}{kv_{ts\sigma}}x'+\frac{k_zv_{zts}}{kv_{ts\sigma}}z' \\
    y=y' \\
    z=\frac{k_zv_{zts}}{kv_{ts\sigma}}x'-\frac{k_xv_{\perp ts\sigma}}{kv_{ts\sigma}}z'
      \end{array}\right.,~~~~~~
    \left\{ \begin{array}{l}
    x'=\frac{k_xv_{\perp ts\sigma}}{kv_{ts\sigma}}x+\frac{k_zv_{zts}}{kv_{ts\sigma}}z\\
    y'=y \\
    z'=\frac{k_zv_{zts}}{kv_{ts\sigma}}x-\frac{k_xv_{\perp ts\sigma}}{kv_{ts\sigma}}z
      \end{array}\right.,
\end{equation}
i.e.,  
\begin{equation}
    \left\{ \begin{array}{l}
    v_x=v_{\perp ts\sigma}\frac{k_xv_{\perp ts\sigma}}{kv_{ts\sigma}}x+v_{\perp ts\sigma}\frac{k_zv_{zts}}{kv_{ts\sigma}}z+v_{dsx}\\
    v_y=v_{\perp ts\sigma}y +v_{dsy}\\
    v_z=v_{zts}\frac{k_zv_{zts}}{kv_{ts\sigma}}x-v_{zts}\frac{k_xv_{\perp ts\sigma}}{kv_{ts\sigma}}z+v_{dsz}
      \end{array}\right.,~~~~~~~~~~dv_xdv_ydv_z={\color{red}-} v_{\perp ts\sigma}^2v_{zts}dxdydz,
\end{equation}
where $kv_{ts\sigma}=\sqrt{k_x^2v_{\perp ts\sigma}^2+k_z^2v_{zts}^2}=v_{zts}\sqrt{k_x^2{\color{red}/\lambda_{Ts\sigma}}+k_z^2}$. {\color{red}To transform the integral to be able to use the $Z$ function, we need $e^{-x^2}$ term be separate. It is not easy to do so, except only when we set $v_{dsr}=0$, i.e., $b_{s\sigma}=0$ and $A_{s\sigma}=1$. At $v_{dsr}=0$ case, we can have $z'^2+(\sqrt{x'^2+y'^2}-b_{s\sigma})^2=x'^2+y'^2+z'^2=x^2+y^2+z^2$. Thus, to make life easy, we set $v_{dsr}=0$ for unmagnetized study\footnote{It is also rare to meet ring beam in unmagnetized case. Thus, using $v_{dsr}=0$ for unmagnetized species will not limit too much to the application of present model. Actually, our present model is much general than Ref.\cite{Muschietti2017} and the non-relativistic case in Ref.\cite{Ruyer2014}.}.}

We define $\zeta_{s\sigma}=\frac{\omega-k_xv_{dsx}-k_zv_{dsz}+i\nu_s}{kv_{ts\sigma}}$, and thus $\omega-k_xv_x-k_zv_z+i\nu_s=kv_{ts\sigma}(\zeta_{s\sigma}-x)$.
The unmagnetized $\sum_{s=u}$ term in Eq.(\ref{eq:esdr0}) is
\begin{eqnarray}\nonumber
   \sum_{s=u}\cdots &=&\sum_{s=u} \frac{\omega_{ps}^2}{k^2}\int dv^3 \frac{-2\sum_{\sigma=a,b}\Big[\frac{k_xx'}{v_{\perp ts\sigma}}+\frac{k_zz'}{v_{zts}} \Big] \frac{r_{s\sigma}}{\pi^{3/2}v_{zts}v_{\perp ts\sigma}^2}e^{-(x^2+y^2+z^2)},}{(\omega-\bm k\cdot\bm v +i\nu_s)}\\\nonumber
   &=&\sum_{s=u} \frac{\omega_{ps}^2}{k^2}2\sum_{\sigma=a,b}\int_{-\infty}^{\infty} dxdydz \frac{\Big[\frac{k^2}{kv_{ts\sigma}}x+\frac{k_xk_z}{kv_{ts\sigma}}\frac{v_{\perp ts\sigma}}{v_{zts}}(\lambda_{Ts\sigma}-1)z \Big]\frac{r_{s\sigma}}{\pi^{3/2}}e^{-(x^2+y^2+z^2)},}{kv_{ts\sigma}(\zeta_\sigma-x)}\\\nonumber
   &=&\sum_{s=u} \frac{\omega_{ps}^2}{k^2}\sum_{\sigma=a,b}\frac{2r_{s\sigma}}{v_{ts\sigma}^2}\frac{1}{\pi^{3/2}}\int_{-\infty}^{\infty} dxdydz \frac{xe^{-(x^2+y^2+z^2)},}{(\zeta_\sigma-x)}\\
   &=&\sum_{s=u} \frac{\omega_{ps}^2}{k^2}\sum_{\sigma=a,b}\frac{2r_{s\sigma}}{v_{ts\sigma}^2}[1+\zeta_{s\sigma} Z(\zeta_{s\sigma})]=\sum_{s=u} \frac{\omega_{ps}^2}{k^2}\sum_{\sigma=a,b}\frac{2r_{s\sigma}}{v_{ts\sigma}^2}Z_1(\zeta_{s\sigma}),
\end{eqnarray}
where we have used $\frac{k_xx'}{v_{\perp ts\sigma}}+\frac{k_zz'}{v_{zts}}=\frac{k_x}{v_{\perp ts\sigma}}[\frac{k_xv_{\perp ts\sigma}}{kv_{ts\sigma}}x+\frac{k_zv_{zts}}{kv_{ts\sigma}}z]+\frac{k_z}{v_{zts}}[\frac{k_zv_{zts}}{kv_{ts\sigma}}x-\frac{k_xv_{\perp ts\sigma}}{kv_{ts\sigma}}z]=\frac{k^2}{kv_{ts\sigma}}x+\frac{k_xk_z}{kv_{ts\sigma}}\frac{v_{zts}}{v_{\perp ts\sigma}}z-\frac{k_xk_z}{kv_{ts\sigma}}\frac{v_{\perp ts\sigma}}{v_{zts}}z=\frac{k^2}{kv_{ts\sigma}}x+\frac{k_xk_z}{kv_{ts\sigma}}\frac{v_{\perp ts\sigma}}{v_{zts}}(\lambda_{Ts\sigma}-1)z$. We note that the $\frac{k_xk_z}{kv_{ts\sigma}}\frac{v_{\perp ts\sigma}}{v_{zts}}(\lambda_{Ts\sigma}-1)z$ term is integralled to be zero which makes the result simplified a lot.

Thus we obtain the final electrostatic dispersion relation
\begin{eqnarray}\label{eq:esdr1}\nonumber
  D(\omega,\bm k)&=&1+\sum_{s=m}\frac{\omega_{ps}^2}{k^2v_{zts}^2}\sum_{n=-\infty}^{\infty} \sum_{\sigma=a,b}r_{s\sigma}\Big\{Z_1(\zeta_{sn})A_{n0\sigma} + \frac{n\omega_{cs} \lambda_{Ts\sigma}}{k_zv_{zts}}Z_0(\zeta_{sn})A_{nbs\sigma}\Big\} \\
  &&+\sum_{s=u} \frac{\omega_{ps}^2}{k^2}\sum_{\sigma=a,b}\frac{2r_{s\sigma}}{v_{ts\sigma}^2}Z_1(\zeta_{s\sigma})=0.
\end{eqnarray}
The above dispersion relation Eq.(\ref{eq:esdr1}) is very general, except that required $v_{dsr}=0$ for unmagnetized species.

\subsection{Electromagnetic dispersion relation}

The electromagnetic case is much complicated than the electrostatic case.

\subsubsection{Unmagnetized terms}
We start from the unmagnetized term firstly. We use the same transformation for $\bm v$ as in the unmagnetized electrostatic case. Again, we assume $v_{dsr}=0$ for unmagnetized species, and calculate term by term:
\begin{eqnarray*}
   \bm k\cdot\frac{\partial f_{s0}}{\partial \bm v}&=&-2\sum_{\sigma=a,b}\Big[\frac{k_xx'}{v_{\perp ts\sigma}}+\frac{k_zz'}{v_{zts}} \Big]\frac{r_{s\sigma}}{\pi^{3/2}v_{zts}v_{\perp ts\sigma}^2}e^{-z'^2-(\sqrt{x'^2+y'^2}-b_{s\sigma})^2}\\
   &=&-2\sum_{\sigma=a,b}\Big[\frac{k^2}{kv_{ts\sigma}}x+\frac{k_xk_z}{kv_{ts\sigma}}\frac{v_{\perp ts\sigma}}{v_{zts}}(\lambda_{Ts\sigma}-1)z \Big]\frac{r_{s\sigma}}{\pi^{3/2}v_{zts}v_{\perp ts\sigma}^2}e^{-(x^2+y^2+z^2)},
\end{eqnarray*}
\begin{eqnarray*}
   \frac{\partial f_{s0}}{\partial v_x}&=&-2(v_x-v_{dsx})\sum_{\sigma=a,b}\frac{r_{s\sigma}f_{s0\sigma}}{v_{\perp ts\sigma}^2}= -2\sum_{\sigma=a,b}\Big[\frac{k_x}{kv_{ts\sigma}}x+\sqrt{\lambda_{Ts\sigma}}\frac{k_z}{kv_{ts\sigma}}z \Big]\frac{r_{s\sigma}}{\pi^{3/2}v_{zts}v_{\perp ts\sigma}^2}e^{-(x^2+y^2+z^2)},\\\nonumber
   \frac{\partial f_{s0}}{\partial v_y}&=&-2(v_y-v_{dsx})\sum_{\sigma=a,b}\frac{r_{s\sigma}f_{s0\sigma}}{v_{\perp ts\sigma}^2}=-2\sum_{\sigma=a,b}\frac{1}{v_{\perp ts\sigma}}y\frac{r_{s\sigma}}{\pi^{3/2}v_{zts}v_{\perp ts\sigma}^2}e^{-(x^2+y^2+z^2)},\\\nonumber
   \frac{\partial f_{s0}}{\partial v_z}&=&-2(v_z-v_{dsz})\sum_{\sigma=a,b}\frac{r_{s\sigma}f_{s0\sigma}}{v_{zts}^2}=-2\sum_{\sigma=a,b}\Big[\frac{k_z}{kv_{ts\sigma}}x-\frac{k_x}{kv_{ts\sigma}}\frac{1}{\sqrt{\lambda_{Ts\sigma}}}z \Big]\frac{r_{s\sigma}}{\pi^{3/2}v_{zts}v_{\perp ts\sigma}^2}e^{-(x^2+y^2+z^2)}.
\end{eqnarray*}
Define $F_{s\sigma}\equiv \frac{r_{s\sigma}}{\pi^{3/2}v_{zts}v_{\perp ts\sigma}^2}e^{-(x^2+y^2+z^2)}$ (note: $A_{s\sigma}=1$), we have
\begin{eqnarray*}
   (\bm v \frac{\partial f_{s0}}{\partial \bm v}) &=& \left( \begin{array}{c}
 v_x \\
v_y \\
 v_z
    \end{array}\right)(\frac{\partial f_{s0}}{\partial v_x},\frac{\partial f_{s0}}{\partial v_y},\frac{\partial f_{s0}}{\partial v_z})\\
    &=&-2\sum_{\sigma=a,b}F_{s\sigma}\left( \begin{array}{c}
 v_x \\
v_y \\
 v_z
    \end{array}\right)(\Big[\frac{k_x}{kv_{ts\sigma}}x+\sqrt{\lambda_{Ts\sigma}}\frac{k_z}{kv_{ts\sigma}}z \Big] ,  \frac{1}{v_{\perp ts\sigma}}y,
\Big[\frac{k_z}{kv_{ts\sigma}}x-\frac{k_x}{kv_{ts\sigma}}\frac{1}{\sqrt{\lambda_{Ts\sigma}}}z \Big])\\
    &\to& -2\sum_{\sigma=a,b}F_{s\sigma}\left( \begin{array}{ccc}
  \frac{k_x^2v_{\perp ts\sigma}^2}{k^2v_{ts\sigma}^2}x^2+\frac{k_xv_{dsx}}{kv_{ts\sigma}}x+\frac{k_z^2v_{zts}^2}{k^2v_{ts\sigma}^2}z^2 & 0 &    \frac{k_xk_zv_{\perp ts\sigma}^2}{k^2v_{ts\sigma}^2}x^2 +\frac{k_zv_{dsx}}{kv_{ts\sigma}}x - \frac{k_xk_zv_{\perp ts\sigma}^2}{k^2v_{ts\sigma}^2}z^2 \\
\frac{k_xv_{dsy}}{kv_{ts\sigma}}x & y^2 & \frac{k_zv_{dsy}}{kv_{ts\sigma}}x\\
     \frac{k_xk_zv_{zts}^2}{k^2v_{ts\sigma}^2}x^2+ \frac{k_xv_{dsz}}{kv_{ts\sigma}}x- \frac{k_z}{kv_{ts\sigma}}\frac{k_xv_{zts}^2}{kv_{ts\sigma}}z^2 & 0 & \frac{k_z^2v_{zts}^2}{k^2v_{ts\sigma}^2}x^2+\frac{k_zv_{dsz}}{kv_{ts\sigma}}x +\frac{k_x^2v_{\perp ts\sigma}^2}{k^2v_{ts\sigma}^2}z^2
    \end{array}\right)
\end{eqnarray*}
Note also:
\begin{eqnarray*}
   (\bm v \frac{\partial f_{s0}}{\partial \bm v}) &\neq& \left( \begin{array}{c}
 \frac{\partial f_{s0}}{\partial v_x} \\
 \frac{\partial f_{s0}}{\partial v_y} \\
 \frac{\partial f_{s0}}{\partial v_z}
    \end{array}\right)(v_x,v_y,v_z)=-2\sum_{\sigma=a,b}F_{s\sigma}\left( \begin{array}{c}
 \Big[\frac{k_x}{kv_{ts\sigma}}x+\sqrt{\lambda_{Ts\sigma}}\frac{k_z}{kv_{ts\sigma}}z \Big] \\
 \frac{1}{v_{\perp ts\sigma}}y \\
\Big[\frac{k_z}{kv_{ts\sigma}}x-\frac{k_x}{kv_{ts\sigma}}\frac{1}{\sqrt{\lambda_{Ts\sigma}}}z \Big]
    \end{array}\right)(v_x,v_y,v_z)\\
    &\to& -2\sum_{\sigma=a,b}F_{s\sigma}\left( \begin{array}{ccc}
  \frac{k_x^2v_{\perp ts\sigma}^2}{k^2v_{ts\sigma}^2}x^2+\frac{k_xv_{dsx}}{kv_{ts\sigma}}x+\frac{k_z^2v_{zts}^2}{k^2v_{ts\sigma}^2}z^2 & \frac{k_xv_{dsy}}{kv_{ts\sigma}}x & \frac{k_xk_zv_{zts}^2}{k^2v_{ts\sigma}^2}x^2+ \frac{k_xv_{dsz}}{kv_{ts\sigma}}x- \frac{k_z}{kv_{ts\sigma}}\frac{k_xv_{zts}^2}{kv_{ts\sigma}}z^2\\
0 & y^2 & 0\\
\frac{k_xk_zv_{\perp ts\sigma}^2}{k^2v_{ts\sigma}^2}x^2 +\frac{k_zv_{dsx}}{kv_{ts\sigma}}x - \frac{k_xk_zv_{\perp ts\sigma}^2}{k^2v_{ts\sigma}^2}z^2 & \frac{k_zv_{dsy}}{kv_{ts\sigma}}x & \frac{k_z^2v_{zts}^2}{k^2v_{ts\sigma}^2}x^2+\frac{k_zv_{dsz}}{kv_{ts\sigma}}x +\frac{k_x^2v_{\perp ts\sigma}^2}{k^2v_{ts\sigma}^2}z^2
    \end{array}\right)
\end{eqnarray*}
The odd function term from $y$ and $z$ can be integralled to vanish, and thus we have omitted them in the above final expressions. Note
\begin{equation}
    \left\{ \begin{array}{l}
    v_x=v_{\perp ts\sigma}\frac{k_xv_{\perp ts\sigma}}{kv_{ts\sigma}}x+v_{\perp ts\sigma}\frac{k_zv_{zts}}{kv_{ts\sigma}}z+v_{dsx}\\
    v_y=v_{\perp ts\sigma}y +v_{dsy}\\
    v_z=v_{zts}\frac{k_zv_{zts}}{kv_{ts\sigma}}x-v_{zts}\frac{k_xv_{\perp ts\sigma}}{kv_{ts\sigma}}z+v_{dsz}
      \end{array}\right.,
\end{equation}
and we have used:
\begin{itemize}

\item 
$[\frac{k_x}{kv_{ts\sigma}}x+\sqrt{\lambda_{Ts\sigma}}\frac{k_z}{kv_{ts\sigma}}z ](v_{\perp ts\sigma}\frac{k_xv_{\perp ts\sigma}}{kv_{ts\sigma}}x+v_{\perp ts\sigma}\frac{k_zv_{zts}}{kv_{ts\sigma}}z+v_{dsx}) \to \frac{k_x}{kv_{ts\sigma}}x(\frac{k_xv_{\perp ts\sigma}^2}{kv_{ts\sigma}}x+v_{dsx})+\frac{k_z^2v_{zts}^2}{k^2v_{ts\sigma}^2}z^2= \frac{k_x^2v_{\perp ts\sigma}^2}{k^2v_{ts\sigma}^2}x^2+\frac{k_xv_{dsx}}{kv_{ts\sigma}}x+\frac{k_z^2v_{zts}^2}{k^2v_{ts\sigma}^2}z^2$,

\item $ \Big[\frac{k_x}{kv_{ts\sigma}}x+\sqrt{\lambda_{Ts\sigma}}\frac{k_z}{kv_{ts\sigma}}z \Big](v_{zts}\frac{k_zv_{zts}}{kv_{ts\sigma}}x-v_{zts}\frac{k_xv_{\perp ts\sigma}}{kv_{ts\sigma}}z+v_{dsz}) \to  \frac{k_x}{kv_{ts\sigma}}x(v_{zts}\frac{k_zv_{zts}}{kv_{ts\sigma}}x+v_{dsz})- \frac{k_z}{kv_{ts\sigma}}\frac{k_xv_{zts}^2}{kv_{ts\sigma}}z^2= \frac{k_xk_zv_{zts}^2}{k^2v_{ts\sigma}^2}x^2+ \frac{k_xv_{dsz}}{kv_{ts\sigma}}x- \frac{k_z}{kv_{ts\sigma}}\frac{k_xv_{zts}^2}{kv_{ts\sigma}}z^2$, 

\item 
$\Big[\frac{k_z}{kv_{ts\sigma}}x-\frac{k_x}{kv_{ts\sigma}}\frac{1}{\sqrt{\lambda_{Ts\sigma}}}z \Big](v_{\perp ts\sigma}\frac{k_xv_{\perp ts\sigma}}{kv_{ts\sigma}}x+v_{\perp ts\sigma}\frac{k_zv_{zts}}{kv_{ts\sigma}}z+v_{dsx}) \to \frac{k_xk_zv_{\perp ts\sigma}^2}{k^2v_{ts\sigma}^2}x^2 +\frac{k_zv_{dsx}}{kv_{ts\sigma}}x - \frac{k_xk_zv_{\perp ts\sigma}^2}{k^2v_{ts\sigma}^2}z^2 $, 

\item 
$\Big[\frac{k_z}{kv_{ts\sigma}}x-\frac{k_x}{kv_{ts\sigma}}\frac{1}{\sqrt{\lambda_{Ts\sigma}}}z \Big](v_{zts}\frac{k_zv_{zts}}{kv_{ts\sigma}}x-v_{zts}\frac{k_xv_{\perp ts\sigma}}{kv_{ts\sigma}}z+v_{dsz}) \to \frac{k_z^2v_{zts}^2}{k^2v_{ts\sigma}^2}x^2+\frac{k_zv_{dsz}}{kv_{ts\sigma}}x +\frac{k_x^2v_{\perp ts\sigma}^2}{k^2v_{ts\sigma}^2}z^2 $.

\end{itemize}

And
\begin{eqnarray*}
   (\bm k\cdot\frac{\partial f_{s0}}{\partial \bm v})(\bm v\bm v) &=&-2\sum_{\sigma=a,b}F_{s\sigma}\Big[\frac{k^2}{kv_{ts\sigma}}x+\frac{k_xk_z}{kv_{ts\sigma}}\frac{v_{\perp ts\sigma}}{v_{zts}}(\lambda_{Ts\sigma}-1)z \Big] \left( \begin{array}{c}
 v_x \\
 v_y \\
 v_z
    \end{array}\right)(v_x,v_y,v_z)
\end{eqnarray*}
with terms in $\bm v\bm v$
\begin{itemize}

\item $v_xv_x \to [v_{\perp ts\sigma}\frac{k_xv_{\perp ts\sigma}}{kv_{ts\sigma}}x+v_{\perp ts\sigma}\frac{k_zv_{zts}}{kv_{ts\sigma}}z+v_{dsx}]^2$.

\item $v_xv_y=v_yv_x \to [v_{\perp ts\sigma}\frac{k_xv_{\perp ts\sigma}}{kv_{ts\sigma}}x+v_{\perp ts\sigma}\frac{k_zv_{zts}}{kv_{ts\sigma}}z+v_{dsx}]v_{dsy}$.

\item $v_xv_z=v_zv_x \to [v_{\perp ts\sigma}\frac{k_xv_{\perp ts\sigma}}{kv_{ts\sigma}}x+v_{\perp ts\sigma}\frac{k_zv_{zts}}{kv_{ts\sigma}}z+v_{dsx}][v_{zts}\frac{k_zv_{zts}}{kv_{ts\sigma}}x-v_{zts}\frac{k_xv_{\perp ts\sigma}}{kv_{ts\sigma}}z+v_{dsz}] $.

\item $v_yv_y\to [v_{\perp ts\sigma}^2y^2 +v_{dsy}^2]$.

\item $v_yv_z=v_zv_y \to  [v_{zts}\frac{k_zv_{zts}}{kv_{ts\sigma}}x-v_{zts}\frac{k_xv_{\perp ts\sigma}}{kv_{ts\sigma}}z+v_{dsz}]v_{dsy}$.

\item $v_zv_z\to [v_{zts}\frac{k_zv_{zts}}{kv_{ts\sigma}}x-v_{zts}\frac{k_xv_{\perp ts\sigma}}{kv_{ts\sigma}}z+v_{dsz}]^2$.

\end{itemize}

Define $\bm Q_s^u=\sum_{\sigma=a,b}r_{s\sigma}\frac{\omega_{ps}^2}{\omega^2}\bm P_s^u$, with $\bm P_s^u=\Big\{-\bm I +\int dv^3 \frac{\Big[-i\nu_s(\bm v \frac{\partial f_{s0\sigma}}{\partial \bm v})+(\bm k\cdot\frac{\partial f_{s0\sigma}}{\partial \bm v})(\bm v\bm v)\Big]}{(\omega-\bm k\cdot\bm v +i\nu_s)}\Big\}$, and $g= \frac{1}{\pi^{3/2} }e^{-(x^2+y^2+z^2)}$. We calculate term by term, similar to 
\begin{itemize}

\item 
$P_{s11}^u=\Big\{-\bm I + \int dv^3 \frac{\Big[-i\nu_s(\bm v \frac{\partial f_{s0\sigma}}{\partial \bm v})+(\bm k\cdot\frac{\partial f_{s0\sigma}}{\partial \bm v})(\bm v\bm v)\Big]}{(\omega-\bm k\cdot\bm v +i\nu_s)}\Big\}_{11}=-1 {\color{red}+} \\ 2 \int dxdydz g\frac{ -i\nu_s( \frac{k_x^2v_{\perp ts\sigma}^2}{k^2v_{ts\sigma}^2}x^2+\frac{k_xv_{dsx}}{kv_{ts\sigma}}x+\frac{k_z^2v_{zts}^2}{k^2v_{ts\sigma}^2}z^2)+\Big[\frac{k^2}{kv_{ts\sigma}}x+\frac{k_xk_z}{kv_{ts\sigma}}\frac{v_{\perp ts\sigma}}{v_{zts}}(\lambda_{Ts\sigma}-1)z \Big]  [v_{\perp ts\sigma}\frac{k_xv_{\perp ts\sigma}}{kv_{ts\sigma}}x+v_{\perp ts\sigma}\frac{k_zv_{zts}}{kv_{ts\sigma}}z+v_{dsx}]^2 }{kv_{ts\sigma}(\zeta_{s\sigma}- x)}=-1+\frac{2}{kv_{ts\sigma}}\Big\{-i\nu_s( \frac{k_x^2v_{\perp ts\sigma}^2}{k^2v_{ts\sigma}^2}Z_2+\frac{k_xv_{dsx}}{kv_{ts\sigma}}Z_1+\frac{k_z^2v_{zts}^2}{2k^2v_{ts\sigma}^2}Z_0)+ \frac{k^2}{kv_{ts\sigma}}(v_{\perp ts\sigma}\frac{k_xv_{\perp ts\sigma}}{kv_{ts\sigma}})^2Z_3+2\frac{k^2}{kv_{ts\sigma}}(v_{\perp ts\sigma}\frac{k_xv_{\perp ts\sigma}}{kv_{ts\sigma}})v_{dsx}Z_2+[\frac{1}{2}\frac{k^2}{kv_{ts\sigma}}(v_{\perp ts\sigma}\frac{k_zv_{zts}}{kv_{ts\sigma}} )^2  +\frac{k^2}{kv_{ts\sigma}}v_{dsx}^2+ \frac{k_xk_z}{kv_{ts\sigma}}\frac{v_{\perp ts\sigma}}{v_{zts}}(\lambda_{Ts\sigma}-1)(v_{\perp ts\sigma}\frac{k_xv_{\perp ts\sigma}}{kv_{ts\sigma}})v_{\perp ts\sigma}\frac{k_zv_{zts}}{kv_{ts\sigma}} ]Z_1 +  \frac{k_xk_z}{kv_{ts\sigma}}\frac{v_{\perp ts\sigma}}{v_{zts}}(\lambda_{Ts\sigma}-1)v_{\perp ts\sigma}\frac{k_zv_{zts}}{kv_{ts\sigma}} v_{dsx}Z_0\Big\}\\
=-1+\frac{2}{kv_{ts\sigma}}\Big\{-i\nu_s( \frac{k_x^2v_{\perp ts\sigma}^2}{k^2v_{ts\sigma}^2}Z_2+\frac{k_xv_{dsx}}{kv_{ts\sigma}}Z_1+\frac{k_z^2v_{zts}^2}{2k^2v_{ts\sigma}^2}Z_0)+ \frac{k^2k_x^2v_{\perp ts\sigma}^4}{k^3v_{ts\sigma}^3}Z_3+2\frac{k^2k_xv_{\perp ts\sigma}^2}{k^2v_{ts\sigma}^2}v_{dsx}Z_2+[\frac{1}{2} v_{\perp ts\sigma}^2\frac{k^2k_z^2v_{zts}^2}{k^3v_{ts\sigma}^3}   +\frac{k^2}{kv_{ts\sigma}}v_{dsx}^2+  (\lambda_{Ts\sigma}-1)\frac{k_x^2k_z^2v_{\perp ts\sigma}^4}{k^3v_{ts\sigma}^3}  ]Z_1 +  \frac{k_xk_z^2}{k^2v_{ts\sigma}^2}(\lambda_{Ts\sigma}-1)v_{\perp ts\sigma}^2 v_{dsx}Z_0\Big\}$,

\end{itemize}
we have others
\begin{itemize}

\item $P_{s12}^u=\Big\{-\bm I + \int dv^3 \frac{\Big[-i\nu_s(\bm v \frac{\partial f_{s0\sigma}}{\partial \bm v})+(\bm k\cdot\frac{\partial f_{s0\sigma}}{\partial \bm v})(\bm v\bm v)\Big]}{(\omega-\bm k\cdot\bm v +i\nu_s)}\Big\}_{12}=0 {\color{red}+} \\ 2 \int dxdydz g\frac{ -i\nu_s 0+\Big[\frac{k^2}{kv_{ts\sigma}}x+\frac{k_xk_z}{kv_{ts\sigma}}\frac{v_{\perp ts\sigma}}{v_{zts}}(\lambda_{Ts\sigma}-1)z \Big]  [v_{\perp ts\sigma}\frac{k_xv_{\perp ts\sigma}}{kv_{ts\sigma}}x+v_{\perp ts\sigma}\frac{k_zv_{zts}}{kv_{ts\sigma}}z+v_{dsx}]v_{dsy} }{kv_{ts\sigma}(\zeta_{s\sigma}- x)}=\frac{2}{kv_{ts\sigma}}\Big\{  \frac{k^2}{kv_{ts\sigma}}\frac{k_xv_{\perp ts\sigma}^2}{kv_{ts\sigma}}v_{dsy}Z_2+\frac{k^2}{kv_{ts\sigma}} v_{dsx}v_{dsy}Z_1+\frac{1}{2}\frac{k_xk_z}{kv_{ts\sigma}}\frac{v_{\perp ts\sigma}}{v_{zts}}(\lambda_{Ts\sigma}-1)v_{\perp ts\sigma}\frac{k_zv_{zts}}{kv_{ts\sigma}}v_{dsy}Z_0\Big\}\\
=\frac{2v_{dsy}}{kv_{ts\sigma}}\Big\{ \frac{k^2k_xv_{\perp ts\sigma}^2}{k^2v_{ts\sigma}^2} Z_2+\frac{k^2}{kv_{ts\sigma}} v_{dsx} Z_1+\frac{1}{2}\frac{k_xk_z^2}{k^2v_{ts\sigma}^2} (\lambda_{Ts\sigma}-1)v_{\perp ts\sigma}^2Z_0\Big\}$,

\item $P_{s21}^u=\Big\{-\bm I + \int dv^3 \frac{\Big[-i\nu_s(\bm v \frac{\partial f_{s0\sigma}}{\partial \bm v})+(\bm k\cdot\frac{\partial f_{s0\sigma}}{\partial \bm v})(\bm v\bm v)\Big]}{(\omega-\bm k\cdot\bm v +i\nu_s)}\Big\}_{21}=0 {\color{red}+} \\ 2 \int dxdydz g\frac{ -i\nu_s \frac{k_xv_{dsy}}{kv_{ts\sigma}}x+\Big[\frac{k^2}{kv_{ts\sigma}}x+\frac{k_xk_z}{kv_{ts\sigma}}\frac{v_{\perp ts\sigma}}{v_{zts}}(\lambda_{Ts\sigma}-1)z \Big]  [v_{\perp ts\sigma}\frac{k_xv_{\perp ts\sigma}}{kv_{ts\sigma}}x+v_{\perp ts\sigma}\frac{k_zv_{zts}}{kv_{ts\sigma}}z+v_{dsx}]v_{dsy} }{kv_{ts\sigma}(\zeta_{s\sigma}- x)}
\\=\frac{2}{kv_{ts\sigma}}\Big\{-i\nu_s\frac{k_xv_{dsy}}{kv_{ts\sigma}}Z_1+  \frac{k^2}{kv_{ts\sigma}}\frac{k_xv_{\perp ts\sigma}^2}{kv_{ts\sigma}}v_{dsy}Z_2+\frac{k^2}{kv_{ts\sigma}} v_{dsx}v_{dsy}Z_1+\frac{1}{2}\frac{k_xk_z}{kv_{ts\sigma}}\frac{v_{\perp ts\sigma}}{v_{zts}}(\lambda_{Ts\sigma}-1)v_{\perp ts\sigma}\frac{k_zv_{zts}}{kv_{ts\sigma}}v_{dsy}Z_0\Big\}\\
=\frac{2v_{dsy}}{kv_{ts\sigma}}\Big\{ -i\nu_s\frac{k_x}{kv_{ts\sigma}}Z_1+  \frac{k^2k_xv_{\perp ts\sigma}^2}{k^2v_{ts\sigma}^2} Z_2+ \frac{k^2}{kv_{ts\sigma}} v_{dsx} Z_1+ \frac{1}{2}\frac{k_xk_z^2}{k^2v_{ts\sigma}^2} (\lambda_{Ts\sigma}-1)v_{\perp ts\sigma}^2Z_0\Big\}$,

\item $P_{s22}^u=\Big\{-\bm I + \int dv^3 \frac{\Big[-i\nu_s(\bm v \frac{\partial f_{s0\sigma}}{\partial \bm v})+(\bm k\cdot\frac{\partial f_{s0\sigma}}{\partial \bm v})(\bm v\bm v)\Big]}{(\omega-\bm k\cdot\bm v +i\nu_s)}\Big\}_{22}=-1{\color{red}+} \\ 2 \int dxdydz g\frac{ -i\nu_s y^2+\Big[\frac{k^2}{kv_{ts\sigma}}x+\frac{k_xk_z}{kv_{ts\sigma}}\frac{v_{\perp ts\sigma}}{v_{zts}}(\lambda_{Ts\sigma}-1)z \Big]  [v_{\perp ts\sigma}^2y^2 +v_{dsy}^2] }{kv_{ts\sigma}(\zeta_{s\sigma}- x)}\\
=-1{\color{red}+} \frac{2}{kv_{ts\sigma}}\Big\{-i\nu_s\frac{1}{2}Z_0+ (\frac{1}{2}v_{\perp ts\sigma}^2+v_{dsy}^2)\frac{k^2}{kv_{ts\sigma}}  Z_1 \Big\}$,

\item 
$P_{s13}^u=\Big\{-\bm I + \int dv^3 \frac{\Big[-i\nu_s(\bm v \frac{\partial f_{s0\sigma}}{\partial \bm v})+(\bm k\cdot\frac{\partial f_{s0\sigma}}{\partial \bm v})(\bm v\bm v)\Big]}{(\omega-\bm k\cdot\bm v +i\nu_s)}\Big\}_{13}=0{\color{red}+} 2 \int dxdydz g\\ \frac{ -i\nu_s( \frac{k_xk_zv_{\perp ts\sigma}^2}{k^2v_{ts\sigma}^2}x^2 +\frac{k_zv_{dsx}}{kv_{ts\sigma}}x - \frac{k_xk_zv_{\perp ts\sigma}^2}{k^2v_{ts\sigma}^2}z^2)+\Big[\frac{k^2}{kv_{ts\sigma}}x+\frac{k_xk_z}{kv_{ts\sigma}}\frac{v_{\perp ts\sigma}}{v_{zts}}(\lambda_{Ts\sigma}-1)z \Big]  [\frac{k_xv_{\perp ts\sigma}^2}{kv_{ts\sigma}}x+v_{\perp ts\sigma}\frac{k_zv_{zts}}{kv_{ts\sigma}}z+v_{dsx}][\frac{k_zv_{zts}^2}{kv_{ts\sigma}}x-v_{zts}\frac{k_xv_{\perp ts\sigma}}{kv_{ts\sigma}}z+v_{dsz}]  }{kv_{ts\sigma}(\zeta_{s\sigma}- x)}\\
= \frac{2}{kv_{ts\sigma}}\Big\{-i\nu_s( \frac{k_xk_zv_{\perp ts\sigma}^2}{k^2v_{ts\sigma}^2}Z_2 +\frac{k_zv_{dsx}}{kv_{ts\sigma}}Z_1 - \frac{k_xk_zv_{\perp ts\sigma}^2}{2k^2v_{ts\sigma}^2}Z_0)+ \frac{k^2}{kv_{ts\sigma}}   \frac{k_xv_{\perp ts\sigma}^2}{kv_{ts\sigma}} \frac{k_zv_{zts}^2}{kv_{ts\sigma}}Z_3 +\frac{k^2}{kv_{ts\sigma}}  [ \frac{k_xv_{\perp ts\sigma}^2}{kv_{ts\sigma}}v_{dsz} +v_{dsx}\frac{k_zv_{zts}^2}{kv_{ts\sigma}}]Z_2 +[\frac{k^2}{kv_{ts\sigma}}  v_{dsx}v_{dsz}-\frac{1}{2}\frac{k^2}{kv_{ts\sigma}}  v_{\perp ts\sigma}\frac{k_zv_{zts}}{kv_{ts\sigma}} v_{zts}\frac{k_xv_{\perp ts\sigma}}{kv_{ts\sigma}}  -\frac{1}{2}\frac{k_xk_z}{kv_{ts\sigma}}\frac{v_{\perp ts\sigma}}{v_{zts}}(\lambda_{Ts\sigma}-1)\frac{k_xv_{\perp ts\sigma}^2}{kv_{ts\sigma}}v_{zts}\frac{k_xv_{\perp ts\sigma}}{kv_{ts\sigma}} +\frac{1}{2}\frac{k_xk_z}{kv_{ts\sigma}}\frac{v_{\perp ts\sigma}}{v_{zts}}(\lambda_{Ts\sigma}-1)v_{\perp ts\sigma}\frac{k_zv_{zts}}{kv_{ts\sigma}} \frac{k_zv_{zts}^2}{kv_{ts\sigma}}]Z_1 +[\frac{1}{2}\frac{k_xk_z}{kv_{ts\sigma}}\frac{v_{\perp ts\sigma}}{v_{zts}}(\lambda_{Ts\sigma}-1)v_{\perp ts\sigma}\frac{k_zv_{zts}}{kv_{ts\sigma}}v_{dsz} -\frac{1}{2}\frac{k_xk_z}{kv_{ts\sigma}}\frac{v_{\perp ts\sigma}}{v_{zts}}(\lambda_{Ts\sigma}-1)v_{dsx}v_{zts}\frac{k_xv_{\perp ts\sigma}}{kv_{ts\sigma}}]Z_0\Big\}\\
= \frac{2}{kv_{ts\sigma}}\Big\{-i\nu_s( \frac{k_xk_zv_{\perp ts\sigma}^2}{k^2v_{ts\sigma}^2}Z_2 +\frac{k_zv_{dsx}}{kv_{ts\sigma}}Z_1 - \frac{k_xk_zv_{\perp ts\sigma}^2}{2k^2v_{ts\sigma}^2}Z_0)
+  \frac{k^2k_xk_zv_{zts}^2v_{\perp ts\sigma}^2}{k^3v_{ts\sigma}^3}Z_3 
+\frac{k^2}{kv_{ts\sigma}}  [ \frac{k_xv_{\perp ts\sigma}^2}{kv_{ts\sigma}}v_{dsz} +v_{dsx}\frac{k_zv_{zts}^2}{kv_{ts\sigma}}]Z_2 
+[\frac{k^2}{kv_{ts\sigma}}  v_{dsx}v_{dsz}
-\frac{1}{2}\frac{k^2k_xk_zv_{zts}^2v_{\perp ts\sigma}^2}{k^3v_{ts\sigma}^3} 
 -\frac{1}{2} (\lambda_{Ts\sigma}-1) \frac{k_x^3k_zv_{\perp ts\sigma}^4}{k^3v_{ts\sigma}^3} 
+\frac{1}{2}(\lambda_{Ts\sigma}-1)\frac{k_xk_z^3v_{zts}^2v_{\perp ts\sigma}^2}{k^3v_{ts\sigma}^3}]Z_1 
+[\frac{1}{2}(\lambda_{Ts\sigma}-1)\frac{k_xk_z^2v_{\perp ts\sigma}^2}{k^2v_{ts\sigma}^2}v_{dsz} -\frac{1}{2} (\lambda_{Ts\sigma}-1)v_{dsx} \frac{k_x^2k_zv_{\perp ts\sigma}^2}{k^2v_{ts\sigma}^2}]Z_0\Big\}$,

\item 
$P_{s31}^u=\Big\{-\bm I + \int dv^3 \frac{\Big[-i\nu_s(\bm v \frac{\partial f_{s0\sigma}}{\partial \bm v})+(\bm k\cdot\frac{\partial f_{s0\sigma}}{\partial \bm v})(\bm v\bm v)\Big]}{(\omega-\bm k\cdot\bm v +i\nu_s)}\Big\}_{31}=0{\color{red}+} 2 \int dxdydz g\\ \frac{ -i\nu_s(  \frac{k_xk_zv_{zts}^2}{k^2v_{ts\sigma}^2}x^2+ \frac{k_xv_{dsz}}{kv_{ts\sigma}}x- \frac{k_xk_zv_{zts}^2}{k^2v_{ts\sigma}^2}z^2)+\Big[\frac{k^2}{kv_{ts\sigma}}x+\frac{k_xk_z}{kv_{ts\sigma}}\frac{v_{\perp ts\sigma}}{v_{zts}}(\lambda_{Ts\sigma}-1)z \Big]  [\frac{k_xv_{\perp ts\sigma}^2}{kv_{ts\sigma}}x+v_{\perp ts\sigma}\frac{k_zv_{zts}}{kv_{ts\sigma}}z+v_{dsx}][\frac{k_zv_{zts}^2}{kv_{ts\sigma}}x-v_{zts}\frac{k_xv_{\perp ts\sigma}}{kv_{ts\sigma}}z+v_{dsz}]  }{kv_{ts\sigma}(\zeta_{s\sigma}- x)}\\
= \frac{2}{kv_{ts\sigma}}\Big\{-i\nu_s(  \frac{k_xk_zv_{zts}^2}{k^2v_{ts\sigma}^2}Z_2+ \frac{k_xv_{dsz}}{kv_{ts\sigma}}Z_1- \frac{k_xk_zv_{zts}^2}{2k^2v_{ts\sigma}^2}Z_0)+ \frac{k^2}{kv_{ts\sigma}}   \frac{k_xv_{\perp ts\sigma}^2}{kv_{ts\sigma}} \frac{k_zv_{zts}^2}{kv_{ts\sigma}}Z_3 +\frac{k^2}{kv_{ts\sigma}}  [ \frac{k_xv_{\perp ts\sigma}^2}{kv_{ts\sigma}}v_{dsz} +v_{dsx}\frac{k_zv_{zts}^2}{kv_{ts\sigma}}]Z_2 +[\frac{k^2}{kv_{ts\sigma}}  v_{dsx}v_{dsz}-\frac{1}{2}\frac{k^2}{kv_{ts\sigma}}  v_{\perp ts\sigma}\frac{k_zv_{zts}}{kv_{ts\sigma}} v_{zts}\frac{k_xv_{\perp ts\sigma}}{kv_{ts\sigma}}  -\frac{1}{2}\frac{k_xk_z}{kv_{ts\sigma}}\frac{v_{\perp ts\sigma}}{v_{zts}}(\lambda_{Ts\sigma}-1)\frac{k_xv_{\perp ts\sigma}^2}{kv_{ts\sigma}}v_{zts}\frac{k_xv_{\perp ts\sigma}}{kv_{ts\sigma}} +\frac{1}{2}\frac{k_xk_z}{kv_{ts\sigma}}\frac{v_{\perp ts\sigma}}{v_{zts}}(\lambda_{Ts\sigma}-1)v_{\perp ts\sigma}\frac{k_zv_{zts}}{kv_{ts\sigma}} \frac{k_zv_{zts}^2}{kv_{ts\sigma}}]Z_1 +[\frac{1}{2}\frac{k_xk_z}{kv_{ts\sigma}}\frac{v_{\perp ts\sigma}}{v_{zts}}(\lambda_{Ts\sigma}-1)v_{\perp ts\sigma}\frac{k_zv_{zts}}{kv_{ts\sigma}}v_{dsz} -\frac{1}{2}\frac{k_xk_z}{kv_{ts\sigma}}\frac{v_{\perp ts\sigma}}{v_{zts}}(\lambda_{Ts\sigma}-1)v_{dsx}v_{zts}\frac{k_xv_{\perp ts\sigma}}{kv_{ts\sigma}}]Z_0\Big\}\\
= \frac{2}{kv_{ts\sigma}}\Big\{-i\nu_s(  \frac{k_xk_zv_{zts}^2}{k^2v_{ts\sigma}^2}Z_2+ \frac{k_xv_{dsz}}{kv_{ts\sigma}}Z_1- \frac{k_xk_zv_{zts}^2}{2k^2v_{ts\sigma}^2}Z_0)+  \frac{k^2k_xk_zv_{zts}^2v_{\perp ts\sigma}^2}{k^3v_{ts\sigma}^3}Z_3 
+\frac{k^2}{kv_{ts\sigma}}  [ \frac{k_xv_{\perp ts\sigma}^2}{kv_{ts\sigma}}v_{dsz} +v_{dsx}\frac{k_zv_{zts}^2}{kv_{ts\sigma}}]Z_2 
+[\frac{k^2}{kv_{ts\sigma}}  v_{dsx}v_{dsz}
-\frac{1}{2}\frac{k^2k_xk_zv_{zts}^2v_{\perp ts\sigma}^2}{k^3v_{ts\sigma}^3} 
 -\frac{1}{2} (\lambda_{Ts\sigma}-1) \frac{k_x^3k_zv_{\perp ts\sigma}^4}{k^3v_{ts\sigma}^3} 
+\frac{1}{2}(\lambda_{Ts\sigma}-1)\frac{k_xk_z^3v_{zts}^2v_{\perp ts\sigma}^2}{k^3v_{ts\sigma}^3}]Z_1 
+[\frac{1}{2}(\lambda_{Ts\sigma}-1)\frac{k_xk_z^2v_{\perp ts\sigma}^2}{k^2v_{ts\sigma}^2}v_{dsz} -\frac{1}{2} (\lambda_{Ts\sigma}-1)v_{dsx} \frac{k_x^2k_zv_{\perp ts\sigma}^2}{k^2v_{ts\sigma}^2}]Z_0\Big\}$,

\item $P_{s23}^u=\Big\{-\bm I + \int dv^3 \frac{\Big[-i\nu_s(\bm v \frac{\partial f_{s0\sigma}}{\partial \bm v})+(\bm k\cdot\frac{\partial f_{s0\sigma}}{\partial \bm v})(\bm v\bm v)\Big]}{(\omega-\bm k\cdot\bm v +i\nu_s)}\Big\}_{23}=0 {\color{red}+} \\ 2 \int dxdydz g\frac{ -i\nu_s \frac{k_zv_{dsy}}{kv_{ts\sigma}}x+\Big[\frac{k^2}{kv_{ts\sigma}}x+\frac{k_xk_z}{kv_{ts\sigma}}\frac{v_{\perp ts\sigma}}{v_{zts}}(\lambda_{Ts\sigma}-1)z \Big]   [v_{zts}\frac{k_zv_{zts}}{kv_{ts\sigma}}x-v_{zts}\frac{k_xv_{\perp ts\sigma}}{kv_{ts\sigma}}z+v_{dsz}]v_{dsy} }{kv_{ts\sigma}(\zeta_{s\sigma}- x)}
\\=\frac{2}{kv_{ts\sigma}}\Big\{-i\nu_s\frac{k_zv_{dsy}}{kv_{ts\sigma}}Z_1+ \frac{k^2}{kv_{ts\sigma}}   v_{zts}\frac{k_zv_{zts}}{kv_{ts\sigma}}v_{dsy}Z_2+\frac{k^2}{kv_{ts\sigma}} v_{dsz}v_{dsy}Z_1-\frac{1}{2}\frac{k_xk_z}{kv_{ts\sigma}}\frac{v_{\perp ts\sigma}}{v_{zts}}(\lambda_{Ts\sigma}-1)v_{zts}\frac{k_xv_{\perp ts\sigma}}{kv_{ts\sigma}}v_{dsy}Z_0\Big\}\\
=\frac{2v_{dsy}}{kv_{ts\sigma}}\Big\{ -i\nu_s\frac{k_z}{kv_{ts\sigma}}Z_1+  \frac{k^2k_zv_{zts}^2}{k^2v_{ts\sigma}^2} Z_2+\frac{k^2}{kv_{ts\sigma}} v_{dsz} Z_1-\frac{1}{2}(\lambda_{Ts\sigma}-1) \frac{k_x^2k_zv_{\perp ts\sigma}^2}{k^2v_{ts\sigma}^2} Z_0\Big\}$,

\item $P_{s32}^u=\Big\{-\bm I + \int dv^3 \frac{\Big[-i\nu_s(\bm v \frac{\partial f_{s0\sigma}}{\partial \bm v})+(\bm k\cdot\frac{\partial f_{s0\sigma}}{\partial \bm v})(\bm v\bm v)\Big]}{(\omega-\bm k\cdot\bm v +i\nu_s)}\Big\}_{32}=0 {\color{red}+} \\ 2 \int dxdydz g\frac{ -i\nu_s 0+\Big[\frac{k^2}{kv_{ts\sigma}}x+\frac{k_xk_z}{kv_{ts\sigma}}\frac{v_{\perp ts\sigma}}{v_{zts}}(\lambda_{Ts\sigma}-1)z \Big]   [v_{zts}\frac{k_zv_{zts}}{kv_{ts\sigma}}x-v_{zts}\frac{k_xv_{\perp ts\sigma}}{kv_{ts\sigma}}z+v_{dsz}]v_{dsy} }{kv_{ts\sigma}(\zeta_{s\sigma}- x)}
\\=\frac{2}{kv_{ts\sigma}}\Big\{ \frac{k^2}{kv_{ts\sigma}}   v_{zts}\frac{k_zv_{zts}}{kv_{ts\sigma}}v_{dsy}Z_2+\frac{k^2}{kv_{ts\sigma}} v_{dsz}v_{dsy}Z_1-\frac{1}{2}\frac{k_xk_z}{kv_{ts\sigma}}\frac{v_{\perp ts\sigma}}{v_{zts}}(\lambda_{Ts\sigma}-1)v_{zts}\frac{k_xv_{\perp ts\sigma}}{kv_{ts\sigma}}v_{dsy}Z_0\Big\}\\
=\frac{2v_{dsy}}{kv_{ts\sigma}}\Big\{   \frac{k^2k_zv_{zts}^2}{k^2v_{ts\sigma}^2}Z_2+\frac{k^2}{kv_{ts\sigma}} v_{dsz}Z_1-\frac{1}{2}(\lambda_{Ts\sigma}-1) \frac{k_x^2k_zv_{\perp ts\sigma}^2}{k^2v_{ts\sigma}^2}Z_0\Big\}$,

\item 
$P_{s33}^u=\Big\{-\bm I + \int dv^3 \frac{\Big[-i\nu_s(\bm v \frac{\partial f_{s0\sigma}}{\partial \bm v})+(\bm k\cdot\frac{\partial f_{s0\sigma}}{\partial \bm v})(\bm v\bm v)\Big]}{(\omega-\bm k\cdot\bm v +i\nu_s)}\Big\}_{33}=-1 {\color{red}+} \\ 2 \int dxdydz g\frac{ -i\nu_s( \frac{k_z^2v_{zts}^2}{k^2v_{ts\sigma}^2}x^2+\frac{k_zv_{dsz}}{kv_{ts\sigma}}x +\frac{k_x^2v_{\perp ts\sigma}^2}{k^2v_{ts\sigma}^2}z^2)+\Big[\frac{k^2}{kv_{ts\sigma}}x+\frac{k_xk_z}{kv_{ts\sigma}}\frac{v_{\perp ts\sigma}}{v_{zts}}(\lambda_{Ts\sigma}-1)z \Big]  [v_{zts}\frac{k_zv_{zts}}{kv_{ts\sigma}}x-v_{zts}\frac{k_xv_{\perp ts\sigma}}{kv_{ts\sigma}}z+v_{dsz}]^2 }{kv_{ts\sigma}(\zeta_{s\sigma}- x)}=-1+\frac{2}{kv_{ts\sigma}}\Big\{-i\nu_s( \frac{k_z^2v_{zts}^2}{k^2v_{ts\sigma}^2}Z_2+\frac{k_zv_{dsz}}{kv_{ts\sigma}}Z_1 +\frac{k_x^2v_{\perp ts\sigma}^2}{2k^2v_{ts\sigma}^2}Z_0)+ \frac{k^2}{kv_{ts\sigma}} (v_{zts}\frac{k_zv_{zts}}{kv_{ts\sigma}})^2Z_3+2\frac{k^2}{kv_{ts\sigma}} v_{zts}\frac{k_zv_{zts}}{kv_{ts\sigma}}v_{dsz}Z_2+[\frac{1}{2}\frac{k^2}{kv_{ts\sigma}}  (v_{zts}\frac{k_xv_{\perp ts\sigma}}{kv_{ts\sigma}})^2  +\frac{k^2}{kv_{ts\sigma}}  v_{dsz}^2- \frac{k_xk_z}{kv_{ts\sigma}}\frac{v_{\perp ts\sigma}}{v_{zts}}(\lambda_{Ts\sigma}-1)v_{zts}\frac{k_zv_{zts}}{kv_{ts\sigma}}v_{zts}\frac{k_xv_{\perp ts\sigma}}{kv_{ts\sigma}}]Z_1 -  \frac{k_xk_z}{kv_{ts\sigma}}\frac{v_{\perp ts\sigma}}{v_{zts}}(\lambda_{Ts\sigma}-1)v_{zts}\frac{k_xv_{\perp ts\sigma}}{kv_{ts\sigma}}v_{dsz}Z_0\Big\}\\
=-1+\frac{2}{kv_{ts\sigma}}\Big\{-i\nu_s( \frac{k_z^2v_{zts}^2}{k^2v_{ts\sigma}^2}Z_2+\frac{k_zv_{dsz}}{kv_{ts\sigma}}Z_1 +\frac{k_x^2v_{\perp ts\sigma}^2}{2k^2v_{ts\sigma}^2}Z_0)+ 
  \frac{k^2k_z^2v_{zts}^4}{k^3v_{ts\sigma}^3}Z_3
+2 \frac{k^2k_zv_{zts}^2}{k^2v_{ts\sigma}^2}v_{dsz}Z_2
+[\frac{1}{2}  \frac{k^2k_x^2v_{zts}^2v_{\perp ts\sigma}^2}{k^3v_{ts\sigma}^3}  +\frac{k^2}{kv_{ts\sigma}}  v_{dsz}^2-  (\lambda_{Ts\sigma}-1) \frac{k_x^2k_z^2v_{zts}^2v_{\perp ts\sigma}^2}{k^3v_{ts\sigma}^3}]Z_1 
-   (\lambda_{Ts\sigma}-1) \frac{k_x^2k_zv_{\perp ts\sigma}^2}{k^2v_{ts\sigma}^2}v_{dsz}Z_0\Big\}$,

\end{itemize}
where the argument for $Z_{0,1,2,3}$ is $\zeta_{s\sigma}$.

Note, with the '$\to$' means that to integral out the $x$, $y$ and $z$ terms, say, $y,z\to0$, $y^3,z^3\to0$, $y^0,z^0\to1$, $y^2,z^2\to\frac{1}{2}$, $x^p\to Z_p(\zeta_{s\sigma})$), we have used
\begin{itemize}

\item $(ax+bz)(cx+dz+e)^2=ac^2x^3+2acdx^2z+2acex^2+ad^2xz^2+2adexz+ae^2x+bc^2x^2z+2bcdxz^2+2bcexz+bd^2z^3+2bdez^2+be^2z\to ac^2Z_3+2aceZ_2+[\frac{1}{2}ad^2  +ae^2+ bcd]Z_1 +  bdeZ_0 $.

\item $(ax+bz)(cx+dz+e)f=acfx^2+adfxz+aefx+bcfxz+bdfz^2+befz \to acfZ_2+aefZ_1+\frac{1}{2}bdfZ_0$.

\item $(ax+bz)(cx-dz+e)f=acfx^2-adfxz+aefx+bcfxz-bdfz^2+befz \to acfZ_2+aefZ_1-\frac{1}{2}bdfZ_0$.

\item $(ax+bz)(c y^2+d)=acxy^2+adx+bcy^2z+bdz\to [\frac{1}{2}ac+ad]Z_1 $. 

\item $(ax+bz)(cx+dz+e)(fx-gz+h)=acfx^3-acgx^2z+achx^2+adfx^2z-adgxz^2+adhxz+aefx^2-aegxz+aehx+bcfx^2z-bcgxz^2+bchxz+bdfxz^2-bdgz^3+bdhz^2+befxz-begz^2+behz\to acfZ_3 +[ach +aef]Z_2 +[aeh-\frac{1}{2}adg  -\frac{1}{2}bcg +\frac{1}{2}bdf]Z_1 +[\frac{1}{2}bdh -\frac{1}{2}beg]Z_0 $.

\item $(ax+bz)(cx-dz+e)^2=ac^2x^3-2acdx^2z+2acex^2+ad^2xz^2-2adexz+ae^2x+bc^2x^2z-2bcdxz^2+2bcexz+bd^2z^3-2bdez^2+be^2z\to ac^2Z_3+2aceZ_2+[\frac{1}{2}ad^2  +ae^2- bcd]Z_1 -  bdeZ_0 $.

\end{itemize}

Although the above expressions are much complicated and need be carefully, they can be solved easily by PASS-K matrix approach. We have also checked that, when $\nu_s=0$, $\lambda_{Ts\sigma}=1$ and $v_{dsz}=v_{dsy}=0$, the result can reduce to exactly the same form as in Ref.\cite{Muschietti2017} .


\subsubsection{Magnetized terms}
Next, we calculate the magnetized terms.

Define $\bm Q_s^m=\sum_{\sigma=a,b}r_{s\sigma}\frac{\omega_{ps}^2}{\omega^2}\bm P_s^m$, with $\bm P_s^m=\sum_{n=-\infty}^{\infty}\int_{-\infty}^{\infty}\int_0^{\infty}  \frac{2\pi v'_\perp dv'_\perp dv'_\parallel}{(\omega-k_\parallel v'_\parallel {-\color{red}k_xv_{dsx}+i\nu_s}-n\omega_{cs})}\bm \Pi_{s\sigma}=\sum_{n=-\infty}^{\infty}\int_{-\infty}^{\infty}\int_0^{\infty}  \frac{2\pi v'_\perp dv'_\perp dv''_\parallel}{(\omega_{sn}-k_z v''_\parallel )}\bm \Pi_{s\sigma}$ and $\omega_{sn}=\omega-k_z v_{dsz}-n\omega_{cs}{-\color{red}k_xv_{dsx}+i\nu_s}$, i.e., $\omega=\omega_{sn}+k_z v_{dsz}+n\omega_{cs}+k_xv_{dsx}-i\nu_s$. We have also
\begin{eqnarray}\nonumber
   \frac{\partial f_{s0\sigma}}{\partial v'_\parallel}&=&-\frac{2(v'_\parallel-v_{dsz})}{v_{zts}^2}f_{s0\sigma}=-2\frac{(v'_\parallel-v_{dsz})}{v_{zts}^2}f_{s0\sigma}=-2\frac{ v''_\parallel }{v_{zts}^2}f_{s0\sigma},\\\nonumber
   \frac{\partial f_{s0\sigma}}{\partial v'_\perp}&=&-2 \frac{ (v'_\perp-v_{dsr})}{v_{\perp ts\sigma}^2}f_{s0\sigma},\\\nonumber
\end{eqnarray}
with $v''_\parallel=v'_\parallel-v_{dsz}$, and $f_{s0\sigma}=\frac{1}{\pi^{3/2}v_{zts}v_{\perp ts\sigma}^2A_{s\sigma}}\exp\Big[-\frac{v''^2_\parallel}{v_{zts}^2}-\frac{(v'-v_{dsr})^2_\perp}{v_{\perp ts}^2}\Big]$, i.e.,  $v'_\parallel=v''_\parallel+v_{dsz}$. Since we have defined $A_n, B_n, C_n$, we do not need transform $v'_\perp$.

Thus,
\begin{itemize}

\item $P_{s11}^{m}=\sum_{n=-\infty}^{\infty}\int_{-\infty}^{\infty}\int_0^{\infty}  \frac{2\pi v'_\perp dv'_\perp dv''_\parallel}{(\omega_{sn}-k_z v''_\parallel )} \Pi_{s\sigma11}=\sum_{n=-\infty}^{\infty} \Big\{[\frac{n\omega_{cs}}{k_x }(n\omega_{cs}+k_xv_{dsx}-i\nu_s)\frac{ A_{nbs\sigma}}{v_{\perp ts\sigma}^2}\frac{Z_0}{k_zv_{zts}}+A_{n0\sigma} (\frac{n\omega_{cs}}{k_x } +v_{dsx})\frac{ Z_1}{v_{zts}^2}]( \frac{n\omega_{cs}}{k_x }+v_{dsx}) \Big\}-{\color{red}\sum_n A_{nbs\sigma}\frac{n\omega_{cs}}{k_x  v_{\perp ts\sigma}^2}( \frac{n\omega_{cs}}{k_x }+v_{dsx})}$.

\item $P_{s12}^{m}=\sum_{n=-\infty}^{\infty}\int_{-\infty}^{\infty}\int_0^{\infty}  \frac{2\pi v'_\perp dv'_\perp dv''_\parallel}{(\omega_{sn}-k_z v''_\parallel )} \Pi_{s\sigma12}=\sum_{n=-\infty}^{\infty} \Big\{ v_{dsy}( \frac{n\omega_{cs}}{k_x }+v_{dsx})(n\omega_{cs}\frac{ A_{nbs\sigma}}{v_{\perp ts\sigma}^2}\frac{Z_0}{k_zv_{zts}}+A_{n0\sigma} \frac{ Z_1}{v_{zts}^2})+i( \frac{n\omega_{cs}}{k_x }+v_{dsx})[(n\omega_{cs}-i\nu_s)\frac{ B_{nbs\sigma}}{v_{\perp ts\sigma}}\frac{Z_0}{k_zv_{zts}}+\frac{B_{n0\sigma} v_{\perp ts\sigma}Z_1}{v_{zts}^2}]\Big\}$.

\item $P_{s21}^{m}=\sum_{n=-\infty}^{\infty}\int_{-\infty}^{\infty}\int_0^{\infty}  \frac{2\pi v'_\perp dv'_\perp dv''_\parallel}{(\omega_{sn}-k_z v''_\parallel )} \Pi_{s\sigma21}=\sum_{n=-\infty}^{\infty} \Big\{ \frac{n\omega_{cs}}{k_x}(n\omega_{cs}+k_xv_{dsx}-i\nu_s)[-iv_{\perp ts\sigma}B_{nbs\sigma}+v_{dsy}A_{nbs\sigma}]\frac{ 1}{v_{\perp ts\sigma}^2}\frac{Z_0}{k_zv_{zts}}+[-iv_{\perp ts\sigma}B_{n0\sigma}+v_{dsy}A_{n0\sigma}] (\frac{n\omega_{cs}}{k_x }+v_{dsx})\frac{Z_1 }{v_{zts}^2}\Big\}$.

\item $P_{s22}^{m}=\sum_{n=-\infty}^{\infty}\int_{-\infty}^{\infty}\int_0^{\infty}  \frac{2\pi v'_\perp dv'_\perp dv''_\parallel}{(\omega_{sn}-k_z v''_\parallel )} \Pi_{s\sigma22}=\sum_{n=-\infty}^{\infty} \Big\{ v_{dsy}[-iv_{\perp ts\sigma} B_{nbs\sigma}+v_{dsy}A_{nbs\sigma}] \frac{n\omega_{cs}}{v_{\perp ts\sigma}^2}\frac{Z_0}{k_zv_{zts}}+v_{dsy}[-iv_{\perp ts\sigma} B_{n0\sigma}+v_{dsy}A_{n0\sigma}] \frac{ Z_1 }{v_{zts}^2}+v_{\perp ts\sigma} [v_{\perp ts\sigma} C_{nbs\sigma}+iv_{dsy}B_{nbs\sigma}]( n\omega_{cs}-i\nu_s)\frac{1}{v_{\perp ts\sigma}^2}\frac{Z_0}{k_zv_{zts}}+v_{\perp ts\sigma} [v_{\perp ts\sigma} C_{n0\sigma}+iv_{dsy}B_{n0\sigma}]\frac{ Z_1 }{v_{zts}^2} \Big\}-{\color{red} \sum_n(C_{nbs\sigma}+i\frac{v_{dsy}}{v_{\perp ts\sigma} }B_{nbs\sigma})}$.

\item $P_{s13}^{m}=\sum_{n=-\infty}^{\infty}\int_{-\infty}^{\infty}\int_0^{\infty}  \frac{2\pi v'_\perp dv'_\perp dv''_\parallel}{(\omega_{sn}-k_z v''_\parallel )} \Pi_{s\sigma13}=\sum_{n=-\infty}^{\infty} \Big\{ ( \frac{n\omega_{cs}}{k_x }+v_{dsx})[n\omega_{cs}  (\frac{Z_1}{k_z}+v_{dsz}\frac{Z_0}{k_zv_{zts}})  \frac{A_{nbs\sigma}}{v_{\perp ts\sigma}^2}+A_{n0\sigma}( v_{zts}Z_2+(k_z v_{dsz}-i\nu_s )\frac{Z_1}{k_z} )\frac{  1}{v_{zts}^2}] \Big\}$.

\item $P_{s31}^{m}=\sum_{n=-\infty}^{\infty}\int_{-\infty}^{\infty}\int_0^{\infty}  \frac{2\pi v'_\perp dv'_\perp dv''_\parallel}{(\omega_{sn}-k_z v''_\parallel )} \Pi_{s\sigma31}=\sum_{n=-\infty}^{\infty} \Big\{ \frac{n\omega_{cs}}{k_x }(n\omega_{cs}+k_xv_{dsx}-i\nu_s)\frac{ A_{nbs\sigma} }{v_{\perp ts\sigma}^2}(\frac{Z_1}{k_z}+v_{dsz}\frac{Z_0}{k_zv_{zts}})+A_{n0\sigma} (\frac{n\omega_{cs}}{k_x } +v_{dsx})(\frac{Z_2}{v_{zts}}+v_{dsz} \frac{Z_1}{v_{zts}^2})\Big\}$.

\item $P_{s23}^{m}=\sum_{n=-\infty}^{\infty}\int_{-\infty}^{\infty}\int_0^{\infty}  \frac{2\pi v'_\perp dv'_\perp dv''_\parallel}{(\omega_{sn}-k_z v''_\parallel )} \Pi_{s\sigma23}=\sum_{n=-\infty}^{\infty} \Big\{ [v_{dsy}A_{nbs\sigma}-iv_{\perp ts\sigma} B_{nbs\sigma}][(\frac{Z_1}{k_z}+\frac{Z_0v_{dsz}}{k_zv_{tsz}})  \frac{ n\omega_{cs} }{v_{\perp ts\sigma}^2}]+ [v_{dsy}A_{n0\sigma}-iv_{\perp ts\sigma} B_{n0\sigma}][v_{zts}Z_2+(k_z v_{dsz}-i\nu_s)\frac{Z_1}{k_z}]\frac{ 1}{v_{zts}^2}  \Big\} $.

\item $P_{s32}^{m}=\sum_{n=-\infty}^{\infty}\int_{-\infty}^{\infty}\int_0^{\infty}  \frac{2\pi v'_\perp dv'_\perp dv''_\parallel}{(\omega_{sn}-k_z v''_\parallel )} \Pi_{s\sigma32}=\sum_{n=-\infty}^{\infty} \Big\{  [ n\omega_{cs} \frac{ A_{nbs\sigma}}{v_{\perp ts\sigma}^2}\frac{Z_1}{k_z}+A_{n0\sigma} \frac{k_z v_{zts}Z_2 }{v_{zts}^2}]v_{dsy}+ [ n\omega_{cs} \frac{ A_{nbs\sigma}}{v_{\perp ts\sigma}^2}\frac{Z_0}{k_zv_{zts}}+A_{n0\sigma} \frac{Z_1 }{v_{zts}^2}]v_{dsy}v_{dsz} +i [(n\omega_{cs} -i\nu_s )\frac{ B_{nbs\sigma}}{v_{\perp ts\sigma}}\frac{Z_1}{k_z}+B_{n0\sigma} \frac{  v_{\perp ts\sigma}Z_2}{v_{zts} }]+i v_{dsz} [(n\omega_{cs} -i\nu_s )\frac{ B_{nbs\sigma}}{v_{\perp ts\sigma}}\frac{Z_0}{k_zv_{zts}}+B_{n0\sigma} \frac{ v_{\perp ts\sigma}Z_1}{v_{zts}^2}]   \Big\} $.

\item $P_{s33}^{m}=\sum_{n=-\infty}^{\infty}\int_{-\infty}^{\infty}\int_0^{\infty}  \frac{2\pi v'_\perp dv'_\perp dv''_\parallel}{(\omega_{sn}-k_z v''_\parallel )} \Pi_{s\sigma33}=\sum_{n=-\infty}^{\infty} \Big\{  [n\omega_{cs}  (\frac{v_{zts}Z_2}{k_z}+v_{dsz}\frac{ Z_1}{k_z})\frac{ A_{nbs\sigma}}{v_{\perp ts\sigma}^2}+A_{n0\sigma}(  Z_3+(k_z v_{dsz}-i\nu_s)\frac{Z_2}{k_zv_{zts}}) ] + v_{dsz}[n\omega_{cs}  (\frac{Z_1}{k_z}+v_{dsz}\frac{Z_0}{k_zv_{zts}})\frac{ A_{nbs\sigma}}{v_{\perp ts\sigma}^2}+A_{n0\sigma}( \frac{Z_2}{v_{zts}}+(k_z v_{dsz}-i\nu_s)\frac{Z_1}{k_zv_{zts}^2})]  \Big\} -{\color{red}\sum_n\frac{1}{2}A_{n0\sigma}}$.

\end{itemize}
where the argument for $Z_{0,1,2,3}$ is $\zeta_{sn}$, and we have used  $\sum_{n=-\infty}^{\infty}J_nJ'_n=0$,  $\sum_{n=-\infty}^{\infty}nJ_n^2=0$,  $\sum_{n=-
\infty}^{\infty}nJ_nJ'_n=0$. If we further use , $\sum_{n=-
\infty}^{\infty}J_n^2=1$,  $\sum_{n=-\infty}^{\infty}(J'_n)^2=\frac{1}{2}$,  $\sum_{n=-\infty}^{\infty}\frac{n^2J_n^2(x)}{x^2}=\frac{1}{2}$, the last $\sum_n$ terms in $P_{s11}^{m}$, $P_{s22}^{m}$ and $P_{s33}^{m}$ yields to $1$. We derive the above equation base on (define $\omega'_{sn}=\omega_{sn}-k_zv''_\parallel$, i.e., $\omega_{sn}=\omega'_{sn}+k_zv''_\parallel$)
\begin{itemize}

\item  $v'_\perp\Pi_{s\sigma11}
= J_n^2[\frac{n\omega_{cs}}{k_x }(\omega-k_zv'_\parallel)\frac{\partial f_{s0\sigma}}{\partial v'_\perp}+v'_\perp k_z(\frac{n\omega_{cs}}{k_x } +v_{dsx})\frac{\partial f_{s0\sigma}}{\partial v'_\parallel}]( \frac{n\omega_{cs}}{k_x }+v_{dsx})\to J_n^2[\frac{n\omega_{cs}}{k_x }(\omega'_{sn}+n\omega_{cs}+k_xv_{dsx}-i\nu_s)\frac{ (v'_\perp-v_{dsr})}{v_{\perp ts\sigma}^2}+v'_\perp k_z(\frac{n\omega_{cs}}{k_x } +v_{dsx})\frac{ v''_\parallel }{v_{zts}^2}]( \frac{n\omega_{cs}}{k_x }+v_{dsx})
\to J_n^2[\frac{n\omega_{cs}}{k_x }(n\omega_{cs}+k_xv_{dsx}-i\nu_s)\frac{ (v'_\perp-v_{dsr})}{v_{\perp ts\sigma}^2}+v'_\perp (\frac{n\omega_{cs}}{k_x } +v_{dsx})\frac{ k_zv''_\parallel }{v_{zts}^2}]( \frac{n\omega_{cs}}{k_x }+v_{dsx})+J_n^2\frac{n\omega_{cs}}{k_x }\frac{ (v'_\perp-v_{dsr})}{v_{\perp ts\sigma}^2}( \frac{n\omega_{cs}}{k_x }+v_{dsx})(\omega'_{sn})$.

\item $v'_\perp\Pi_{s\sigma12}
= J_n^2v_{dsy}( \frac{n\omega_{cs}}{k_x }+v_{dsx})(n\omega_{cs}\frac{\partial f_{s0\sigma}}{\partial v'_\perp}+v'_\perp k_z\frac{\partial f_{s0\sigma}}{\partial v'_\parallel})+iv'_\perp J_nJ'_n( \frac{n\omega_{cs}}{k_x }+v_{dsx})[(\omega-k_zv'_\parallel-{\color{red}k_xv_{dsx}})\frac{\partial f_{s0\sigma}}{\partial v'_\perp}+k_zv'_\perp \frac{\partial f_{s0\sigma}}{\partial v'_\parallel}]
\to  J_n^2v_{dsy}( \frac{n\omega_{cs}}{k_x }+v_{dsx})(n\omega_{cs}\frac{ (v'_\perp-v_{dsr})}{v_{\perp ts\sigma}^2}+v'_\perp k_z\frac{ v''_\parallel }{v_{zts}^2})+iv'_\perp J_nJ'_n( \frac{n\omega_{cs}}{k_x }+v_{dsx})[(\omega'_{sn}+n\omega_{cs}-i\nu_s)\frac{ (v'_\perp-v_{dsr})}{v_{\perp ts\sigma}^2}+k_zv'_\perp \frac{ v''_\parallel }{v_{zts}^2}]
\to  J_n^2v_{dsy}( \frac{n\omega_{cs}}{k_x }+v_{dsx})(n\omega_{cs}\frac{ (v'_\perp-v_{dsr})}{v_{\perp ts\sigma}^2}+v'_\perp \frac{ k_zv''_\parallel }{v_{zts}^2})+iv'_\perp J_nJ'_n( \frac{n\omega_{cs}}{k_x }+v_{dsx})[(n\omega_{cs}-i\nu_s)\frac{ (v'_\perp-v_{dsr})}{v_{\perp ts\sigma}^2}+v'_\perp \frac{k_z v''_\parallel }{v_{zts}^2}]+iv'_\perp J_nJ'_n( \frac{n\omega_{cs}}{k_x }+v_{dsx})\frac{ (v'_\perp-v_{dsr})}{v_{\perp ts\sigma}^2}(\omega'_{sn})$. 

\item $v'_\perp\Pi_{s\sigma21}
= (-iv'_\perp J_nJ'_n+v_{dsy}J_n^2)[\frac{n\omega_{cs}}{k_x}(\omega-k_zv'_\parallel)\frac{\partial f_{s0\sigma}}{\partial v'_\perp}+v'_\perp\frac{n\omega_{cs}}{k_x }k_z \frac{\partial f_{s0\sigma}}{\partial v'_\parallel}+v'_\perp k_zv_{dsx}\frac{\partial f_{s0\sigma}}{\partial v'_\parallel}]
\to (-iv'_\perp J_nJ'_n+v_{dsy}J_n^2)[\frac{n\omega_{cs}}{k_x}(\omega'_{sn}+n\omega_{cs}+k_xv_{dsx}-i\nu_s)\frac{ (v'_\perp-v_{dsr})}{v_{\perp ts\sigma}^2}+v'_\perp k_z(\frac{n\omega_{cs}}{k_x }+v_{dsx})\frac{ v''_\parallel }{v_{zts}^2}]
\to (-iv'_\perp J_nJ'_n+v_{dsy}J_n^2)[\frac{n\omega_{cs}}{k_x}(n\omega_{cs}+k_xv_{dsx}-i\nu_s)\frac{ (v'_\perp-v_{dsr})}{v_{\perp ts\sigma}^2}+v'_\perp (\frac{n\omega_{cs}}{k_x }+v_{dsx})\frac{k_z v''_\parallel }{v_{zts}^2}]+(-iv'_\perp J_nJ'_n+v_{dsy}J_n^2)\frac{n\omega_{cs}}{k_x}\frac{ (v'_\perp-v_{dsr})}{v_{\perp ts\sigma}^2}(\omega'_{sn}) $. 

\item $v'_\perp\Pi_{s\sigma22}
= J_nv_{dsy}(-iv'_\perp J'_n+v_{dsy}J_n)( n\omega_{cs} \frac{\partial f_{s0\sigma}}{\partial v'_\perp}+v'_\perp k_z\frac{\partial f_{s0\sigma}}{\partial v'_\parallel})+iv'_\perp J'_n(-iv'_\perp J'_n+v_{dsy}J_n)[(\omega-k_zv'_\parallel-{\color{red}k_xv_{dsx}})\frac{\partial f_{s0\sigma}}{\partial v'_\perp}+k_zv'_\perp \frac{\partial f_{s0}}{\partial v'_\parallel}]
\to J_nv_{dsy}(-iv'_\perp J'_n+v_{dsy}J_n)( n\omega_{cs} \frac{ (v'_\perp-v_{dsr})}{v_{\perp ts\sigma}^2}+v'_\perp k_z\frac{ v''_\parallel }{v_{zts}^2})+iv'_\perp J'_n(-iv'_\perp J'_n+v_{dsy}J_n)[(\omega'_{sn}+n\omega_{cs}-i\nu_s)\frac{ (v'_\perp-v_{dsr})}{v_{\perp ts\sigma}^2}+k_zv'_\perp \frac{ v''_\parallel }{v_{zts}^2}]
\to J_nv_{dsy}(-iv'_\perp J'_n+v_{dsy}J_n)( n\omega_{cs} \frac{ (v'_\perp-v_{dsr})}{v_{\perp ts\sigma}^2}+v'_\perp \frac{ k_zv''_\parallel }{v_{zts}^2})+iv'_\perp J'_n(-iv'_\perp J'_n+v_{dsy}J_n)[( n\omega_{cs}-i\nu_s)\frac{ (v'_\perp-v_{dsr})}{v_{\perp ts\sigma}^2}+v'_\perp \frac{ k_zv''_\parallel }{v_{zts}^2}]+iv'_\perp J'_n(-iv'_\perp J'_n+v_{dsy}J_n)\frac{ (v'_\perp-v_{dsr})}{v_{\perp ts\sigma}^2}(\omega'_{sn})$. 

\item $v'_\perp \Pi_{s\sigma13}
=J_n^2( \frac{n\omega_{cs}}{k_x }+v_{dsx})[n\omega_{cs}  v'_\parallel  \frac{\partial f_{s0\sigma}}{\partial v'_\perp}+v'_\perp(\omega-n\omega_{cs} -{\color{red}k_xv_{dsx}})\frac{\partial f_{s0\sigma}}{\partial v'_\parallel}]
\to J_n^2( \frac{n\omega_{cs}}{k_x }+v_{dsx})[n\omega_{cs}  (v''_\parallel+v_{dsz})  \frac{ (v'_\perp-v_{dsr})}{v_{\perp ts\sigma}^2}+v'_\perp(\omega'_{sn}+k_zv''_\parallel+k_z v_{dsz}-i\nu_s )\frac{ v''_\parallel }{v_{zts}^2}]
\to J_n^2( \frac{n\omega_{cs}}{k_x }+v_{dsx})[n\omega_{cs}  (v''_\parallel+v_{dsz})  \frac{ (v'_\perp-v_{dsr})}{v_{\perp ts\sigma}^2}+v'_\perp( k_zv''_\parallel+k_z v_{dsz}-i\nu_s )\frac{ v''_\parallel }{v_{zts}^2}]+J_n^2( \frac{n\omega_{cs}}{k_x }+v_{dsx})v'_\perp\frac{ v''_\parallel }{v_{zts}^2}(\omega'_{sn} )$. 

\item $v'_\perp\Pi_{s\sigma31}
= v'_\parallel J_n^2[\frac{n\omega_{cs}}{k_x }(\omega-k_zv'_\parallel)\frac{\partial f_{s0\sigma}}{\partial v'_\perp}+v'_\perp\frac{n\omega_{cs}}{k_x }k_z \frac{\partial f_{s0\sigma}}{\partial v'_\parallel}+v'_\perp k_zv_{dsx}\frac{\partial f_{s0\sigma}}{\partial v'_\parallel}]
\to (v''_\parallel +v_{dsz})J_n^2[\frac{n\omega_{cs}}{k_x }(\omega'_{sn}+n\omega_{cs}+k_xv_{dsx}-i\nu_s)\frac{ (v'_\perp-v_{dsr})}{v_{\perp ts\sigma}^2}+v'_\perp k_z(\frac{n\omega_{cs}}{k_x } +v_{dsx})\frac{ v''_\parallel }{v_{zts}^2}]
\to (v''_\parallel +v_{dsz}) J_n^2[\frac{n\omega_{cs}}{k_x }(n\omega_{cs}+k_xv_{dsx}-i\nu_s)\frac{ (v'_\perp-v_{dsr})}{v_{\perp ts\sigma}^2}+v'_\perp (\frac{n\omega_{cs}}{k_x } +v_{dsx})\frac{k_z v''_\parallel }{v_{zts}^2}]+(v''_\parallel +v_{dsz})J_n^2\frac{n\omega_{cs}}{k_x }\frac{ (v'_\perp-v_{dsr})}{v_{\perp ts\sigma}^2}(\omega'_{sn})$.

\item $v'_\perp\Pi_{s\sigma23}
=(v_{dsy}J_n^2-iv'_\perp J_nJ'_n)[n\omega_{cs} v'_\parallel  \frac{\partial f_{s0\sigma}}{\partial v'_\perp}+v'_\perp(\omega-n\omega_{cs}-{\color{red}k_xv_{dsx}})\frac{\partial f_{s0\sigma}}{\partial v'_\parallel}]
\to (v_{dsy}J_n^2-iv'_\perp J_nJ'_n)[n\omega_{cs} (v''_\parallel+v_{dsz})  \frac{ (v'_\perp-v_{dsr})}{v_{\perp ts\sigma}^2}+v'_\perp(\omega'_{sn}+k_zv''_\parallel+k_z v_{dsz}-i\nu_s)\frac{ v''_\parallel }{v_{zts}^2}]
\to (v_{dsy}J_n^2-iv'_\perp J_nJ'_n)[n\omega_{cs} (v''_\parallel+v_{dsz})  \frac{ (v'_\perp-v_{dsr})}{v_{\perp ts\sigma}^2}+v'_\perp(k_zv''_\parallel+k_z v_{dsz}-i\nu_s)\frac{ v''_\parallel }{v_{zts}^2}]+(v_{dsy}J_n^2-iv'_\perp J_nJ'_n)v'_\perp\frac{ v''_\parallel }{v_{zts}^2}(\omega'_{sn})$.

\item $v'_\perp\Pi_{s\sigma32}=v'_\parallel J_n^2( n\omega_{cs} \frac{\partial f_{s0\sigma}}{\partial v'_\perp}+v'_\perp k_z\frac{\partial f_{s0}}{\partial v'_\parallel})v_{dsy}+iv'_\parallel v'_\perp J_nJ'_n[(\omega-k_zv'_\parallel-{\color{red}k_xv_{dsx}})\frac{\partial f_{s0}}{\partial v'_\perp}+k_zv'_\perp \frac{\partial f_{s0\sigma}}{\partial v'_\parallel}]
\to (v''_\parallel+v_{dsz}) J_n^2( n\omega_{cs} \frac{ (v'_\perp-v_{dsr})}{v_{\perp ts\sigma}^2}+v'_\perp k_z\frac{ v''_\parallel }{v_{zts}^2})v_{dsy}+i(v''_\parallel+v_{dsz}) v'_\perp J_nJ'_n[(\omega'_{sn}+n\omega_{cs} -i\nu_s )\frac{ (v'_\perp-v_{dsr})}{v_{\perp ts\sigma}^2}+k_zv'_\perp \frac{ v''_\parallel }{v_{zts}^2}]
\to (v''_\parallel+v_{dsz}) J_n^2( n\omega_{cs} \frac{ (v'_\perp-v_{dsr})}{v_{\perp ts\sigma}^2}+v'_\perp \frac{ k_zv''_\parallel }{v_{zts}^2})v_{dsy}+i(v''_\parallel+v_{dsz}) v'_\perp J_nJ'_n[(n\omega_{cs} -i\nu_s )\frac{ (v'_\perp-v_{dsr})}{v_{\perp ts\sigma}^2}+v'_\perp \frac{k_z v''_\parallel }{v_{zts}^2}]+i(v''_\parallel+v_{dsz}) v'_\perp J_nJ'_n\frac{ (v'_\perp-v_{dsr})}{v_{\perp ts\sigma}^2} (\omega'_{sn} )$. 

\item $v'_\perp\Pi_{s\sigma33}=v'_\parallel J_n^2[n\omega_{cs}  v'_\parallel \frac{\partial f_{s0\sigma}}{\partial v'_\perp}+v'_\perp(\omega-n\omega_{cs}-{\color{red}k_xv_{dsx}})\frac{\partial f_{s0\sigma}}{\partial v'_\parallel}]
\to (v''_\parallel+v_{dsz}) J_n^2[n\omega_{cs}  (v''_\parallel+v_{dsz})\frac{ (v'_\perp-v_{dsr})}{v_{\perp ts\sigma}^2}+v'_\perp(\omega'_{sn}+k_zv''_\parallel+k_z v_{dsz}-i\nu_s)\frac{ v''_\parallel }{v_{zts}^2}]
\to (v''_\parallel+v_{dsz}) J_n^2[n\omega_{cs}  (v''_\parallel+v_{dsz})\frac{ (v'_\perp-v_{dsr})}{v_{\perp ts\sigma}^2}+v'_\perp( k_zv''_\parallel+k_z v_{dsz}-i\nu_s)\frac{ v''_\parallel }{v_{zts}^2}]+{\color{red}v'_\perp J_n^2\frac{(v''_\parallel +v_{dsz}) v''_\parallel }{v_{zts}^2}(\omega'_{sn})}$. 

\end{itemize}
where '$\to$' means we have omitted the coefficient '$-2f_{s0\sigma}$'.

We find by set $\nu_s=0$, $v_{dsx}=v_{dsy}=0$, these $P_{sij}^m$ reduce exactly the same one as in Ref.\cite{Umeda2012} ring beam case. By set $\nu_s=0$, $v_{dsy}=v_{dsr}=0$, we have also checked that they can reduce to the one in Ref.\cite{Umeda2018} for drift across magnetic field case.

\subsubsection{Final Form}

The final form of the electromagnetic dispersion relation is the combine of the above unmagnetized $P_s^u$ and magnetized $P_s^m$ terms to $\bm Q$  in Eq.(\ref{eq:emdrQ}), and then to Eq.(\ref{eq:emdr0}).

\section{Transform to PASS-K matrix Equation}\label{sec:passk_dr}

The conventional root finding approach to solve the above dispersion relations can only give one solution at one time and heavily depends on initial guess. The Cauchy contour integral approach \cite{Kravanja2000} can locate all the solutions in a selected complex domain, however which still can not give all the important solutions and is also difficult for complicated dispersion relation. To solve the dispersion relation using PASS-K matrix approach \cite{Xie2016}, which can give all the important solution at one time, we need two further steps: 
\begin{itemize}
\item 

(1) Do $J$-pole expansion of $Z$ function, i.e.,
\begin{equation}
Z(\zeta)\simeq Z_J(\zeta)=\sum_{j=1}^{J}\frac{b_j}{\zeta-c_j},
\end{equation}
where $b_j$ and $c_j$ are constants for given $J$, as given in Ref.\cite{Xie2016} for $k_{ts}>0$; 

\item (2)  Do linear transformation to a equivalent matrix eigenvalue problem. The standard eigenvalue library can solve all the eigenvalues of a matrix.

\end{itemize}
The first step with $J=8$ has been used well for more than thirty years in WHAMP \cite{Ronnmark1982} code;  The second step is firstly developed in the first version of PASS-K/PDRK code \cite{Xie2016}.
We derive the corresponding equations step by step.

We note: $\sum_jb_j=-1$,
$\sum_jb_jc_j=0$, $\sum_jb_jc_j^2=-1/2$ and $\sum_{j=1}^{J}b_jc_j^3=0$ \cite{Ronnmark1983}. And also
\begin{equation*}
\frac{1}{\omega}\frac{b}{\omega-c}=\frac{b}{c}\Big(\frac{1}{\omega-c}-\frac{1}{\omega}\Big).
\end{equation*}
For  $\zeta_s=\frac{\omega-k_{cs}}{k_{ts}}$, we have
\begin{eqnarray*}\nonumber
Z_0(\zeta_s,k_{ts})=  \left\{ \begin{array}{l}
    Z(\zeta_s)\simeq k_{ts}\sum_{j=1}^{J}\frac{b_j}{\omega-k_{cs}-k_{ts}c_{j}}\\
    -\frac{k_{ts}}{\omega-k_{cs}}=k_{ts}\sum_{j=1}^{J}\frac{b_j}{\omega-k_{cs}}\\
    -Z(-\zeta_s)\simeq k_{ts}\sum_{j=1}^{J}\frac{b_j}{\omega-k_{cs}+k_{ts}c_{j}}
      \end{array}\right.=k_{ts}\sum_{j=1}^{J}\frac{b_j}{\omega-k_{cs}-|k_{ts}|c_{j}}=\sum_{j=1}^{J}\frac{k_{ts}b_j}{\omega-c_{sj}},
\end{eqnarray*}
which is written to one compact form for both $k_{ts}>0$ and $k_{ts}\leq0$, with $c_{sj}=k_{cs}+|k_{ts}|c_{j}$. And
\begin{eqnarray*}\nonumber
Z_1(\zeta_s)=1+\zeta_s Z_0(\zeta_s)&\simeq&1+\zeta_s\sum_{j=1}^{J}\frac{b_j}{\zeta_s-c_j}=1+\sum_{j=1}^{J}\Big[b_j+\frac{b_jc_j}{\zeta_s-c_j}\Big]=\sum_{j=1}^{J}\frac{b_j|k_{ts}|c_j}{\omega-c_{sj}},
\end{eqnarray*}
where we have  used $\sum_{j=1}^{J}b_j=-1$.  And
\begin{eqnarray*}\nonumber
Z_2(\zeta_s)&=&\zeta_s(1+\zeta_s Z_0)\simeq\frac{\omega-k_{cs}}{k_{ts}}\sum_{j=1}^{J}\frac{b_j|k_{ts}|c_j}{\omega-c_{sj}}=\frac{|k_{ts}|}{k_{ts}}\sum_{j=1}^{J}\frac{b_jc_j(\omega-k_{cs})}{\omega-c_{sj}}\\
&=&\frac{|k_{ts}|}{k_{ts}}\sum_{j=1}^{J}\Big[b_jc_j+\frac{b_jc_j(c_{sj}-k_{cs})}{\omega-c_{sj}}\Big]=\frac{1}{k_{ts}}\sum_{j=1}^{J}\frac{b_j|k_{ts}|^2c_j^2}{\omega-c_{sj}},
\end{eqnarray*}
where we have used  $\sum_{j=1}^{J}b_jc_j=0$.  And
\begin{eqnarray*}\nonumber
Z_3(\zeta_s)&=&\frac{1}{2}+\zeta_s^2(1+\zeta_s Z_0)\simeq\frac{1}{2}+\frac{\omega-k_{cs}}{k_{ts}}k_{ts}\sum_{j=1}^{J}\frac{b_jc_j^2}{\omega-c_{sj}}=\frac{1}{2}+\sum_{j=1}^{J}\frac{b_jc_j^2(\omega-k_{cs})}{\omega-c_{sj}}\\
&=&\frac{1}{2}+\sum_{j=1}^{J}\Big[b_jc_j^2+\frac{b_jc_j^2(c_{sj}-k_{cs})}{\omega-c_{sj}}\Big]=\frac{1}{k_{ts}^2}\sum_{j=1}^{J}\frac{b_j|k_{ts}|^3c_j^3}{\omega-c_{sj}},
\end{eqnarray*}
where we have used  $\sum_{j=1}^{J}b_jc_j^2=-1/2$. In the above $Z_{0,1,2,3}$ and $c_{sj}$, we find for $k_{ts}<0$ it could be simply by doing the variables change with: $|k_{ts}|\to k_{ts}$ and $c_j\to -c_j$. Thus, in the later usage, to simplify the notation, all $|k_{ts}|$ is changed to be $k_{ts}$, and the meaning of $c_j$ for $k_{ts}<0$ is the $-c_j$ in the default $Z_J$. Thus for both $k_{ts}>0$ and $k_{ts}\leq0$, we have a single compact form
\begin{eqnarray*}\nonumber
Z_p(\zeta_s)\simeq k_{ts}\sum_{j=1}^{J}\frac{b_jc_j^p}{\omega-c_{sj}},
\end{eqnarray*}
with $c_{sj}=k_{cs}+k_{ts}c_{j}$. And the only change is that $c_j=c_{j0}$ for $k_{ts}\geq0$ and $c_j=-c_{j0}$ for $k_{ts}<0$ , with $c_{j0}$ be the $c_j$ for $k_{ts}>0$.

Typically, for magnetized species $\zeta_{sn}=\frac{\omega-k_zv_{dsz}-n\omega_{cs}{-\color{red}k_xv_{dsx}+i\nu_s}}{k_zv_{zts}}$ and unmagnetized species $\zeta_{s\sigma}=\frac{\omega-k_xv_{dsx}-k_zv_{dsz}+i\nu_s}{kv_{ts\sigma}}$, with $kv_{ts\sigma}=\sqrt{k_x^2v_{\perp ts\sigma}^2+k_z^2v_{zts}^2}$ (which always $\geq0$), we have corresponding $k_{ts}=k_zv_{zts}$ and $kv_{ts\sigma}$, respectively. For $c_{sj}$: $c_{snj}=k_zv_{dsz}+n\omega_{cs}+k_xv_{dsx}-i\nu_s+k_zv_{zts}c_j$, and $c_{sj\sigma}=k_zv_{dsz}+k_xv_{dsx}-i\nu_s+kv_{ts\sigma}c_j$.

\subsection{The Electrostatic case}

We solve Eq.(\ref{eq:esdr1}) for electrostatic case. The corresponding linear transformation is straightforward and simple. 
\begin{eqnarray}\nonumber
  D(\omega,\bm k)&\simeq&1+\sum_{s=m}\frac{\omega_{ps}^2}{k^2v_{zts}^2}\sum_{n=-\infty}^{\infty} \sum_{\sigma=a,b}r_{s\sigma}\Big\{[1+\zeta_{sn}\sum_{j=1}^{J}\frac{b_j}{\zeta_{sn}-c_j}]A_{n0\sigma} + \frac{n\omega_{cs} \lambda_{Ts\sigma}}{k_zv_{zts}}\sum_{j=1}^{J}\frac{b_j}{\zeta_{sn}-c_j}A_{nbs\sigma}\Big\}\\\nonumber
  &&+\sum_{s=u} \frac{\omega_{ps}^2}{k^2}\sum_{\sigma=a,b}\frac{2r_{s\sigma}}{v_{ts\sigma}^2}[1+\zeta_{s\sigma} \sum_{j=1}^{J}\frac{b_j}{\zeta_{s\sigma}-c_j}]\\\nonumber
  &=&1+\sum_{s=m}\frac{\omega_{ps}^2}{k^2v_{zts}^2}\sum_{n=-\infty}^{\infty} \sum_{\sigma=a,b}r_{s\sigma}\sum_{j=1}^{J}\frac{A_{n0\sigma}b_jc_j+\frac{A_{nbs\sigma}n\omega_{cs} \lambda_{Ts\sigma}}{k_zv_{zts}}b_j}{\zeta_{sn}-c_j} +\sum_{s=u} \frac{\omega_{ps}^2}{k^2}\sum_{\sigma=a,b}\frac{2r_{s\sigma}}{v_{ts\sigma}^2}\sum_{j=1}^{J}\frac{b_jc_j}{\zeta_{s\sigma}-c_j}\\\nonumber
  &=&1+\sum_{sn\sigma j}^{m}\frac{r_{s\sigma}\omega_{ps}^2}{k^2v_{zts}^2}\frac{A_{n0\sigma}k_zv_{zts}b_jc_j+{A_{nbs\sigma}n\omega_{cs} \lambda_{Ts\sigma}}b_j}{\omega-c_{snj}} +\sum_{sj\sigma}^{u} \frac{2r_{s\sigma}\omega_{ps}^2}{kv_{ts\sigma}}\frac{b_jc_j}{\omega-c_{sj\sigma}}\\
  &=&1+\sum_{snj}^{m}\frac{b_{snj}}{\omega-c_{snj}} +\sum_{sj\sigma}^{u}\frac{b_{sj\sigma}}{\omega-c_{sj\sigma}}=0.
\end{eqnarray}

Notation: $\zeta_{sn}=\frac{\omega-k_zv_{dsz}-n\omega_{cs}{-\color{red}k_xv_{dsx}+i\nu_s}}{k_zv_{zts}}$, $\zeta_{s\sigma}=\frac{\omega-k_xv_{dsx}-k_zv_{dsz}+i\nu_s}{kv_{ts\sigma}}$, $kv_{ts\sigma}=\sqrt{k_x^2v_{\perp ts\sigma}^2+k_z^2v_{zts}^2}=v_{zts}\sqrt{k_x^2{\color{red}/\lambda_{Ts\sigma}}+k_z^2}$.

Define: $c_{snj}=k_zv_{dsz}+n\omega_{cs}+k_xv_{dsx}-i\nu_s+k_zv_{zts}c_j$, $b_{snj}=\sum_{\sigma}\frac{r_{s\sigma}\omega_{ps}^2}{k^2v_{zts}^2}(A_{n0\sigma}k_zv_{zts}b_jc_j+{A_{nbs\sigma}n\omega_{cs} \lambda_{Ts\sigma}}b_j)$, $c_{sj\sigma}=k_zv_{dsz}+k_xv_{dsx}-i\nu_s+kv_{ts\sigma}c_j$, and $b_{sj\sigma}=\frac{2r_{s\sigma}\omega_{ps}^2}{kv_{ts\sigma}}b_jc_j$.

The equivalent linear system can be
\begin{eqnarray}\nonumber
\omega n_{sj}&=&c_{sj}n_{sj}+b_{sj}E,\\
E&=&-\sum_{sj}n_{sj},
\end{eqnarray}
or sparse matrix one
\begin{eqnarray}\nonumber
\omega n_{sj}&=&c_{sj}n_{sj}+b_{sj}E,\\
\omega E&=&-\sum_{sj}c_{sj}n_{sj}-\sum_{sj}b_{sj}E,
\end{eqnarray}
where $b_{sj}$ and $c_{sj}$ is short notation for both unmagnetized and magnetized species $b_{sj\sigma,snj}$ and $c_{sj\sigma,snj}$.
We find the only singularity in the above final form occurs at $k=0$, which requires $k_x=k_z=0$. And thus the final form can be applied for arbitrary $k\neq0$. Some other solvers in literature may meet singularity for $k_z=0$ or $k_x=0$, and may be incorrect for $k_z\leq0$.

It is also obvious that the magnetized species can not reduce to unmagnetized species by set $\omega_{cs}=0$. The major difference is $k_zv_{zts}$ in magnetized species and $kv_{ts\sigma}$ in unmagnetized species.

\subsection{The Electromagnetic case}

The electromagnetic case is much complicated. However, the linear transformation for $\bm Q^m(\omega,\bm k)$ is still similar to the original PASS-K/PDRK derivation.

To seek an equivalent linear system, the Maxwell’s equations
\begin{subequations} \label{eq:em3dmaxw}
\begin{eqnarray}
  & \partial_t {\bm E} = c^2\nabla\times{\bm B}-{\bm J}/\epsilon_0,\\
  & \partial_t {\bm B} = -\nabla\times{\bm E},
\end{eqnarray}
\end{subequations}
do not need to be changed. We only need to seek a new linear system for
{\color{red}${\bm J}={\bm J^m}+{\bm J^u}=({\bm{\sigma^m}+\bm{\sigma^u}})\cdot{\bm E}=\bm{\sigma}\cdot{\bm E}$}.

\subsubsection{The unmagnetized terms}

Considering the defination $\bm \sigma_s^u=-i\epsilon_0\omega\bm Q_s^u=-i\epsilon_0\sum_{\sigma=a,b}r_{s\sigma}\frac{\omega_{ps}^2}{\omega}\bm P_{s\sigma}^u$, after $J$-pole expansion, we have 
\begin{itemize}

\item 
$P_{s\sigma11}^u
=-1+\frac{2}{kv_{ts\sigma}}\Big\{-i\nu_s( \frac{k_x^2v_{\perp ts\sigma}^2}{k^2v_{ts\sigma}^2}Z_2+\frac{k_xv_{dsx}}{kv_{ts\sigma}}Z_1+\frac{k_z^2v_{zts}^2}{2k^2v_{ts\sigma}^2}Z_0)+ \frac{k^2k_x^2v_{\perp ts\sigma}^4}{k^3v_{ts\sigma}^3}Z_3+2\frac{k^2k_xv_{\perp ts\sigma}^2}{k^2v_{ts\sigma}^2}v_{dsx}Z_2+[\frac{1}{2} v_{\perp ts\sigma}^2\frac{k^2k_z^2v_{zts}^2}{k^3v_{ts\sigma}^3}   +\frac{k^2}{kv_{ts\sigma}}v_{dsx}^2+  (\lambda_{Ts\sigma}-1)\frac{k_x^2k_z^2v_{\perp ts\sigma}^4}{k^3v_{ts\sigma}^3}  ]Z_1 +  \frac{k_xk_z^2}{k^2v_{ts\sigma}^2}(\lambda_{Ts\sigma}-1)v_{\perp ts\sigma}^2 v_{dsx}Z_0\Big\}
\simeq-1+2\sum_{j=1}^J\frac{b_j}{\omega-c_{sj\sigma}}\Big\{-i\nu_s( \frac{k_x^2v_{\perp ts\sigma}^2}{k^2v_{ts\sigma}^2}c_j^2+\frac{k_xv_{dsx}}{kv_{ts\sigma}}c_j+\frac{k_z^2v_{zts}^2}{2k^2v_{ts\sigma}^2})+ \frac{k^2k_x^2v_{\perp ts\sigma}^4}{k^3v_{ts\sigma}^3}c_j^3+2\frac{k^2k_xv_{\perp ts\sigma}^2}{k^2v_{ts\sigma}^2}v_{dsx}c_j^2+[\frac{1}{2} v_{\perp ts\sigma}^2\frac{k^2k_z^2v_{zts}^2}{k^3v_{ts\sigma}^3}   +\frac{k^2}{kv_{ts\sigma}}v_{dsx}^2+  (\lambda_{Ts\sigma}-1)\frac{k_x^2k_z^2v_{\perp ts\sigma}^4}{k^3v_{ts\sigma}^3}  ]c_j +  \frac{k_xk_z^2}{k^2v_{ts\sigma}^2}(\lambda_{Ts\sigma}-1)v_{\perp ts\sigma}^2 v_{dsx}\Big\}=-1+\sum_{j=1}^J\frac{p_{11sj\sigma}}{\omega-c_{sj\sigma}}$,

\item $P_{s\sigma12}^u
=\frac{2v_{dsy}}{kv_{ts\sigma}}\Big\{ \frac{k^2k_xv_{\perp ts\sigma}^2}{k^2v_{ts\sigma}^2} Z_2+\frac{k^2}{kv_{ts\sigma}} v_{dsx} Z_1+\frac{1}{2}\frac{k_xk_z^2}{k^2v_{ts\sigma}^2} (\lambda_{Ts\sigma}-1)v_{\perp ts\sigma}^2Z_0\Big\}
\simeq2v_{dsy}\sum_{j=1}^J\frac{b_j}{\omega-c_{sj\sigma}}\Big\{ \frac{k^2k_xv_{\perp ts\sigma}^2}{k^2v_{ts\sigma}^2} c_j^2+\frac{k^2v_{dsx}}{kv_{ts\sigma}}  c_j+\frac{1}{2}\frac{k_xk_z^2}{k^2v_{ts\sigma}^2} (\lambda_{Ts\sigma}-1)v_{\perp ts\sigma}^2\Big\}=\sum_{j=1}^J\frac{p_{12sj\sigma}}{\omega-c_{sj\sigma}}$,

\item $P_{s\sigma21}^u
=\frac{2v_{dsy}}{kv_{ts\sigma}}\Big\{ -i\nu_s\frac{k_x}{kv_{ts\sigma}}Z_1+  \frac{k^2k_xv_{\perp ts\sigma}^2}{k^2v_{ts\sigma}^2} Z_2+ \frac{k^2}{kv_{ts\sigma}} v_{dsx} Z_1+ \frac{1}{2}\frac{k_xk_z^2}{k^2v_{ts\sigma}^2} (\lambda_{Ts\sigma}-1)v_{\perp ts\sigma}^2Z_0\Big\}
\simeq2v_{dsy}\sum_{j=1}^J\frac{b_j}{\omega-c_{sj\sigma}}\Big\{ -i\nu_s\frac{k_x}{kv_{ts\sigma}}c_j+  \frac{k^2k_xv_{\perp ts\sigma}^2}{k^2v_{ts\sigma}^2} c_j^2+ \frac{k^2}{kv_{ts\sigma}} v_{dsx} c_j+ \frac{1}{2}\frac{k_xk_z^2}{k^2v_{ts\sigma}^2} (\lambda_{Ts\sigma}-1)v_{\perp ts\sigma}^2 \Big\}=\sum_{j=1}^J\frac{p_{21sj\sigma}}{\omega-c_{sj\sigma}}$,

\item $P_{s\sigma22}^u
=-1{\color{red}+} \frac{2}{kv_{ts\sigma}}\Big\{-i\nu_s\frac{1}{2}Z_0+ (\frac{1}{2}v_{\perp ts\sigma}^2+v_{dsy}^2)\frac{k^2}{kv_{ts\sigma}}  Z_1 \Big\}
\simeq-1{\color{red}+} 2\sum_{j=1}^J\frac{b_j}{\omega-c_{sj\sigma}}\Big\{-i\nu_s\frac{1}{2}+ (\frac{1}{2}v_{\perp ts\sigma}^2+v_{dsy}^2)\frac{k^2}{kv_{ts\sigma}}  c_j \Big\}=-1+\sum_{j=1}^J\frac{p_{22sj\sigma}}{\omega-c_{sj\sigma}}$,

\item 
$P_{s\sigma13}^u
= \frac{2}{kv_{ts\sigma}}\Big\{-i\nu_s( \frac{k_xk_zv_{\perp ts\sigma}^2}{k^2v_{ts\sigma}^2}Z_2 +\frac{k_zv_{dsx}}{kv_{ts\sigma}}Z_1 - \frac{k_xk_zv_{\perp ts\sigma}^2}{2k^2v_{ts\sigma}^2}Z_0)
+  \frac{k^2k_xk_zv_{zts}^2v_{\perp ts\sigma}^2}{k^3v_{ts\sigma}^3}Z_3 
+\frac{k^2}{kv_{ts\sigma}}  [ \frac{k_xv_{\perp ts\sigma}^2}{kv_{ts\sigma}}v_{dsz} +v_{dsx}\frac{k_zv_{zts}^2}{kv_{ts\sigma}}]Z_2 
+[\frac{k^2}{kv_{ts\sigma}}  v_{dsx}v_{dsz}
-\frac{1}{2}\frac{k^2k_xk_zv_{zts}^2v_{\perp ts\sigma}^2}{k^3v_{ts\sigma}^3} 
 -\frac{1}{2} (\lambda_{Ts\sigma}-1) \frac{k_x^3k_zv_{\perp ts\sigma}^4}{k^3v_{ts\sigma}^3} 
+\frac{1}{2}(\lambda_{Ts\sigma}-1)\frac{k_xk_z^3v_{zts}^2v_{\perp ts\sigma}^2}{k^3v_{ts\sigma}^3}]Z_1 
+[\frac{1}{2}(\lambda_{Ts\sigma}-1)\frac{k_xk_z^2v_{\perp ts\sigma}^2}{k^2v_{ts\sigma}^2}v_{dsz} -\frac{1}{2} (\lambda_{Ts\sigma}-1)v_{dsx} \frac{k_x^2k_zv_{\perp ts\sigma}^2}{k^2v_{ts\sigma}^2}]Z_0\Big\}
\simeq 2\sum_{j=1}^J\frac{b_j}{\omega-c_{sj\sigma}}\Big\{-i\nu_s( \frac{k_xk_zv_{\perp ts\sigma}^2}{k^2v_{ts\sigma}^2}c_j^2 +\frac{k_zv_{dsx}}{kv_{ts\sigma}}c_j - \frac{k_xk_zv_{\perp ts\sigma}^2}{2k^2v_{ts\sigma}^2})
+  \frac{k^2k_xk_zv_{zts}^2v_{\perp ts\sigma}^2}{k^3v_{ts\sigma}^3}c_j^3
+\frac{k^2}{kv_{ts\sigma}}  [ \frac{k_xv_{\perp ts\sigma}^2}{kv_{ts\sigma}}v_{dsz} +v_{dsx}\frac{k_zv_{zts}^2}{kv_{ts\sigma}}]c_j^2 
+[\frac{k^2}{kv_{ts\sigma}}  v_{dsx}v_{dsz}
-\frac{1}{2}\frac{k^2k_xk_zv_{zts}^2v_{\perp ts\sigma}^2}{k^3v_{ts\sigma}^3} 
 -\frac{1}{2} (\lambda_{Ts\sigma}-1) \frac{k_x^3k_zv_{\perp ts\sigma}^4}{k^3v_{ts\sigma}^3} 
+\frac{1}{2}(\lambda_{Ts\sigma}-1)\frac{k_xk_z^3v_{zts}^2v_{\perp ts\sigma}^2}{k^3v_{ts\sigma}^3}]c_j
+[\frac{1}{2}(\lambda_{Ts\sigma}-1)\frac{k_xk_z^2v_{\perp ts\sigma}^2}{k^2v_{ts\sigma}^2}v_{dsz} -\frac{1}{2} (\lambda_{Ts\sigma}-1)v_{dsx} \frac{k_x^2k_zv_{\perp ts\sigma}^2}{k^2v_{ts\sigma}^2}] \Big\}=\sum_{j=1}^J\frac{p_{13sj\sigma}}{\omega-c_{sj\sigma}}$,

\item 
$P_{s\sigma31}^u
= \frac{2}{kv_{ts\sigma}}\Big\{-i\nu_s(  \frac{k_xk_zv_{zts}^2}{k^2v_{ts\sigma}^2}Z_2+ \frac{k_xv_{dsz}}{kv_{ts\sigma}}Z_1- \frac{k_xk_zv_{zts}^2}{2k^2v_{ts\sigma}^2}Z_0)+  \frac{k^2k_xk_zv_{zts}^2v_{\perp ts\sigma}^2}{k^3v_{ts\sigma}^3}Z_3 
+\frac{k^2}{kv_{ts\sigma}}  [ \frac{k_xv_{\perp ts\sigma}^2}{kv_{ts\sigma}}v_{dsz} +v_{dsx}\frac{k_zv_{zts}^2}{kv_{ts\sigma}}]Z_2 
+[\frac{k^2}{kv_{ts\sigma}}  v_{dsx}v_{dsz}
-\frac{1}{2}\frac{k^2k_xk_zv_{zts}^2v_{\perp ts\sigma}^2}{k^3v_{ts\sigma}^3} 
 -\frac{1}{2} (\lambda_{Ts\sigma}-1) \frac{k_x^3k_zv_{\perp ts\sigma}^4}{k^3v_{ts\sigma}^3} 
+\frac{1}{2}(\lambda_{Ts\sigma}-1)\frac{k_xk_z^3v_{zts}^2v_{\perp ts\sigma}^2}{k^3v_{ts\sigma}^3}]Z_1 
+[\frac{1}{2}(\lambda_{Ts\sigma}-1)\frac{k_xk_z^2v_{\perp ts\sigma}^2}{k^2v_{ts\sigma}^2}v_{dsz} -\frac{1}{2} (\lambda_{Ts\sigma}-1)v_{dsx} \frac{k_x^2k_zv_{\perp ts\sigma}^2}{k^2v_{ts\sigma}^2}]Z_0\Big\}
\simeq 2\sum_{j=1}^J\frac{b_j}{\omega-c_{sj\sigma}}\Big\{-i\nu_s(  \frac{k_xk_zv_{zts}^2}{k^2v_{ts\sigma}^2}c_j^2+ \frac{k_xv_{dsz}}{kv_{ts\sigma}}c_j- \frac{k_xk_zv_{zts}^2}{2k^2v_{ts\sigma}^2} )+  \frac{k^2k_xk_zv_{zts}^2v_{\perp ts\sigma}^2}{k^3v_{ts\sigma}^3}c_j^3
+\frac{k^2}{kv_{ts\sigma}}  [ \frac{k_xv_{\perp ts\sigma}^2}{kv_{ts\sigma}}v_{dsz} +v_{dsx}\frac{k_zv_{zts}^2}{kv_{ts\sigma}}]c_j^2 
+[\frac{k^2}{kv_{ts\sigma}}  v_{dsx}v_{dsz}
-\frac{1}{2}\frac{k^2k_xk_zv_{zts}^2v_{\perp ts\sigma}^2}{k^3v_{ts\sigma}^3} 
 -\frac{1}{2} (\lambda_{Ts\sigma}-1) \frac{k_x^3k_zv_{\perp ts\sigma}^4}{k^3v_{ts\sigma}^3} 
+\frac{1}{2}(\lambda_{Ts\sigma}-1)\frac{k_xk_z^3v_{zts}^2v_{\perp ts\sigma}^2}{k^3v_{ts\sigma}^3}]c_j 
+[\frac{1}{2}(\lambda_{Ts\sigma}-1)\frac{k_xk_z^2v_{\perp ts\sigma}^2}{k^2v_{ts\sigma}^2}v_{dsz} -\frac{1}{2} (\lambda_{Ts\sigma}-1)v_{dsx} \frac{k_x^2k_zv_{\perp ts\sigma}^2}{k^2v_{ts\sigma}^2}] \Big\}=\sum_{j=1}^J\frac{p_{31sj\sigma}}{\omega-c_{sj\sigma}}$,

\item $P_{s\sigma23}^u
=\frac{2v_{dsy}}{kv_{ts\sigma}}\Big\{ -i\nu_s\frac{k_z}{kv_{ts\sigma}}Z_1+  \frac{k^2k_zv_{zts}^2}{k^2v_{ts\sigma}^2} Z_2+\frac{k^2}{kv_{ts\sigma}} v_{dsz} Z_1-\frac{1}{2}(\lambda_{Ts\sigma}-1) \frac{k_x^2k_zv_{\perp ts\sigma}^2}{k^2v_{ts\sigma}^2} Z_0\Big\}
\simeq2v_{dsy}\sum_{j=1}^J\frac{b_j}{\omega-c_{sj\sigma}}\Big\{ -i\nu_s\frac{k_z}{kv_{ts\sigma}}c_j+  \frac{k^2k_zv_{zts}^2}{k^2v_{ts\sigma}^2} c_j^2+\frac{k^2}{kv_{ts\sigma}} v_{dsz} c_j-\frac{1}{2}(\lambda_{Ts\sigma}-1) \frac{k_x^2k_zv_{\perp ts\sigma}^2}{k^2v_{ts\sigma}^2} \Big\}=\sum_{j=1}^J\frac{p_{23sj\sigma}}{\omega-c_{sj\sigma}}$,

\item $P_{s\sigma32}^u
=\frac{2v_{dsy}}{kv_{ts\sigma}}\Big\{   \frac{k^2k_zv_{zts}^2}{k^2v_{ts\sigma}^2}Z_2+\frac{k^2}{kv_{ts\sigma}} v_{dsz}Z_1-\frac{1}{2}(\lambda_{Ts\sigma}-1) \frac{k_x^2k_zv_{\perp ts\sigma}^2}{k^2v_{ts\sigma}^2}Z_0\Big\}
\simeq2v_{dsy}\sum_{j=1}^J\frac{b_j}{\omega-c_{sj\sigma}}\Big\{   \frac{k^2k_zv_{zts}^2}{k^2v_{ts\sigma}^2}c_j^2+\frac{k^2}{kv_{ts\sigma}} v_{dsz}c_j-\frac{1}{2}(\lambda_{Ts\sigma}-1) \frac{k_x^2k_zv_{\perp ts\sigma}^2}{k^2v_{ts\sigma}^2} \Big\}=\sum_{j=1}^J\frac{p_{32sj\sigma}}{\omega-c_{sj\sigma}}$,

\item 
$P_{s\sigma33}^u
=-1+\frac{2}{kv_{ts\sigma}}\Big\{-i\nu_s( \frac{k_z^2v_{zts}^2}{k^2v_{ts\sigma}^2}Z_2+\frac{k_zv_{dsz}}{kv_{ts\sigma}}Z_1 +\frac{k_x^2v_{\perp ts\sigma}^2}{2k^2v_{ts\sigma}^2}Z_0)+ 
  \frac{k^2k_z^2v_{zts}^4}{k^3v_{ts\sigma}^3}Z_3
+2 \frac{k^2k_zv_{zts}^2}{k^2v_{ts\sigma}^2}v_{dsz}Z_2
+[\frac{1}{2}  \frac{k^2k_x^2v_{zts}^2v_{\perp ts\sigma}^2}{k^3v_{ts\sigma}^3}  +\frac{k^2}{kv_{ts\sigma}}  v_{dsz}^2-  (\lambda_{Ts\sigma}-1) \frac{k_x^2k_z^2v_{zts}^2v_{\perp ts\sigma}^2}{k^3v_{ts\sigma}^3}]Z_1 
-   (\lambda_{Ts\sigma}-1) \frac{k_x^2k_zv_{\perp ts\sigma}^2}{k^2v_{ts\sigma}^2}v_{dsz}Z_0\Big\}
\simeq-1+2\sum_{j=1}^J\frac{b_j}{\omega-c_{sj\sigma}}\Big\{-i\nu_s( \frac{k_z^2v_{zts}^2}{k^2v_{ts\sigma}^2}c_j^2+\frac{k_zv_{dsz}}{kv_{ts\sigma}}c_j +\frac{k_x^2v_{\perp ts\sigma}^2}{2k^2v_{ts\sigma}^2} )+ 
  \frac{k^2k_z^2v_{zts}^4}{k^3v_{ts\sigma}^3}c_j^3
+2 \frac{k^2k_zv_{zts}^2}{k^2v_{ts\sigma}^2}v_{dsz}c_j^2
+[\frac{1}{2}  \frac{k^2k_x^2v_{zts}^2v_{\perp ts\sigma}^2}{k^3v_{ts\sigma}^3}  +\frac{k^2}{kv_{ts\sigma}}  v_{dsz}^2-  (\lambda_{Ts\sigma}-1) \frac{k_x^2k_z^2v_{zts}^2v_{\perp ts\sigma}^2}{k^3v_{ts\sigma}^3}]c_j 
-   (\lambda_{Ts\sigma}-1) \frac{k_x^2k_zv_{\perp ts\sigma}^2}{k^2v_{ts\sigma}^2}v_{dsz} \Big\}=-1+\sum_{j=1}^J\frac{p_{33sj\sigma}}{\omega-c_{sj\sigma}}$.

\end{itemize}
In the above, for example, $p_{11sj\sigma}=2b_j\Big\{-i\nu_s( \frac{k_x^2v_{\perp ts\sigma}^2}{k^2v_{ts\sigma}^2}c_j^2+\frac{k_xv_{dsx}}{kv_{ts\sigma}}c_j+\frac{k_z^2v_{zts}^2}{2k^2v_{ts\sigma}^2})+ \frac{k^2k_x^2v_{\perp ts\sigma}^4}{k^3v_{ts\sigma}^3}c_j^3+2\frac{k^2k_xv_{\perp ts\sigma}^2}{k^2v_{ts\sigma}^2}v_{dsx}c_j^2+[\frac{1}{2} v_{\perp ts\sigma}^2\frac{k^2k_z^2v_{zts}^2}{k^3v_{ts\sigma}^3}   +\frac{k^2}{kv_{ts\sigma}}v_{dsx}^2+  (\lambda_{Ts\sigma}-1)\frac{k_x^2k_z^2v_{\perp ts\sigma}^4}{k^3v_{ts\sigma}^3}  ]c_j +  \frac{k_xk_z^2}{k^2v_{ts\sigma}^2}(\lambda_{Ts\sigma}-1)v_{\perp ts\sigma}^2 v_{dsx}\Big\}$, and others are similar and thus we have not written them out explicitly.

{\color{red}We find the result is very simple by use $Z_{0,1,2,3}$, which would also make the EM3D-M case be much simpler than the previous PASS-K \cite{Xie2016} derivation. This is also why we use $Z_{0,1}$ in the ES3D case, which gives a more compact form.}

Thus, we obtain the relations between $\bm J^u$ and $\bm E$, which has the following form (with $\sum_{s=u}$)
\begin{equation}\label{eq:JuE}
 \left( \begin{array}{c}J_x^u \\ J_y^u \\ J_z^u\end{array}\right)
    =-i\epsilon_0\left( \begin{array}{ccc}
    \frac{b_{11}^u}{\omega}+\sum_{sj\sigma}\frac{b_{sj\sigma11}}{\omega-c_{sj\sigma}} & \frac{b_{12}^u}{\omega}+\sum_{sj\sigma}\frac{b_{sj\sigma12}}{\omega-c_{sj\sigma}}
    & \frac{b_{13}^u}{\omega}+\sum_{sj\sigma}\frac{b_{sj\sigma13}}{\omega-c_{sj\sigma}}\\
    \frac{b_{21}^u}{\omega}+\sum_{sj\sigma}\frac{b_{sj\sigma21}}{\omega-c_{sj\sigma}} & \frac{b_{22}^u}{\omega}+\sum_{sj\sigma}\frac{b_{sj\sigma22}}{\omega-c_{sj\sigma}}
    & \frac{b_{23}^u}{\omega}+\sum_{sj\sigma}\frac{b_{sj\sigma23}}{\omega-c_{sj\sigma}} \\
    \frac{b_{31}^u}{\omega}+\sum_{sj\sigma}\frac{b_{sj\sigma31}}{\omega-c_{sj\sigma}} & \frac{b_{32}^u}{\omega}+\sum_{sj\sigma}\frac{b_{sj\sigma32}}{\omega-c_{sj\sigma}}
    & \frac{b_{33}^u}{\omega}+\sum_{sj\sigma}\frac{b_{sj\sigma33}}{\omega-c_{sj\sigma}}
    \end{array}\right) \left( \begin{array}{c}E_x \\ E_y \\
    E_z\end{array}\right),
\end{equation}
with the coefficients
\begin{equation}\label{eq:JEmatu}
    \left\{ \begin{array}{ccc}
   && b_{sj\sigma11} =  r_{s\sigma}\omega_{ps}^2p_{11sj\sigma}/c_{sj\sigma},~~~~~~~~
    b_{11}^u = - \sum_{s}^{s=u}\omega_{ps}^2[1+ \sum_{\sigma}r_{s\sigma}\sum_{j}p_{11sj\sigma}/c_{sj\sigma}],\\
    
    &&b_{sj\sigma12} =  r_{s\sigma}\omega_{ps}^2p_{12sj\sigma}/c_{sj\sigma},~~~~~~~~
    b_{12}^u = - \sum_{s}^{s=u}\omega_{ps}^2[\sum_{\sigma}r_{s\sigma}\sum_{j}p_{12sj\sigma}/c_{sj\sigma}],\\

   && b_{sj\sigma21} =  r_{s\sigma}\omega_{ps}^2p_{21sj\sigma}/c_{sj\sigma},~~~~~~~~
    b_{21}^u = - \sum_{s}^{s=u}\omega_{ps}^2[\sum_{\sigma}r_{s\sigma}\sum_{j}p_{21sj\sigma}/c_{sj\sigma}],\\

   && b_{sj\sigma22} =  r_{s\sigma}\omega_{ps}^2p_{22sj\sigma}/c_{sj\sigma},~~~~~~~~
    b_{22}^u = - \sum_{s}^{s=u}\omega_{ps}^2[1+ \sum_{\sigma}r_{s\sigma}\sum_{j}p_{22sj\sigma}/c_{sj\sigma}],\\

   && b_{sj\sigma13} =  r_{s\sigma}\omega_{ps}^2p_{13sj\sigma}/c_{sj\sigma},~~~~~~~~
    b_{13}^u = - \sum_{s}^{s=u}\omega_{ps}^2[\sum_{\sigma}r_{s\sigma}\sum_{j}p_{13sj\sigma}/c_{sj\sigma}],\\

   && b_{sj\sigma31} =  r_{s\sigma}\omega_{ps}^2p_{31sj\sigma}/c_{sj\sigma},~~~~~~~~
    b_{31}^u = - \sum_{s}^{s=u}\omega_{ps}^2[\sum_{\sigma}r_{s\sigma}\sum_{j}p_{31sj\sigma}/c_{sj\sigma}],\\

  &&  b_{sj\sigma23} =  r_{s\sigma}\omega_{ps}^2p_{23sj\sigma}/c_{sj\sigma},~~~~~~~~
    b_{23}^u = - \sum_{s}^{s=u}\omega_{ps}^2[\sum_{\sigma}r_{s\sigma}\sum_{j}p_{23sj\sigma}/c_{sj\sigma}],\\

   && b_{sj\sigma32} =  r_{s\sigma}\omega_{ps}^2p_{32sj\sigma}/c_{sj\sigma},~~~~~~~~
    b_{32}^u = - \sum_{s}^{s=u}\omega_{ps}^2[\sum_{\sigma}r_{s\sigma}\sum_{j}p_{32sj\sigma}/c_{sj\sigma}],\\

   && b_{sj\sigma33} =  r_{s\sigma}\omega_{ps}^2p_{33sj\sigma}/c_{sj\sigma},~~~~~~~~
    b_{33}^u = - \sum_{s}^{s=u}\omega_{ps}^2[1+ \sum_{\sigma}r_{s\sigma}\sum_{j}p_{33sj\sigma}/c_{sj\sigma}],\\

    &&c_{sj\sigma} = k_zv_{dsz}+k_xv_{dsx}-i\nu_s+kv_{ts\sigma}c_j,
    \end{array}\right.
\end{equation}

Considering that $c_j$ are complex number and $k_x$, $k_z$ and $\nu_s$ are real number, the singularity of the above form is also only $k=0$.

\subsubsection{The magnetized terms}

Similarly to the unmagnetized case, considering the defination $\bm \sigma_s^m=-i\epsilon_0\omega\bm Q_s^m=-i\epsilon_0\sum_{\sigma=a,b}r_{s\sigma}\frac{\omega_{ps}^2}{\omega}\bm P_{s\sigma}^m$, after $J$-pole expansion, we have
\begin{itemize}

\item $P_{s\sigma11}^{m}\simeq\sum_{n=-\infty}^{\infty} \sum_{j=1}^J\frac{k_zv_{zts}b_j}{\omega-c_{snj}}\Big\{[\frac{n\omega_{cs}}{k_x }(n\omega_{cs}+k_xv_{dsx}-i\nu_s)\frac{ A_{nbs\sigma}}{v_{\perp ts\sigma}^2}\frac{1}{k_zv_{zts}}+A_{n0\sigma} (\frac{n\omega_{cs}}{k_x } +v_{dsx})\frac{ c_j}{v_{zts}^2}]( \frac{n\omega_{cs}}{k_x }+v_{dsx}) \Big\}-{\color{red}1}=-1+\sum_n\sum_j\frac{p_{11snj}}{\omega-c_{snj}}$.

\item $P_{s\sigma12}^{m}\simeq\sum_{n=-\infty}^{\infty}\sum_{j=1}^J\frac{k_zv_{zts}b_j}{\omega-c_{snj}} \Big\{ v_{dsy}( \frac{n\omega_{cs}}{k_x }+v_{dsx})(n\omega_{cs}\frac{ A_{nbs\sigma}}{v_{\perp ts\sigma}^2}\frac{1}{k_zv_{zts}}+A_{n0\sigma} \frac{c_j}{v_{zts}^2})+i( \frac{n\omega_{cs}}{k_x }+v_{dsx})[(n\omega_{cs}-i\nu_s)\frac{ B_{nbs\sigma}}{v_{\perp ts\sigma}}\frac{1}{k_zv_{zts}}+\frac{B_{n0\sigma} v_{\perp ts\sigma}c_j}{v_{zts}^2}]\Big\}= \sum_n\sum_j\frac{p_{12snj}}{\omega-c_{snj}}$.

\item $P_{s\sigma21}^{m}\simeq\sum_{n=-\infty}^{\infty}\sum_{j=1}^J\frac{k_zv_{zts}b_j}{\omega-c_{snj}} \Big\{ \frac{n\omega_{cs}}{k_x}(n\omega_{cs}+k_xv_{dsx}-i\nu_s)[-iv_{\perp ts\sigma}B_{nbs\sigma}+v_{dsy}A_{nbs\sigma}]\frac{ 1}{v_{\perp ts\sigma}^2}\frac{1}{k_zv_{zts}}+[-iv_{\perp ts\sigma}B_{n0\sigma}+v_{dsy}A_{n0\sigma}] (\frac{n\omega_{cs}}{k_x }+v_{dsx})\frac{c_j}{v_{zts}^2}\Big\}=\sum_n\sum_j\frac{p_{21snj}}{\omega-c_{snj}}$.

\item $P_{s\sigma22}^{m}\simeq\sum_{n=-\infty}^{\infty}\sum_{j=1}^J\frac{k_zv_{zts}b_j}{\omega-c_{snj}} \Big\{ v_{dsy}[-iv_{\perp ts\sigma} B_{nbs\sigma}+v_{dsy}A_{nbs\sigma}] \frac{n\omega_{cs}}{v_{\perp ts\sigma}^2}\frac{1}{k_zv_{zts}}+v_{dsy}[-iv_{\perp ts\sigma} B_{n0\sigma}+v_{dsy}A_{n0\sigma}] \frac{ c_j}{v_{zts}^2}+v_{\perp ts\sigma} [v_{\perp ts\sigma} C_{nbs\sigma}+iv_{dsy}B_{nbs\sigma}]( n\omega_{cs}-i\nu_s)\frac{1}{v_{\perp ts\sigma}^2}\frac{1}{k_zv_{zts}}+v_{\perp ts\sigma} [v_{\perp ts\sigma} C_{n0\sigma}+iv_{dsy}B_{n0\sigma}]\frac{ c_j }{v_{zts}^2} \Big\}-{\color{red}1}=-1+\sum_n\sum_j\frac{p_{22snj}}{\omega-c_{snj}}$.

\item $P_{s\sigma13}^{m}\simeq\sum_{n=-\infty}^{\infty}\sum_{j=1}^J\frac{k_zv_{zts}b_j}{\omega-c_{snj}} \Big\{ ( \frac{n\omega_{cs}}{k_x }+v_{dsx})[n\omega_{cs}  (\frac{c_j}{k_z}+v_{dsz}\frac{1}{k_zv_{zts}})  \frac{A_{nbs\sigma}}{v_{\perp ts\sigma}^2}+A_{n0\sigma}( v_{zts}c_j^2+(k_z v_{dsz}-i\nu_s )\frac{c_j}{k_z} )\frac{  1}{v_{zts}^2}] \Big\}= \sum_n\sum_j\frac{p_{13snj}}{\omega-c_{snj}}$.

\item $P_{s\sigma31}^{m}\simeq\sum_{n=-\infty}^{\infty}\sum_{j=1}^J\frac{k_zv_{zts}b_j}{\omega-c_{snj}} \Big\{ \frac{n\omega_{cs}}{k_x }(n\omega_{cs}+k_xv_{dsx}-i\nu_s)\frac{ A_{nbs\sigma} }{v_{\perp ts\sigma}^2}(\frac{c_j}{k_z}+v_{dsz}\frac{1}{k_zv_{zts}})+A_{n0\sigma} (\frac{n\omega_{cs}}{k_x } +v_{dsx})(\frac{c_j^2}{v_{zts}}+v_{dsz} \frac{c_j}{v_{zts}^2})\Big\}= \sum_n\sum_j\frac{p_{31snj}}{\omega-c_{snj}}$.

\item $P_{s\sigma23}^{m}\simeq\sum_{n=-\infty}^{\infty}\sum_{j=1}^J\frac{k_zv_{zts}b_j}{\omega-c_{snj}} \Big\{ [v_{dsy}A_{nbs\sigma}-iv_{\perp ts\sigma} B_{nbs\sigma}][(\frac{c_j}{k_z}+\frac{ v_{dsz}}{k_zv_{tsz}})  \frac{ n\omega_{cs} }{v_{\perp ts\sigma}^2}]+ [v_{dsy}A_{n0\sigma}-iv_{\perp ts\sigma} B_{n0\sigma}][v_{zts}c_j^2+(k_z v_{dsz}-i\nu_s)\frac{c_j}{k_z}]\frac{ 1}{v_{zts}^2}  \Big\} = \sum_n\sum_j\frac{p_{23snj}}{\omega-c_{snj}}$.

\item $P_{s\sigma32}^{m}\simeq\sum_{n=-\infty}^{\infty}\sum_{j=1}^J\frac{k_zv_{zts}b_j}{\omega-c_{snj}} \Big\{  [ n\omega_{cs} \frac{ A_{nbs\sigma}}{v_{\perp ts\sigma}^2}\frac{c_j}{k_z}+A_{n0\sigma} \frac{k_z v_{zts}c_j^2 }{v_{zts}^2}]v_{dsy}+ [ n\omega_{cs} \frac{ A_{nbs\sigma}}{v_{\perp ts\sigma}^2}\frac{1}{k_zv_{zts}}+A_{n0\sigma} \frac{c_j }{v_{zts}^2}]v_{dsy}v_{dsz} +i [(n\omega_{cs} -i\nu_s )\frac{ B_{nbs\sigma}}{v_{\perp ts\sigma}}\frac{c_j}{k_z}+B_{n0\sigma} \frac{  v_{\perp ts\sigma}c_j^2}{v_{zts} }]+i v_{dsz} [(n\omega_{cs} -i\nu_s )\frac{ B_{nbs\sigma}}{v_{\perp ts\sigma}}\frac{1}{k_zv_{zts}}+B_{n0\sigma} \frac{ v_{\perp ts\sigma}c_j}{v_{zts}^2}]   \Big\} = \sum_n\sum_j\frac{p_{32snj}}{\omega-c_{snj}}$.

\item $P_{s\sigma33}^{m}\simeq\sum_{n=-\infty}^{\infty}\sum_{j=1}^J\frac{k_zv_{zts}b_j}{\omega-c_{snj}} \Big\{  [n\omega_{cs}  (\frac{v_{zts}c_j^2}{k_z}+v_{dsz}\frac{ c_j}{k_z})\frac{ A_{nbs\sigma}}{v_{\perp ts\sigma}^2}+A_{n0\sigma}(  c_j^3+(k_z v_{dsz}-i\nu_s)\frac{c_j^2}{k_zv_{zts}}) ] + v_{dsz}[n\omega_{cs}  (\frac{c_j}{k_z}+v_{dsz}\frac{1}{k_zv_{zts}})\frac{ A_{nbs\sigma}}{v_{\perp ts\sigma}^2}+A_{n0\sigma}( \frac{c_j^2}{v_{zts}}+(k_z v_{dsz}-i\nu_s)\frac{c_j}{k_zv_{zts}^2})]  \Big\} -{\color{red}1}=-1+\sum_n\sum_j\frac{p_{33snj}}{\omega-c_{snj}}$.

\end{itemize}
In the above, say, $p_{11snj}= k_zv_{zts}b_j\Big\{[\frac{n\omega_{cs}}{k_x }(n\omega_{cs}+k_xv_{dsx}-i\nu_s)\frac{ A_{nbs\sigma}}{v_{\perp ts\sigma}^2}\frac{1}{k_zv_{zts}}+A_{n0\sigma} (\frac{n\omega_{cs}}{k_x } +v_{dsx})\frac{ c_j}{v_{zts}^2}]( \frac{n\omega_{cs}}{k_x }+v_{dsx}) \Big\}$, and other terms are similar.

It is thus easy to find that after $J$-pole  expansion, the relations between $\bm J^m$ and $\bm E$ has the following form (with $\sum_{s=m}$)
\begin{equation}\label{eq:JmE}
    \left( \begin{array}{c}J_x^m \\ J_y^m \\ J_z^m\end{array}\right)
    =-i\epsilon_0\left( \begin{array}{ccc}
    \frac{b_{11}^m}{\omega}+\sum_{snj}\frac{b_{snj11}}{\omega-c_{snj}} & \frac{b_{12}^m}{\omega}+\sum_{snj}\frac{b_{snj12}}{\omega-c_{snj}}
    & \frac{b_{13}^m}{\omega}+\sum_{snj}\frac{b_{snj13}}{\omega-c_{snj}}\\
    \frac{b_{21}^m}{\omega}+\sum_{snj}\frac{b_{snj21}}{\omega-c_{snj}} & \frac{b_{22}^m}{\omega}+\sum_{snj}\frac{b_{snj22}}{\omega-c_{snj}}
    & \frac{b_{23}^m}{\omega}+\sum_{snj}\frac{b_{snj23}}{\omega-c_{snj}} \\
    \frac{b_{31}^m}{\omega}+\sum_{snj}\frac{b_{snj31}}{\omega-c_{snj}} & \frac{b_{32}^m}{\omega}+\sum_{snj}\frac{b_{snj32}}{\omega-c_{snj}}
    & \frac{b_{33}^m}{\omega}+\sum_{snj}\frac{b_{snj33}}{\omega-c_{snj}}
    \end{array}\right) \left( \begin{array}{c}E_x \\ E_y \\
    E_z\end{array}\right).
\end{equation}
with the coefficients
\begin{equation}\label{eq:JEmatm}
    \left\{ \begin{array}{ccc}
   b_{snj11} =  \sum_{\sigma}r_{s\sigma}\omega_{ps}^2p_{11snj}/c_{snj},~
    b_{11}^m = - \sum_{s}^{s=m}\omega_{ps}^2\sum_{\sigma}r_{s\sigma}[\sum_n A_{nbs\sigma}\frac{n\omega_{cs}}{k_x  v_{\perp ts\sigma}^2}( \frac{n\omega_{cs}}{k_x }+v_{dsx})+ \sum_{nj}p_{11snj}/c_{snj}],\\
    
    b_{snj12} =  \sum_{\sigma}r_{s\sigma}\omega_{ps}^2p_{12snj}/c_{snj},~~~~~~
    b_{12}^m = - \sum_{s}^{s=m}\omega_{ps}^2[\sum_{\sigma}r_{s\sigma}\sum_{nj}p_{12snj}/c_{snj}],\\

    b_{snj21} =  \sum_{\sigma}r_{s\sigma}\omega_{ps}^2p_{21snj}/c_{snj},~~~~~~
    b_{21}^m = - \sum_{s}^{s=m}\omega_{ps}^2[\sum_{\sigma}r_{s\sigma}\sum_{nj}p_{21snj}/c_{snj}],\\

   b_{snj22} =  \sum_{\sigma}r_{s\sigma}\omega_{ps}^2p_{22snj}/c_{snj},~~
    b_{22}^m = - \sum_{s}^{s=m}\omega_{ps}^2\sum_{\sigma}r_{s\sigma}[\sum_n(C_{nbs\sigma}+i\frac{v_{dsy}}{v_{\perp ts\sigma} }B_{nbs\sigma})+\sum_{nj}p_{22snj}/c_{snj}],\\

   b_{snj13} =  \sum_{\sigma}r_{s\sigma}\omega_{ps}^2p_{13snj}/c_{snj},~~~~~~
    b_{13}^m = - \sum_{s}^{s=m}\omega_{ps}^2[\sum_{\sigma}r_{s\sigma}\sum_{nj}p_{13snj}/c_{snj}],\\

    b_{snj31} =  \sum_{\sigma}r_{s\sigma}\omega_{ps}^2p_{31snj}/c_{snj},~~~~~~
    b_{31}^m = - \sum_{s}^{s=m}\omega_{ps}^2[\sum_{\sigma}r_{s\sigma}\sum_{nj}p_{31snj}/c_{snj}],\\

   b_{snj23} =  \sum_{\sigma}r_{s\sigma}\omega_{ps}^2p_{23snj}/c_{snj},~~~~~~
    b_{23}^m = - \sum_{s}^{s=m}\omega_{ps}^2[\sum_{\sigma}r_{s\sigma}\sum_{nj}p_{23snj}/c_{snj}],\\

   b_{snj32} =  \sum_{\sigma}r_{s\sigma}\omega_{ps}^2p_{32snj}/c_{snj},~~~~~~
    b_{32}^m = - \sum_{s}^{s=m}\omega_{ps}^2[\sum_{\sigma}r_{s\sigma}\sum_{nj}p_{32snj}/c_{snj}],\\

  b_{snj33} =  \sum_{\sigma}r_{s\sigma}\omega_{ps}^2p_{33snj}/c_{snj},~~~~~~
    b_{33}^m = - \sum_{s}^{s=m}\omega_{ps}^2 \sum_{\sigma}r_{s\sigma}[\sum_n\frac{1}{2}A_{n0\sigma}+\sum_{nj}p_{33snj}/c_{snj}],\\

  c_{snj} = k_zv_{dsz}+n\omega_{cs}+k_xv_{dsx}-i\nu_s+k_zv_{zts}c_j.
    \end{array}\right.
\end{equation}
In numerical test, considering the cut of summation $n$, i.e., $\sum_n=\sum_{n=-N}^N$ with $N\neq\infty$, we find the above original form of $b_{11}^m$,  $b_{22}^m$ and  $b_{33}^m$, are better than the below form [to undetstand]
\begin{equation}\label{eq:JEmatm}
    \left\{ \begin{array}{ccc}
   &&b_{11}^m = - \sum_{s}^{s=m}\omega_{ps}^2[1+ \sum_{\sigma}r_{s\sigma}\sum_{nj}p_{11snj}/c_{snj}],\\
    
   &&  b_{22}^m = - \sum_{s}^{s=m}\omega_{ps}^2[1+ \sum_{\sigma}r_{s\sigma}\sum_{nj}p_{22snj}/c_{snj}],\\

   && b_{33}^m = - \sum_{s}^{s=m}\omega_{ps}^2[1+ \sum_{\sigma}r_{s\sigma}\sum_{nj}p_{33snj}/c_{snj}].
        \end{array}\right.
\end{equation}
It is readily to see that all the singularities from $\frac{1}{k_z}$ in $\bm P_{s\sigma}^m$ are removable. The $\frac{n\omega_{cs}}{k_x}$ singularities at $k_x=0$ in $P_{s\sigma11}^m$,  $P_{s\sigma12}^m$,  $P_{s\sigma21}^m$,  $P_{s\sigma13}^m$, $P_{s\sigma31}^m$ are also removable. Thus, the overall equations have no singularity and will not meet numerical difficulty. In the solver, to short the code, if $k_x\rho_s<k_{\delta}$ we set $k_x\rho_s=k_{\delta}$ for magnetized species in EM version. For example, we can set $k_{\delta}=10^{-30}$.
 
Combining Eqs. (\ref{eq:em3dmaxw}),  (\ref{eq:JuE}) and (\ref{eq:JmE}), the equivalent linear system for electromagnetic dispersion relation can be obtained as
\begin{equation}\label{eq:passkem}
    \left\{ \begin{array}{ccc}
    \omega v_{snjx}^{s=m} &=& c_{snj} v_{snjx} + b_{snj11} E_x + b_{snj12} E_y + b_{snj13} E_z, \\
    \omega v_{sj\sigma x}^{s=u} &=&  c_{sj\sigma } v_{sj\sigma x} + b_{sj\sigma 11} E_x + b_{sj\sigma 12} E_y + b_{sj\sigma 13} E_z, \\
    \omega j_x &=&  b_{11} E_x + b_{12} E_y + b_{13} E_z, \\
    i J_x\epsilon_0 &=& j_x+\sum_{snj}^{s=m}v_{snjx}+ \sum_{sj\sigma }^{s=u}v_{sj\sigma x}, \\
    \omega v_{snjy}^{s=m} &=& c_{snj} v_{snjy} + b_{snj21} E_x + b_{snj22} E_y + b_{snj23} E_z, \\
    \omega v_{sj\sigma y}^{s=u} &=& c_{sj\sigma } v_{sj\sigma y} + b_{sj\sigma 21} E_x + b_{sj\sigma 22} E_y + b_{sj\sigma 23} E_z, \\
    \omega j_y &=&  b_{21} E_x + b_{22} E_y + b_{23} E_z, \\
    iJ_y/\epsilon_0 &=& j_y+\sum_{snj}^{s=m}v_{snjy}+ \sum_{sj\sigma }^{s=u}v_{sj\sigma y}, \\
    \omega v_{snjz}^{s=m} &=& c_{snj} v_{snjz} + b_{snj31} E_x + b_{snj32} E_y + b_{snj33} E_z, \\
    \omega v_{sj\sigma z}^{s=u} &=& c_{sj\sigma } v_{sj\sigma z} + b_{sj\sigma 31} E_x + b_{sj\sigma 32} E_y + b_{sj\sigma 33} E_z, \\
    \omega j_z &=&  b_{31} E_x + b_{32} E_y + b_{33} E_z, \\
    iJ_z/\epsilon_0 &=& j_z+\sum_{snj}^{s=m}v_{snjz}+\sum_{sj\sigma }^{s=u}v_{sj\sigma z}, \\
    \omega E_x &=& c^2k_z B_y-iJ_x/\epsilon_0, \\
    \omega E_y &=& -c^2k_z B_x +c^2k_x B_z-iJ_y/\epsilon_0,\\
    \omega E_z &=& -c^2k_x B_y-iJ_z/\epsilon_0, \\
    \omega B_x &=& -k_z E_y, \\
    \omega B_y &=& k_z E_x - k_x E_z, \\
    \omega B_z &=& k_x E_y,
    \end{array}\right.
\end{equation}
which yields a sparse matrix eigenvalue problem, where $b_{11}=b_{11}^m+b_{11}^u$ and so on. The symbols $v_{snjx}$, $j_{x,y,z}$ and $J_{x,y,z}$ used here do not have direct physical meanings but are analogy to the perturbed velocity and current density in the fluid derivations of plasma waves. The elements of the eigenvector
$(E_x, E_y, E_z, B_x, B_y, B_z)$ still represent the original electric and magnetic fields. Thus, the polarization of the solutions can also be obtained in a straightforward manner. The dimension of the matrix is {\color{red}$N_N=3\times (N_{SmNJ}+N_{SuJ}+1)+6=3\times
\{[S_m\times(2\times N+1)+S_u\times2]\times J+1\}+6$}. And another good aspect of the final PASS-K matrix equation is that it is valid for arbitrary real number of $k_x$ and $k_z$, i.e., $\theta\in[0,2\pi]$ and the only requirement is $k\neq0$.

\subsection{The Darwin case}

Based on the electromagnetic result, the Darwin model case is straightforward, where the linear system for
{\color{red}${\bm J}={\bm J^m}+{\bm J^u}=({\bm{\sigma^m}+\bm{\sigma^u}})\cdot{\bm E}=\bm{\sigma}\cdot{\bm E}$} is the same to electromagnetic case. We only need to modify the the linear system of the Maxwell’s equations, which are also straightforward
\begin{subequations}
\begin{eqnarray}
  & \omega{\color{red}(\frac{\bm k\bm k}{k^2})}\cdot {\bm E} = -c^2\bm k\times{\bm B}-i{\bm J}/\epsilon_0,\\
  & \omega \bm I \cdot\bm B = \bm k\times{\bm E},
\end{eqnarray}
\end{subequations}
and the matrix eigenvalue problem becomes $\omega\bm M_B\cdot\bm X=\bm M_A\cdot\bm X$, where $\bm M_A$ is still the same as the electromagnetic one from Eq.(\ref{eq:passkem}) and $\bm M_B$ changes from unit matrix $\bm I_{N_N\times N_N}$ to 
\begin{equation}
\bm M_B=\left( \begin{array}{ccc}
 \bm I_{(N_N-6)\times (N_N-6)}  &  &     \\
 & (\frac{\bm k\bm k}{k^2})_{3\times3} &  \\
     &  & \bm I_{3\times3}
    \end{array}\right).
\end{equation}
Though $ (\frac{\bm k\bm k}{k^2})_{3\times3}$ may not be full rank matrix, the standard eigenvalue library, such as 'eig()' in Matlab, can solve the eigenvalue problem well. 

\subsection{The polarizations}

The matrix solver can obtain $(E_x,E_y,E_z,B_x,B_y,B_z)$ directly\footnote{In principle, PASS-K matrix can also obtain $(E_x,E_y,E_z)$ as in standard $3\times3$ matrix $\bm D\cdot\bm E=0$. To obtain group velocity $\bm v_g=d\omega/d\bm k$ or do ray tracing, we may also need $\partial D/\partial \omega$ and $\partial D/\partial \bm k$.} from the matrix eigenvalue problem. Considered that  the magnitude of the wave has no meaning for a linear system, we should do normalizations. We set $|E|=1 mV/m$ and $E_x=Re(E_x)$. 


Some other useful: electric field energy $U_E=\frac{\epsilon_0}{2}\bm E\cdot\bm E^*$, magnetic field energy $U_B=\frac{1}{2\mu_0}\bm B\cdot\bm B^*$, energy flux Poynting vector $\bm S=\frac{1}{\mu_0}\bm E\times\bm B^*$. $P=E_y/iE_x$.

\section{Benchmark}\label{sec:benchmark}

There exists numerous applications of this newly developed updated version PASS-K tool, we only show some typical benchmarks to the reader to get a flavor of it.

\begin{figure}[htbp]
\begin{center}
\includegraphics[width=16cm]{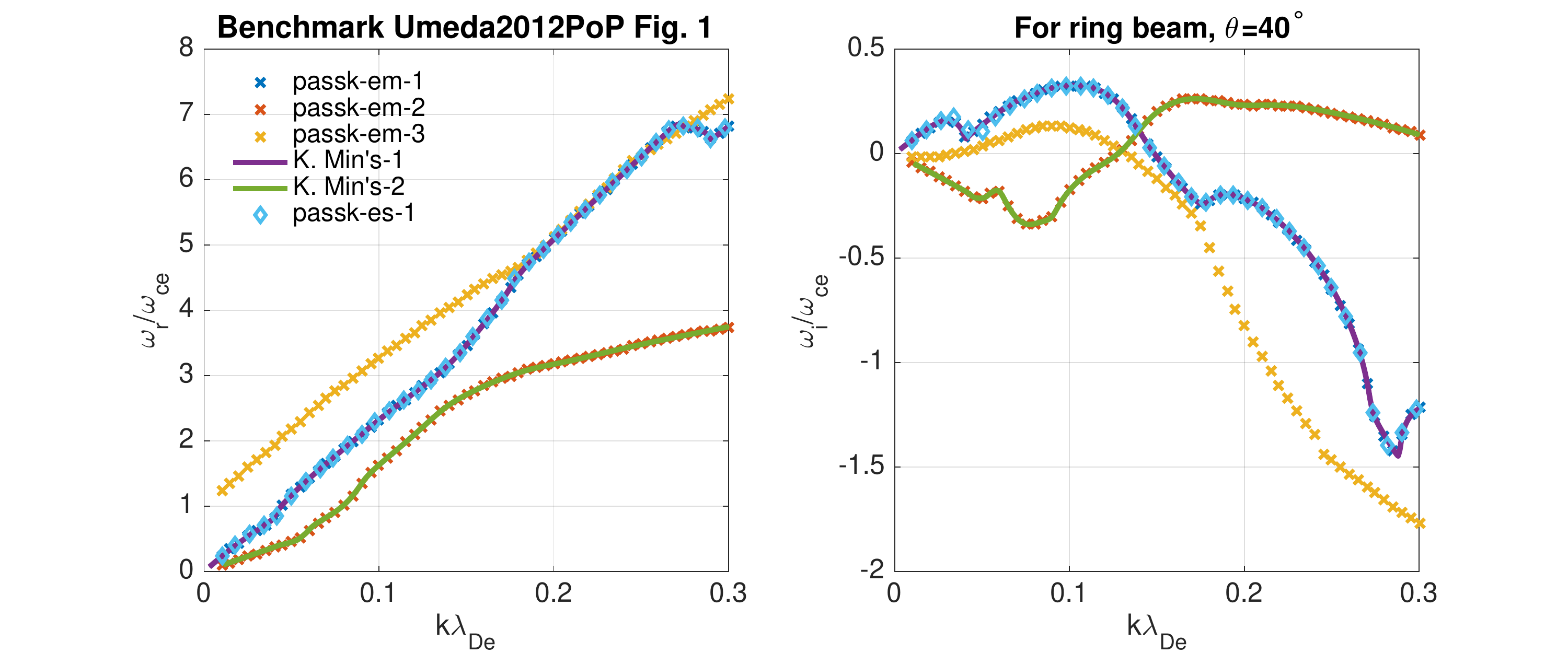}
\caption{Benchmark the electromagnetic ring beam dispersion relation between PASS-K and Min's code \cite{Min2015} with the parameters in Ref.\cite{Umeda2012} Fig.1 for $\theta=40^\circ$, which shows very good agreement. PASS-K gives all the three unstable branches, whereas the third branch 'passk-em-3' could easily be missed by the iterative root finding approaches used in Min's \cite{Min2015}  or Umeda's \cite{Umeda2012} codes. One branch of the electrostatic PASS-K solution is also shown for reference, which is close to the electromagnetic one implies that this branch is essentially an electrostatic mode. The  results also agree well with the PIC simulation result in Ref.\cite{Umeda2012}.}
\label{fig:benchmark_ring_beam}
\end{center}
\end{figure}

\begin{figure}[htbp]
\begin{center}
\includegraphics[width=16cm]{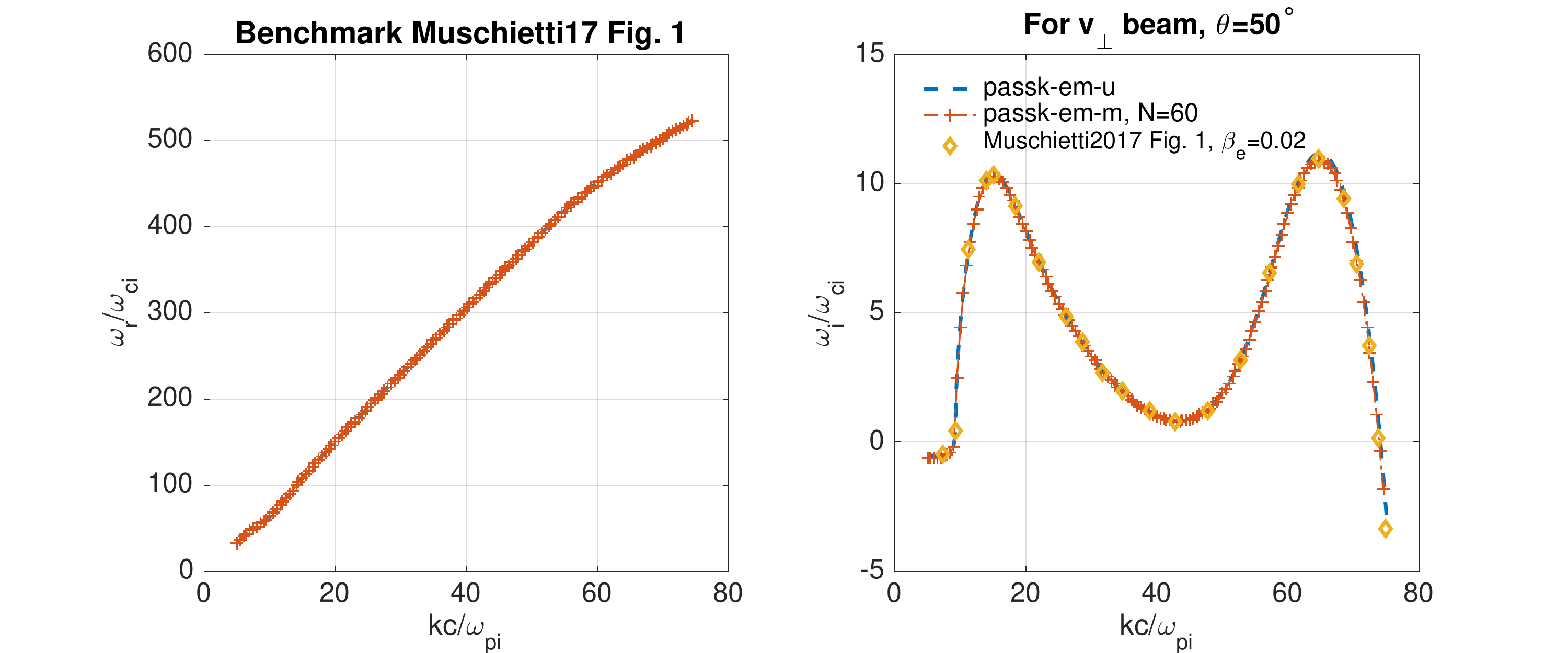}
\caption{Benchmark the electromagnetic penpendicular beam dispersion relation between PASS-K and Ref.\cite{Muschietti2017} Fig. 1 with the parameters $\theta=50^\circ$ and $\beta_e=0.02$, which shows very good agreement.  Ref.\cite{Muschietti2017} assumes unmagnetized ions and magnetized electron. 'passk-em-u' uses the same assumption; whereas 'passk-em-m' uses also magnetized ions with $N=60$ where the $\sum_n$  has convergent. The agreement between '-U' and '-M' versions also implies that the unmagnetized ions is a valid assumption for this case.}
\label{fig:passk_vs_Muschietti17_vdperp_theta=50}
\end{center}
\end{figure}

\begin{figure}[htbp]
\begin{center}
\includegraphics[width=16cm]{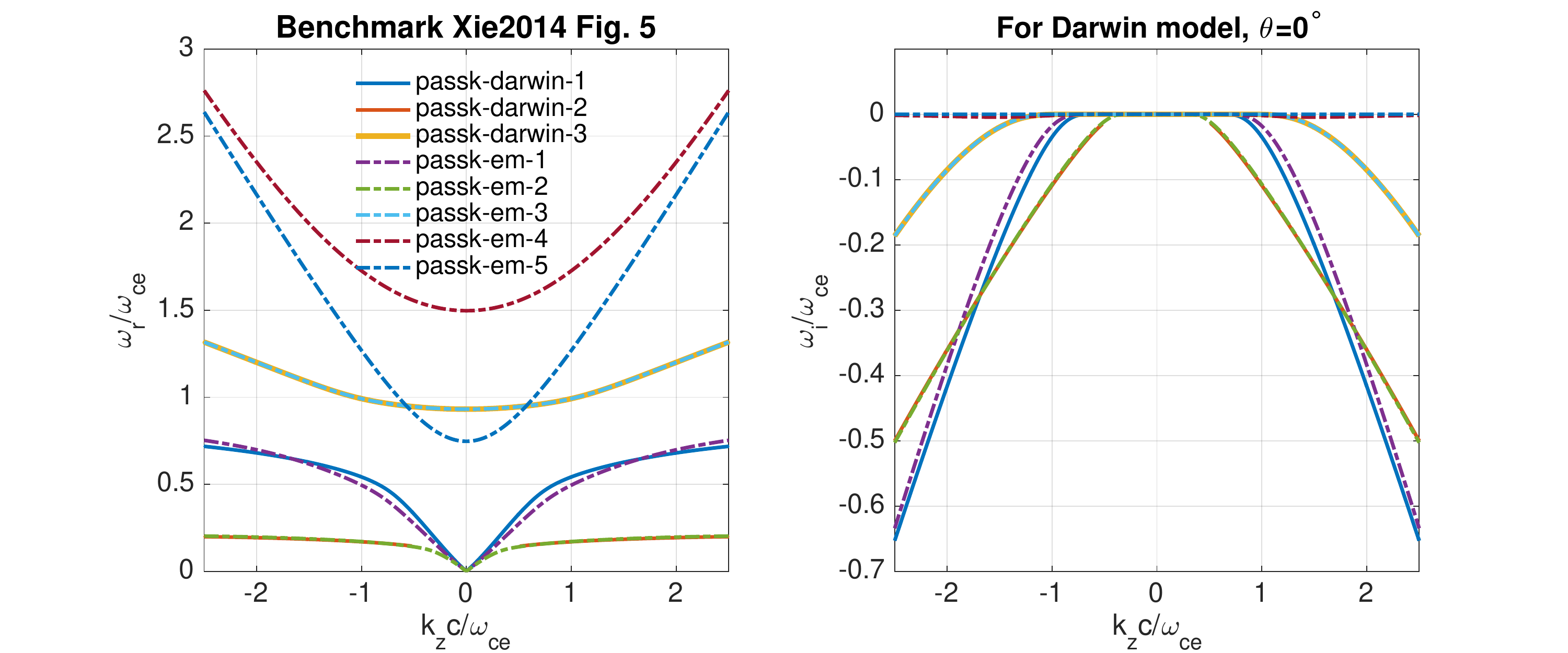}
\caption{Compare the electromagnetic model and Darwin model in PASS-K with the parameters in Ref.\cite{Xie2014} Fig. 5, $\theta=0^\circ$ and $\omega_{ce}/\omega_{pe}=1.2$, $c/v_{zte}=5/\sqrt{2}$, $m_i/m_e=4$, $T_i=T_e$. The $k_z<0$ solutions are also shown, which shows good symmetry to the $k_z>0$ solutions. The results show that Darwin model is a good approximate model in this case, and the slight deviations agree with the theoretical expectation in Ref.\cite{Xie2014}.}
\label{fig:passk_vs_Xie14_Darwin_theta=0}
\end{center}
\end{figure}

The first benchmark is the ring beam case in Ref.\cite{Umeda2012}, which is to make sure the function $A_n$, $B_n$ and $C_n$ are treated correctly in our model. The results are shown in Fig.\ref{fig:benchmark_ring_beam}, with very good agreement with Min's  \cite{Min2015}  code. The second benchmark shown in Fig.\ref{fig:passk_vs_Muschietti17_vdperp_theta=50} is the mixed of magnetized and unmagnetized species in Ref.\cite{Muschietti2017} for the instabilities driven by perpendicular beam in shock, which also show good agreement. And, the treatment of ion to be magnetized species also shows close result to the unmagnetized ion model, which implies that the unmagnetized ion assumption is valid in that case. This also gives us confidence of the validity of our magnetized model, since that the equations are totally different but yield similar solution as should be. The third benchmark shown in Fig.\ref{fig:passk_vs_Xie14_Darwin_theta=0} is for the Darwin model in Ref.\cite{Xie2014}, which is the same as the one solved using accurate $Z$ function with conventional iterative root finding in Fig.5 of Ref.\cite{Xie2014}. Here, we have also shown other branches and $k_z<0$ branches. The symmetry between $k_z<0$ and $k_z>0$ solutions implies that the $Z$ function is treated correctly for $k_z<0$ in this new solver.

The benchmark parameters for above cases are listed below.
\begin{itemize}
\item 
Defaultly we use SI unit: $c=2.99792458e8$, $\epsilon_0=8.854187817e-12$, $k_B=1.38064852e-23$, $q_e=1.60217662e-19$, $m_p=1.6726219e-27$. 

\item Parameters for Ref.\cite{Umeda2012} benchmark: $B_0=96.24E-9$. $q_s=[-1,-1]$, $m_s=[5.447e-4,5.447e-4]$, $n_s=[1e5,9e5]$, $T_{zs}=T_{\perp s}=[5.1,5.1]$, $v_{dsz}/c=[0.1,0]$, $v_{dsr}/c=[0.05,0]$.

\item Parameters for Ref.\cite{Muschietti2017} benchmark:  $q_s=[1,1,-1]$, $m_s=[1,1,1.1111e-3]$, $n_s=[0.8,0.2,1.0]$, $T_{zs}=T_{\perp s}=[0.111,0.111,0.01]$, $v_{dsx}/c=[-8.333e-3,3.333e-2,0]$. $B_0=1$, $c=300$, $\mu_0=1.0$, $k_B=1.0$, $q_e=1.0$, $m_p=1.0$.

\end{itemize}

The purpose of the present work is to provide the foundation of this new tool. And thus, the applications to new examples would be discussed in other works.

\section{Summary and Discussion}\label{sec:discussion}

In summary, a powerful new kinetic dispersion relation tool is developed, which largely extend both the physical models and numerical capacilities of other works in literature.
The advantages of this new version of PASS-K tool is that it contents many new features (anisotropic temperature/loss cone/drift in arbitrary direction/ring beam/collision, unmagnetized/magnetized, electrostatic/electromagnetic/Darwin, etc) and can be widely applied. And compared to some other solvers, the $k_z=0$ ($\theta=\pi/2$) and $k_x=0$ ($\theta=0$) cases are not singular in PASS-K. What is more, the $k_z<0$ modes are also correctly treated. The most attractive feature is that it does not require initial guess for root finding and thus will not miss solutions. Thus, we think that this is a unified tool what exactly dreamed of by the plasma community. 

The limitation of the present PASS-K approach is that it can only used for cases when the 2D velocity integral are decoupled, i.e., usually required $f_{s0}(\bm v)=f_z(v'_\parallel)f_\perp(v'_\perp)$. Typically, the present PASS-K approach can not be used to oblique propagation $\kappa$-distribution function \cite{Summers1994, Astfalk2015} and relativistic \cite{Stix1992} or other arbitrary distribution\cite{Verscharen2018} cases. However, extensions are possible. For example, for $\kappa$-distribution, we can use the similar $J$-pole for corresponding $Z_\kappa$ function\footnote{The $\kappa$-distribution $Z_\kappa(\zeta)=\sum_{j=1}^{J}\frac{b_j}{(\zeta-c_j)^j}$, which is slightly different from the standard one  $Z(\zeta)=\sum_{j=1}^{J}\frac{b_j}{\zeta-c_j}$ and thus the corresponding transformation is also slightly different.} of parallel velocity integral and use Gaussian quadrature for perpendicular velocity integral, which will yield a 2D $J$-pole expansion and then can be transformed to a equivalent linear matrix eigenvalue problem, which is solvable though the corresponding matrix dimension could be much larger than the Maxwellian one, depends on how many nodes are used for perpendicular velocity integral. And thus, with 2D $J$-pole expansion, the PASS-K approach can also be used for other distributions and non-uniform magnetic field drift wave case such as in gyrokinetic model \cite{Xie2017}. The discussion of these extensions are out of the range of the present work. Besides the conventional iterative root finding approaches, it is still required to develop a new powerful algorithm, say, similar to PASS-K approach, to solve the more challenged relativistic kinetic plasma dispersion relation generally.

\section{Acknowledgments}

The author thanks the many suggestions, discussions and benchmarks from Richard Denton, Xin Tao, Jin-song Zhao, Zhong-wei Yang  (helped the benchmark of Fig.\ref{fig:passk_vs_Muschietti17_vdperp_theta=50}), Chao-jie Zhang (helped the benchmark of unmagnetized version), Can Huang, Wen-ya Li, Liang Wang, Kyungguk Min (provided the benchmark data in Fig.\ref{fig:benchmark_ring_beam}), Yang Li, and many others. This work, which is particular useful, say for examples to study the loss-cone and LHDI instabilities in magnetic confinement fusion and beam instabilities in inertial fusion, is supported by the compact fusion project in ENN group.

\appendix

\section{The PASS-K Solver}\label{sec:manu}
The update version of PASS-K/PDRK code includes more features, but still consists of two parts: the main program and the input data
file. The input file ``passk.in" has the follow structure (the case in Fig.\ref{fig:passk_vs_Muschietti17_vdperp_theta=50})
{{\scriptsize
\begin {verbatim}
qs(e)   ms(mp)      ns(m^-3)    Tzs(eV)    Tps(eV)    alphas   Deltas    vdsz/c    vdsx/c    vdsy/c    vdsr/c    nu_s    m_or_u(1/0)
1       1.0         0.8e0       0.111      0.111      1.0      1.0       0.0       -8.333e-3 0.0       0.0       0.0     0
1       1.0         0.2e0       0.111      0.111      1.0      1.0       0.0       3.333e-2  0.0       0.0       0.0     0
-1      1.1111e-3   1.0e0       0.010      0.010      1.0      1.0       0.0       0.0       0.0       0.0       0.0     1
\end {verbatim}
}}One can add the corresponding parameters for additional species. Here, 'qs(e)',   ' ms(mp) ',     'ns(m\^-3)',    'Tzs(eV) ',   'Tps(eV)  ',  'alphas',   'Deltas',    'vdsz/c',    'vdsx/c ',  ' vdsy/c',   ' vdsr/c ',   'nu\_s ',   'm\_or\_u(1/0) ', are the charge in electron charge unit $q_s/e$, mass in proton mass unit $m_s/m_p$, density $n_{s0}$, parallel temperature $T_{zs}$, perpendicular temperature $T_{\perp s}$, loss cone parameters $\alpha_s$ and $\Delta_s$, drift velocities $v_{dsz}/c$, $v_{dsx}/c$, $v_{dsy}/c$, ring beam velocity $v_{dsr}/c$, collision frequency $\nu_s$, and whether the species is magnetized, respectively.
Normalizations and other parameters can be set in PASS-K main program  (more information can be found at: http://hsxie.me/codes/pdrk/ or https://github.com/hsxie/pdrk/).

\section{Relation to the dispersion relation of drift instabilities in inhomogeneous plasma}
In conventional derivation \cite{Stix1992,Gary1993,Swanson2003} of drift instabilities in inhomogeneous magnetized plasma, the distribution function is assumed to depend on space inhomogeneous. We take the electrostatic density inhomogeneous case as in Chap. 4 of Ref.\cite{Gary1993} for example to compare with our model. The conventional derivation, i.e., Eqs.(4.2.1) and (4.2.2) in Ref.\cite{Gary1993}, which ignores all terms of order $\epsilon_n^2$ and valids only for $\epsilon_n\ll k$ and $\epsilon_n \rho_{cs}\ll1$, gives
\begin{eqnarray}
f_{s0}&=&\frac{1}{\sqrt{\pi} v_{ts}^3}\Big[1+\epsilon_n\Big(x+\frac{v_y-v_{gs}}{\omega_{cs}}\Big)\Big]e^{-\frac{v_z^2}{v_{ts^2}}}e^{-\frac{(v_y-v_{gs})^2+v_x^2}{v_{ts^2}}},~~\bm k=(0,k_y,k_z),\\\nonumber
D(\omega,\bm k)&=&1+\sum_{s}\frac{\omega_{ps}^2}{2k^2v_{zts}^2} \Big\{1+\sum_{n=-\infty}^{\infty}\Big[\frac{\omega-k_y(v_{ns}+v_{gs})}{|k_z|v_{ts}} + \frac{n\omega}{|k_z|v_{zts}}\frac{\epsilon_n}{k_y}\Big]Z(\zeta_{sn})\Gamma_n(\frac{a_s^2}{2})\Big\}\\\label{eq:dwdrgary}
&\simeq&1+\sum_{s}\frac{\omega_{ps}^2}{2k^2v_{zts}^2} \Big\{1+\sum_{n=-\infty}^{\infty}\frac{\omega-k_\perp(v_{ns}+v_{gs})}{|k_z|v_{ts}}Z(\zeta_{sn})\Gamma_n(\frac{a_s^2}{2})\Big\}=0,
\end{eqnarray}
with $v_{ns}=\frac{\epsilon_n v_{ts}^2}{\omega_{cs}}$, $v_{gs}=\frac{g}{\omega_{cs}}$ and $\zeta_{sn}=\frac{\omega-k_\perp v_{gs}-n\omega_{cs}}{|k_z|v_{ts}}$, and where we have ignored the $\frac{\epsilon_n}{k_y}\ll1$ term. Here, $\epsilon_n$ is parameter for space inhomogeneous and $g=|\bm g|$ is for force such as gravity. The corresponding dispersion relation in our model is
\begin{eqnarray}
f_{s0}&=&\frac{1}{\sqrt{\pi} v_{ts}^3}e^{-\frac{v_z^2}{v_{ts^2}}}e^{-\frac{(v_x-v_{dsx})^2+v_y^2}{v_{ts^2}}},~~\bm k=(k_x,0,k_z),\\\nonumber
D(\omega,\bm k)&=&1+\sum_{s}\frac{\omega_{ps}^2}{2k^2v_{zts}^2}\sum_{n=-\infty}^{\infty} \Big[Z_1(\zeta_{sn}) + \frac{n\omega_{cs} }{k_zv_{zts}}Z_0(\zeta_{sn})\Big]\Gamma_n(\frac{a_s^2}{2})\\\label{eq:dwdrpassk}
&=&1+\sum_{s}\frac{\omega_{ps}^2}{2k^2v_{zts}^2}\Big\{1+\sum_{n=-\infty}^{\infty}\frac{\omega-k_\perp v_{dsx}}{|k_z|v_{ts}}Z(\zeta_{sn})\Gamma_n(\frac{a_s^2}{2})\Big\}=0,
\end{eqnarray}
with $\zeta_{sn}=\frac{\omega-k_\perp v_{dsx}-n\omega_{cs}}{|k_z|v_{ts}}$, and we have used $\sum_{n=-\infty}^{\infty}\Gamma_n=1$. If we consider the perpendicular drift in our model is due to space inhomogeneous $-\frac{\nabla p}{n_{s0}}$ and gravity $m\bm g$, with $p=n_{s0}(y)k_BT_{s0}$, $n_{s0}(y)=n_{s0}(1-\epsilon_ny)$ and $\bm g=g{\hat y}$, by using force balance and $\bm F_s=m_s\bm{g}-\frac{\nabla (n_{s0}k_BT_{s0})}{n_{s0}}=(m_sg+\epsilon_n m_s v_{ts}^2){\bm\hat{y}}$, we have $v_{dsx}=\frac{F_s}{q_sB_0}=v_{gs}+\frac{\epsilon_n v_{ts}^2}{\omega_{cs}}=v_{gs}+v_{ns}$. We noticed that Eqs.(\ref{eq:dwdrgary}) and (\ref{eq:dwdrpassk}) are exactly the same, except that in the conventional derivation (\ref{eq:dwdrgary}) the density inhomogeneous drift velocity $v_{ns}$ in not included in the argument of $Z$ function $\zeta_{sn}$.

That is to say, our model can be used to study the drift modes in inhomogeneous plasma although we assume the derivation is under homogeneous plasma assumption. It is also not easy to say which of the the two derivations, i.e., the conventional one and the present one, is more accurate to describe the drift modes in inhomogeneous plasma, because that in our derivation the inhomogeneity contributes to the drift velocity and all other steps are rigorous, whereas the conventional derivation ignores high order terms in more than one steps to obtain the final dispersion relation. 


\section{More $J$-pole coefficients of $Z(\zeta)$ function}
Ref.\cite{Xie2016} has provided the typical $J$-pole coefficients of $Z(\zeta)$ function with $J=4,8,12$. Here, following Ref.\cite{Xie2016}, we update them to more accurate data and even beyond the double  precision (16 digit number) number with the help of symbolic computation. In the Table \ref{tab:NpoleZ}, $I$ means we keep $I$ equations for $\zeta\to0$, and $K$ means we keep $K$ equations for $\zeta\to \infty$, as described in Ref.\cite{Xie2016}.

\begin{table}
\begin{center}
\caption{\label{tab:NpoleZ} The more accurate coefficients  $c_j$ and $b_j$ for
$J=4$, $J=8$, $J=12$, $J=16$ and $J=24$ under $J$-pole Pade approximations of $Z(\zeta)$,
where the asterisk denotes complex conjugation, and $\delta_Z=|Z_J(\zeta)-Z(\zeta)|$ is the maximum error between $Z_J(\zeta)$ and $Z(\zeta)$ for $\zeta=x+iy$ with $x\in [0,50]$ and $y=-0.1$. }
{\scriptsize
\begin{tabular}{cccc} \hline\hline
  & $b_1$=- 1.0467968598346571444 - 2.1018525680357924343i & $c_1$=0.37861161238699661757 - 1.3509435854325440801i &\\
   $J=4$ & $b_2$=0.54679685983465714437 + 0.037196505239893094435i & $c_2$=1.2358876534356917085 - 1.2149821325576149719i & $\delta_Z=3\times10^{-3}$ \\
 $(I=5,K=3)$ & $b(3:4)$=$b^*(1:2)$ &  $c(3:4)$=$-c^*(1:2)$ & \\\hline
  & $b_1$=- 5.5833741816150427087 - 11.208550459628098648i & $c_1$=0.27393621805538084727 - 1.9417870375760945628i &  \\
  & $b_2$=- 0.7399178112200519477 - 0.83951828462027428396i & $c_2$=- 1.4652340919391423883 - 1.7896202996033145873i &  \\
   $J=8$ & $b_3$=- 0.017340112270400811857 - 0.04630643962629377424i & $c_3$=2.2376877251342932158 - 1.6259410241203623422i & $\delta_Z=6\times10^{-6}$ \\
  $(I=10,K=6)$  & $b_4$=5.8406321051054954683 - 0.95360275132203964347i & $c_4$=- 0.83925396636792203153 - 1.8919952115314256943i&  \\
  & $b(5:8)$=$b^*(1:4)$ &  $c(5:8)$=$-c^*(1:4)$ &  \\\hline
  & $b_1$=- 47.913598578418315281 - 106.98699311451399461i & $c_1$=0.22536708628380726987 - 2.4862558428460328565i  & \\
  & $b_2$=- 20.148858425809293248 + 12.874749056250453631i & $c_2$=1.1590491549279069691 - 2.4061921257040740764i & \\
  & $b_3$=- 4.5311004339957471789E-3 + 6.3311756354943215316E-4i & $c_3$=2.9785703941315209704 - 2.0490809954949754985i & \\
   $J=12$ & $b_4$=0.2150040123642351701 + 0.20042340981056393122i & $c_4$=2.2568587892309227294 - 2.2080229126485700572i & $\delta_Z=8\times10^{-9}$  \\
  $(I=16,K=8)$ & $b_5$=0.43131038679231352184 - 4.1505366661190555077i & $c_5$=1.6738373878120108271 - 2.3235155478934783777i  & \\
  & $b_6$=66.920673705505055584 + 20.747375125403268524i & $c_6$=0.68229440981712468 - 2.4598334422617114946i & \\
  & $b(7:12)$=$b^*(1:6)$ &  $c(7:12)$=$-c^*(1:6)$ & \\\hline
  & $b_1$=- 10.020983259474214017 - 14.728932929429874883i & $c_1$=0.22660012611958088508 - 2.0716877594897791206i  & \\
  & $b_2$=- 0.58878169153449514493 + 0.19067303610080007359i & $c_2$=- 1.7002921516300350075 - 1.882247422161272446i & \\
  & $b_3$=- 0.27475707659732384029 + 3.617920717493884482i & $c_3$=1.1713932508560117853 - 1.9772503319208541098i & \\
   $J=12$ & $b_4$=4.5713742777499515344E-4 + 2.7155393843737098852E-4i & $c_4$=3.0666201126826972102 - 1.5900208259325997176i & $\delta_Z=5\times10^{-8}$  \\
  $(I=12,K=12)$ & $b_5$=0.017940627032508378515 - 0.036436053276701248142i & $c_5$=2.3073274904105782764 - 1.7546732543728200654i & \\
  & $b_6$=10.366124263145749629 - 2.5069048649816145967i & $c_6$=0.68720052490601906567 - 2.0402885259758440187i & \\
  & $b(7:12)$=$b^*(1:6)$ &  $c(7:12)$=$-c^*(1:6)$ &   \\\hline
  & $b_1$=- 86.416592794839804566 - 147.57960545984972964i & $c_1$=0.19664397441136646085 - 2.5854046363167904821i  & \\
  & $b_2$=- 22.962540986214500398 + 46.211318219085729914i & $c_2$=1.0004276870893045112 - 2.5277610669350594581i & \\
  & $b_3$=- 8.8757833558787660662 - 11.561957978688249474i & $c_3$=1.4263380087098663429 - 2.4694803409658086505i & \\
   $J=16$ & $b_4$=- 0.025134802434111256483 + 0.19730442150379382482i & $c_4$=2.382753075769737514 - 2.2903917960623787648i & $\delta_Z=3\times10^{-11}$  \\
  $(I=18,K=14)$ & $b_5$=- 5.6462830661756538039E-3 - 2.7884991898011769583E-3i & $c_5$=2.9566517643704010427 - 2.1658992556376956217i & \\
  & $b_6$=2.8262945845046458372E-5 + 2.6335348714810255537E-5i & $c_6$=- 3.6699741330155866185 - 2.0087276133120462601i & \\
  & $b_7$=2.3290098166119338312 - 0.57238325918028725167i & $c_7$=1.8818356204685089975 - 2.3907395820644127768i & \\
  & $b_8$=115.45666014287557906 - 2.8617578808752183449i & $c_8$=0.59330036294742852232 - 2.5662607006180515205i & \\
  & $b(9:16)$=$b^*(1:8)$ &  $c(9:16)$=$-c^*(1:8)$ &     \\\hline
  & $b_1$=- 579.77656932346560644 - 844.01436313629880827i & $c_1$=0.16167711630587375808 - 2.9424665391729649011i  & \\
  & $b_2$=- 179.52530851977905732 - 86.660002027244731382i & $c_2$=1.1509135876493567245 - 2.874554296549015316i & \\
  & $b_3$=- 52.107235029274485215 + 453.3246806707749413i & $c_3$=0.81513635269214329287 - 2.9085569383176322447i& \\
  & $b_4$=- 2.1607927691932962178 + 0.63681255371973499384i & $c_4$=2.2362950589041724111 - 2.7033607074680388479i & \\
  & $b_5$=- 0.018283386874895507814 - 0.21941582055233427677i & $c_5$=2.6403561313404041541 - 2.6228400297078984517i & \\
   $J=24$ & $b_6$=- 6.819511737162705016E-5 + 3.2026091897256872621E-4i & $c_6$=3.5620497451197056658 - 2.4245607245823420556i& $\delta_Z<2\times10^{-13}$  \\
  $(I=24,K=24)$ & $b_7$=- 2.8986123310445793648E-6 - 9.9510625011385493369E-7i & $c_7$=4.1169251257106753931 - 2.3036541720854573609i & \\
  & $b_8$=2.3382228949223867744E-9 - 4.0404517369565098657E-9i & $c_8$=4.8034117493360317933 - 2.1592490859689535413i & \\
  & $b_9$=0.01221466589423530596 + 0.00097890737323377354166i & $c_9$=3.0778922349246567316 - 2.5301774598854448463i & \\
  & $b_{10}$=7.3718296773233126912 - 12.575687057120635407i & $c_{10}$=- 1.8572088635240765004 - 2.7720571884094886584i & \\
  & $b_{11}$=44.078424019374375065 - 46.322124026599601416i & $c_{11}$=1.496988132246689338 - 2.8290855580900544693i & \\
  & $b_{12}$=761.62579175738689742 + 185.11797721443392707i & $c_{12}$=- 0.48636891219330428093 - 2.9311741817223824196i & \\
  & $b(13:24)$=$b^*(1:12)$ &  $c(13:24)$=$-c^*(1:12)$ &  \\\hline\hline
\end{tabular}
}
\end{center}
\end{table}

\section{The Possible Numerical Inaccuracy}

Since PASS-K solves the dispersion relation use matrix eigenvalue approach, the computational time to obtain all solutions for a $N_N\times N_N$ dimensions matrix is $O(N_N^\alpha)$, with usually $\alpha\sim2.7$. Thus, the numerical $\sum_n$ with cut off to $n\in[-N,N]$ is crucial. For larger $N$, the matrix dimension could be large and the computation will slower. Usually $N\leq10$ is sufficient for many cases and the computation is very fast, which can yield all solutions for a typical single $k$ run in less than 1 second. For larger $N$, say $N>50$, the sparse matrix approach to search solutions around some initial guesses are still fast and thus the PASS-K tool has not limitation on this.

However, there exists several possible numerical inaccuracies, which should be careful of:
\begin{itemize}
\item (1) The $J$-pole expansion of $Z(\zeta)$ may give some artificial growing mode with growth rate $\gamma\sim10^{-6}$ for $J=8$, which however can be distinguished and reduced by using $J=12,16,24$, etc.
\item (2) The cut off in $\sum_{n=-\infty}^{\infty}\to\sum_{n=-N}^{N}$, which should be checked by using larger $N$ to make sure the results are convergent.
\item (3) Rand off error of eigen solver library, due to that the default numerical data is double precision (16 digit number) in the solver. This in principle can be solved by using high precision digit data.
\item (4) Others, such as singularity in Darwin matrix, degree of accuracy of functions $I_n$, $A_n$, $B_n$ and $C_n$, etc.
\end{itemize}
The above inaccuracies mainly occur at extremely small $k$ and large $k$. The third inaccuracy is not easy to distinguish at this moment, and however, we have not met this problem for most benchmarks.

\end{document}